{}

\documentclass[12pt,a4paper]{article}
\pdfoutput=1

\newif\ifpublic\publicfalse
\newif\ifniklas\niklastrue

\usepackage{ifpdf}
\ifniklas\ifpdf\publictrue\else\publicfalse
\PassOptionsToPackage{hypertex}{hyperref}
\PassOptionsToPackage{draft}{graphicx}
\fi\fi
\usepackage{amsmath,amssymb}
\usepackage[bookmarks=true,hyperfigures=true]{hyperref}
\usepackage{graphicx}
\usepackage[nosort]{cite}
\usepackage[bulletsep]{collref}
\usepackage{graphicx}
\usepackage[usenames,dvipsnames]{color}

\ifniklas

\def\showkeysrefformat#1{{\normalfont\tiny\ttfamily#1}}
\makeatletter
\def\SK@@ref#1>#2\SK@{%
 {\@inlabelfalse\leavevmode\vbox to\z@{%
 \vss\SK@refcolor\rlap{\vrule\raise .75em%
  \hbox{\showkeysrefformat{#2}}}}}}
\makeatother
\fi

\usepackage[a4paper,text={160mm,247mm},centering]{geometry}

\allowdisplaybreaks[3]

\numberwithin{equation}{section}

\usepackage[font=small,labelfont=bf,width=0.85\textwidth]{caption}

\expandafter\def\expandafter\bfseries\expandafter{\bfseries\ifmmode\else\boldmath\fi}
\expandafter\def\expandafter\mdseries\expandafter{\mdseries\ifmmode\else\unboldmath\fi}
\expandafter\def\expandafter\normalfont\expandafter{\normalfont\ifmmode\else\unboldmath\fi}

\RequirePackage{verbatim}

\makeatletter
\newwrite\bibinl@out
\newenvironment{bibtex}[1][\jobname]{%
  \immediate\openout\bibinl@out #1.bib
  \immediate\write\bibinl@out{\@percentchar generated from `\jobname' starting line \the\inputlineno^^J}%
  \def\verbatim@processline{\immediate\write\bibinl@out{\the\verbatim@line}}%
  \@bsphack\let\do\@makeother\dospecials\catcode`\^^M\active\verbatim@start
}%
{\immediate\closeout\bibinl@out\@esphack}
\makeatother

\RequirePackage{verbatim}

\makeatletter
\newwrite\mpi@out
\def\mpi@write#1{\immediate\write\mpi@out{#1}}
\immediate\openout\mpi@out\jobname.mp
\mpi@write{\@percentchar generated from `\jobname'}%
\AtEndDocument{\immediate\closeout\mpi@out}

\newcommand{\mpi@putlineno}{%
  \mpi@write{\@percentchar---------------------------------------}%
  \mpi@write{\@percentchar l.\the\inputlineno}%
}

\newcommand{\mpi@verbatim}{
  \@bsphack
  \let\do\@makeother\dospecials
  \catcode`\^^M\active
  \def\verbatim@processline{\mpi@write{\the\verbatim@line}}%
  \verbatim@start
}

{\mpi@write{}\@esphack}

{\mpi@write{endfig;}\@esphack}

\newcommand{\includegraphicsex}[2][]{%
  \xdef\mpi@tmp{#2}%
  \IfFileExists{\mpi@tmp}%
    {\includegraphics[#1]{\mpi@tmp}}%
    {\textbf{??}\typeout{file \mpi@tmp{} missing}}%
}

\makeatother

\ifx\genfrac\sdflkaj\else\fi
\newcommand{\sfrac}[2]{{\textstyle\frac{#1}{#2}}}
\newcommand{\half}{\sfrac{1}{2}}

\newcommand{\Complex}{\mathbb{C}}
\newcommand{\Integer}{\mathbb{Z}}

\newcommand{\hopf}[1]{\mathrm{#1}}
\newcommand{\yang}{\hopf{Y}}

\newcommand{\alg}[1]{\mathfrak{#1}}
\newcommand{\grp}[1]{\mathrm{#1}}
\newcommand{\gen}[1]{\mathfrak{#1}}
\newcommand{\genyang}[1]{\widehat{\gen{#1}}}
\newcommand{\rep}{\rho}
\newcommand{\copro}{\mathrm{\Delta}}

\newcommand{\fund}{\mathrm{F}}

\newcommand{\repfund}{\rep^\fund}

\newcommand{\bigbrk}[1]{\bigl(#1\bigr)}

\newcommand{\gcomm}[2]{[#1,#2\}}

\newcommand{\sbra}[1]{\langle #1|}

\newcommand{\cket}[1]{|#1]}

\newcommand{\sprods}[2]{\langle#1#2\rangle}

\newcommand{\cprods}[2]{[#1#2]}

\DeclareMathOperator{\End}{End}

\newcommand{\superN}{\mathcal{N}}

\newcommand{\nn}{\nonumber}
\newcommand{\nln}{\nonumber\\}

\def\[{\begin{equation}}
\def\]{\end{equation}}

\providecommand{\href}[2]{#2}

\makeatletter
\def\mr@ignsp#1 {\ifx\:#1\@empty\else #1\expandafter\mr@ignsp\fi}%
\newcommand{\multiref}[1]{\begingroup
\xdef\mr@no@sparg{\expandafter\mr@ignsp#1 \: }%
\def\mr@comma{}%
\@for\mr@refs:=\mr@no@sparg\do{\mr@comma\def\mr@comma{,}\ref{\mr@refs}}%
\endgroup}
\renewcommand{\eqref}[1]{(\multiref{#1})}
\makeatother

\makeatletter
\newcommand{\namedref}[2]{\hyperref[#2]{#1~\ref*{#2}}}
\newcommand{\secref}{\@ifstar{\namedref{Section}}{\namedref{sec.}}}
\newcommand{\subsecref}{\@ifstar{\namedref{Subsection}}{\namedref{subsec.}}}
\newcommand{\appref}{\@ifstar{\namedref{Appendix}}{\namedref{app.}}}
\newcommand{\tabref}{\@ifstar{\namedref{Table}}{\namedref{tab.}}}
\newcommand{\figref}{\@ifstar{\namedref{Figure}}{\namedref{fig.}}}
\makeatother

\providecommand{\hypersetup}[1]{}
\providecommand{\texorpdfstring}[2]{#1}

\hypersetup{plainpages=false}
\hypersetup{pdfpagemode=UseNone}
\hypersetup{bookmarksnumbered=true}
\hypersetup{pdfstartview=FitH}
\hypersetup{colorlinks=false}
\hypersetup{citebordercolor={.5 1 .5}}
\hypersetup{urlbordercolor={.5 1 1}}
\hypersetup{linkbordercolor={1 .7 .7}}

\makeatletter
\let\@keywords\@empty
\let\@subject\@empty
\providecommand{\keywords}[1]{\gdef\@keywords{#1}}
\providecommand{\subject}[1]{\gdef\@subject{#1}}
\def\thetitle{\@title}
\def\theauthor{\@author}
\def\thesubject{\@subject}
\def\thedate{\@date}
\def\thekeywords{\@keywords}
\makeatother
\AtBeginDocument{
\hypersetup{pdftitle={\thetitle}}%
\hypersetup{pdfauthor={\theauthor}}%
\hypersetup{pdfsubject={\thesubject}}%
\hypersetup{pdfkeywords={\thekeywords}}%
}

\makeatletter
\newsavebox{\apb@box}\newlength{\apb@width}
\newcommand{\autoparbox}[2][c]{\sbox{\apb@box}{#2}%
 \settowidth{\apb@width}{\usebox{\apb@box}}%
 \parbox[#1]{\apb@width}{\usebox{\apb@box}}}

\newcommand{\includegraphicsboxex}[2][]{\autoparbox{\includegraphicsex[#1]{#2}}}
\makeatother

\newif\ifmrnote 
\mrnotetrue

\newcommand{\mtw}{\mathcal{W}}
\usepackage[active]{srcltx}
\usepackage{color}

\newcommand{\rop}{\mathrm{R}}
\newcommand{\lop}{\mathrm{L}}
\newcommand{\dl}[1]{\delta_{#1}}
\newcommand{\dlb}[1]{\delta_{\text{-}#1}}
\newcommand{\msYM}{$\superN=4$ sYM }

\newif\ifjbnote 
\jbnotetrue

\newcommand{\emptyperm}{
  \begin{permutation}
    \cdot & \cdot & \cdots & \cdot\\
    \downarrow & \downarrow &&\downarrow\\
    \cdot & \cdot & \cdots & \cdot\\
  \end{permutation}
}
\newenvironment{permutation}{\Big(\begin{smallmatrix}}{\end{smallmatrix}\Big)}
\newcommand{\dblpar}[1]{(\!(#1)\!)}

\title{A dictionary between \texorpdfstring{$\rop$}{R}-operators, on-shell
graphs and Yangian algebras}
\author{Johannes Broedel, Marius de Leeuw and Matteo Rosso}

\begin{document}

\iftrue

\pdfbookmark[1]{Title Page}{title}
\thispagestyle{empty}


\vspace*{2cm}
\begin{center}%
\begingroup\Large\bfseries\thetitle\par\endgroup
\vspace{1cm}

\begingroup\scshape\theauthor\par\endgroup
\vspace{5mm}%

\begingroup\itshape
Institut f\"ur Theoretische Physik,\\
Eidgen\"ossische Technische Hochschule Z\"urich\\
Wolfgang-Pauli-Strasse 27, 8093 Z\"urich, Switzerland
\par\endgroup
\vspace{5mm}

\begingroup\ttfamily
\verb+{+jbroedel,deleeuwm,mrosso\verb+}+@itp.phys.ethz.ch
\par\endgroup

\vfill

\textbf{Abstract}\vspace{5mm}

\begin{minipage}{12.7cm}
  We translate between different formulations of Yangian invariants relevant
  for the computation of tree-level scattering amplitudes in $\superN=4$
  super-Yang--Mills theory.  While the $\rop$-operator formulation allows to
  relate scattering amplitudes to structures well known from integrability, it
  can equally well be connected to the permutations encoded by on-shell graphs. 
\end{minipage}

\vspace*{4cm}

\end{center}

\newpage

\fi

\tableofcontents

\section{Introduction and outline}
\label{sec:intro}

$\superN=4$ super-Yang--Mills (sYM) theory is the maximally supersymmetric
four-dimen\-sional gauge theory not including gravity
\cite{Brink:1976bc,Gliozzi:1976qd}. It is a gauge theory with gauge group
$\grp{SU}(N_c)$; the spectrum includes a single $\superN=4$ massless multiplet,
consisting of one gauge field, four Weyl fermions and six real scalars, all
transforming in the adjoint representation of the gauge group. One of the most
remarkable properties of this theory is that it is superconformally invariant
even at the quantum level
\cite{Howe:1983sr,Brink:1982pd,Brink:1982wv,Mandelstam1982} -- its symmetry
group effectively being $\grp{PSU}(2,2|4)$.

In recent years, the study of scattering amplitudes in $\superN=4$ sYM theory
unveiled a rich underlying structure. One of the most striking discoveries is
the hidden dual superconformal symmetry of tree-level amplitudes
\cite{Drummond:2008vq}; the closure of the ordinary superconformal algebra and
this dual superconformal symmetry is the Yangian algebra
$\yang[\alg{psu}(2,2|4)]$ \cite{Drummond:2009fd}. Although the tree-level
S-matrix enjoys this symmetry, Yangian invariance is broken at loop level due to
IR divergences.

An important tool to investigate the rich structure arising from amplitudes in
$\superN=4$ sYM theory is the so-called Grassmannian formalism
\cite{ArkaniHamed:2009dn,ArkaniHamed:2008gz,ArkaniHamed:2010kv,ArkaniHamed:2009vw,ArkaniHamed:2009dg}. While
the original formulation allowed to express leading singularities of amplitudes
in terms of contour integrals over a suitable Grassmannian manifold, it was
subsequently shown that it is possible to identify the correct contours leading
to the Britto--Cachazo--Feng--Witten (BCFW) decomposition of tree- and
loop-level amplitudes. The approach was later generalised to the study of
\emph{on-shell graphs} (or diagrams) \cite{ArkaniHamed:2012nw}, planar bicolored
graphs that correspond to Yangian invariants. These diagrams were studied and
generalised in refs.~\cite{Ferro:2012xw,Ferro:2013dga} (see also
\cite{Beisert:2014qba}), where it was shown that it is possible to deform the
external helicities to complex values while preserving Yangian invariance.

Recently, a new method for the study of Yangian invariants related to
scattering amplitudes was proposed in
refs.~\cite{Chicherin:2013ora,Chicherin:2013sqa}. The authors employ an
algebraic approach to construct Yangian invariants by defining a set of
operators acting on a suitable vacuum.  Their construction is manifestly
Yangian invariant at each step and is shown to yield the correct form of
(deformed) tree-level MHV scattering amplitudes. The authors argue that the
same approach can be used to construct all tree-level amplitudes in a
manifestly Yangian-invariant way by building single channels via the
inverse-soft-limit construction \cite{Nandan:2012rk}. A similar construction
arose in the context of the Bethe-ansatz approach in
ref.~\cite{Frassek:2013xza}.
\medskip

The aim of this paper is to study the algebraic approach for the construction
of Yangian invariants and tree-level scattering amplitudes and relate it to the
known formulations of Yangian invariants in $\superN=4$ sYM theory. In
\secref{sec:ampl} we review some of the properties of scattering amplitudes in
$\superN=4$ sYM theory, focusing mainly on the symmetries of the tree-level
S-matrix. An important part of the review concerns the Grassmannian formalism
for scattering amplitudes
\cite{ArkaniHamed:2009dn,ArkaniHamed:2008gz,ArkaniHamed:2010kv,ArkaniHamed:2009vw,ArkaniHamed:2009dg}
in terms of on-shell diagrams \cite{ArkaniHamed:2012nw}. One of the most
important results for the current article is the correspondence between the
Yangian-invariant leading singularities of amplitudes and \emph{decorated
permutations}. We will see that a similar combinatorial construction arises
also in the language of refs.~\cite{Chicherin:2013ora,Chicherin:2013sqa}.

In \secref{sec:algebra} we discuss the algebraic framework underlying the
construction of refs.~\cite{Chicherin:2013ora,Chicherin:2013sqa}. In these
papers the authors show how to construct invariants under the Yangian of
$\alg{gl}(4|4)$ via some suitable operators. These operators (called
$\rop$-operators) act on generalised functions defined on on-shell superspace
as
\begin{equation}
  \label{eq:r_act_1} 
  \rop_{ab}(u)
  f(\lambda_a,\tilde{\lambda}_a,\tilde{\eta}_a,\lambda_b,\tilde{\lambda}_b,\tilde{\eta}_b)
  := \int\frac{\mathrm{d}
    z}{z^{1+u}}f(\lambda_a-z\lambda_b,\tilde{\lambda}_a,\tilde{\eta}_a,\lambda_b,\tilde{\lambda}_b+
  z \tilde{\lambda}_a,\tilde{\eta}_b+z \tilde{\eta}_a) \ .
\end{equation}
The ``vacuum state'' used as starting point of the construction is a combination
of superconformally invariant delta functions. The operator $\rop_{a b}$ is
introduced as an intertwiner of representations of the Yangian algebra. It is
evident that its action has a clear ``physical'' interpretation: it performs a
BCFW shift. A function defined in this way on on-shell superspace is a Yangian
invariant if it is an eigenfunction of the monodromy matrix naturally
arising in this context.

We compute and analyse in detail the low-multiplicity Yangian invariants arising
from this construction. We show that it is possible to construct the single
channels of the BCFW decomposition of the six-point NMHV amplitude starting from
a single invariant, which we will subsequently show to be equivalent to the
top-cell on-shell diagram of ref.~\cite{ArkaniHamed:2012nw}. Analysing the
symmetries of the Yangian invariants in detail allows us to associate a
permutation to each of them in a natural way. 

\secref*{sec:onshell} relates the algebraic construction of Yangian invariants
with the on-shell diagram formalism. We will show that the three ways of
encoding a Yangian invariant discussed -- $\rop$-operator approach, associated
permutation, on-shell diagrams -- can be actually translated one into the other.
\begin{figure}[h]
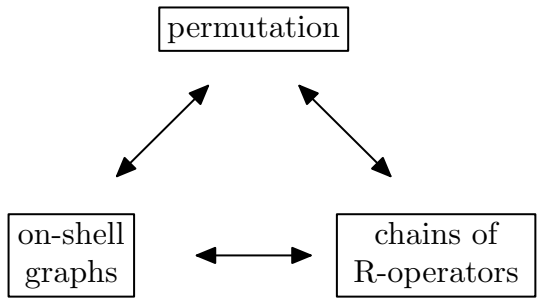

  \begin{center}
  \includegraphicsboxex[scale=1.5]{TripleDescription.mps}
  \end{center}
  \caption{Three equivalent ways of describing Yangian invariants.}
  \label{fig:tripledescription}
\end{figure}
After showing how to associate an on-shell diagram to a Yangian-invariant chain
of $\rop$-operators, we demonstrate how to associate a Yangian-invariant
$\rop$-chain to a permutation and vice versa, thus completing the three-way
correspondence. Parity and dihedral symmetries can be nicely interpreted in
terms of permutations. The section is concluded with the discussion of the six-
and seven-point NMHV Yangian invariants.  Finally, we summarise the results and
consider some possible future directions of inquiry in the concluding section.

\textbf{Note:} In the process of preparing this article for publication, we
learnt about the paper~\cite{Kanning:2014maa}, which shares some conclusions
with our present project. We thank the authors for providing us with a draft of
their paper.


\section{Superamplitudes in \texorpdfstring{$\superN=4$}{N=4} super-Yang--Mills
  theory}
\label{sec:ampl}
 
In this section we will review the main features of scattering amplitudes in
$\superN=4$ sYM theory and some of the tools available to compute them. We will
focus on some more recent developments in the study of the underlying symmetries
of the (planar) S-matrix.

For the study of scattering amplitudes in $\superN=4$ sYM it is convenient to
introduce the so-called on-shell superspace variables
$(\lambda^\alpha,\tilde{\lambda}_{\dot{\alpha}},\tilde{\eta}^A)$
\cite{Nair:1988bq}, which are the supersymmetric extension of the ordinary
spinor-helicity variables. Here, Greek and upper case Latin indices are indices
of the fundamental representation of $\grp{SL}(2)$ and $\grp{SU}(4)$,
respectively.

Since the $\superN=4$ multiplet is CPT self-conjugate, it can be expressed as a
single superfield defined on the on-shell superspace
\begin{equation}
  \label{eq:superfield}
  \Phi ( \lambda, \tilde{\lambda}, \tilde{\eta} )
  :=
  g^+ + \tilde{\eta}^A \psi_A + \,\frac{1}{2} \tilde{\eta}^A \tilde{\eta}^B \,\phi_{A B}
  + \frac{1}{3!} \epsilon_{A B C D} \,\tilde{\eta}^A \tilde{\eta}^B \tilde{\eta}^C \,\bar{\psi}^D
  + \frac{1}{4!} \epsilon_{A B C D}\, \tilde{\eta}^A \tilde{\eta}^B \tilde{\eta}^C \tilde{\eta}^D\, g^- \ .
\end{equation}
The colour-ordered tree-level scattering amplitudes of the full supermultiplet
can be expressed as functions on $n$ copies of the on-shell superspace as\footnote{%
We will refer to superamplitudes simply as ``amplitudes'' below for convenience.} %
\begin{equation}
  \label{eq:superampl_def}
  \mathcal{A}(\Phi_1,\dots,\Phi_n) := \mathcal{A}_{n;2} \, \mathcal{P}_n \ ,
\end{equation}
where
\begin{equation}
  \label{eq:mhv_npt}
  \mathcal{A}_{n;2} = \frac{\delta^4 \bigl( \sum_{i=1}^n \lambda_i
    \tilde{\lambda}_i \bigr) \delta^{0|8} \bigl( \sum_{i=1}^n \lambda_i
    \tilde{\eta}_i \bigr) }{\sprods{1}{2} \sprods{2}{3} \dots \sprods{n}{1} } 
\end{equation}
is the supersymmetric version of the MHV gluon scattering amplitude
\cite{Nair:1988bq}, and our conventions for spinor brackets are
\[
\begin{aligned}
 \sprods{i}{j}&=\lambda_i^{\alpha}\lambda_{j\alpha} \ ,&\qquad
 \cprods{i}{j}&=\tilde{\lambda}_{i\dot{\alpha}}\tilde{\lambda}_j^{\dot{\alpha}} \ ,\\
 \lambda_\alpha = \epsilon_{\alpha\beta}
 \lambda^\beta \ ,\qquad \lambda^\beta&=\epsilon^{\beta\gamma}\lambda_\gamma \ ,&\qquad
 \tilde{\lambda}_{\dot{\alpha}} &= \epsilon_{\dot{\alpha}\dot{\beta}}
 \tilde{\lambda}^{\dot{\beta}} \ ,\qquad\tilde{\lambda}^{\dot{\beta}}=
 \epsilon^{\dot{\beta}\dot{\gamma}}\tilde{\lambda}_{\dot{\gamma}}
\end{aligned}
\nn\]
with $\epsilon_{12}=\epsilon_{\dot{1}\dot{2}}=-1$.

The function $\mathcal{P}_n$ is the sum
\begin{equation}
  \label{eq:P_n}
  \mathcal{P}_n = \mathcal{P}_{n;0} + \mathcal{P}_{n;1} + \dots + \mathcal{P}_{n;n-4} \ ,
\end{equation}
where each ${\cal P}_{n;k-2}$ is a function of homogeneous Grassmann degree $4(k-2)$.
The quantity $k-2$ determines the $\mathrm{MHV}$ level of the amplitude. The
$n$-point $\mathrm{N}^{k-2}\mathrm{MHV}$ amplitude is the term
$\mathcal{A}_{n;k} := \mathcal{A}_{n;2} \mathcal{P}_{n;k-2}$; obviously,
$\mathcal{P}_{n;0}=1$.


\subsection{Symmetries of tree-level scattering amplitudes}
\label{subsec:symmetries}

Scattering amplitudes in \msYM are invariant under the action of the generators
of $\alg{psu}(2,2|4)$ (the representation of these generators on the space of
functions on on-shell superspace was derived in ref.~\cite{Witten:2003nn}). The
additional requirement of physicality external legs amounts to imposing the
invariance of the amplitude under the action of the central charge
\begin{equation}
  \label{eq:cc}
  C_i = \lambda_i^\alpha \frac{\partial}{\partial \lambda_i^\alpha} -
  \tilde{\lambda}_i^{\dot{\alpha}} \frac{\partial}{\partial \tilde{\lambda}_i^{\dot{\alpha}}}
  - \tilde{\eta}_i^A \frac{\partial}{\partial \tilde{\eta}_i^A} + 2\ ,
\end{equation}
where the index $i=1,\dots,n$ labels the external legs. We will henceforth
consider invariance under the centrally extended algebra $\alg{su}(2,2|4)$.

A striking property of tree-level amplitudes in \msYM is that, in addition to
the ordinary invariance under $\grp{PSU}(2,2|4)$, they are covariant under a
\emph{dual} superconformal symmetry, acting on the coordinates of the dual space
\cite{Drummond:2008vq}. In ref.~\cite{Drummond:2009fd} it was shown that the
closure of the realisations of these two copies of $\alg{psu}(2,2|4)$ is the
Yangian algebra $\yang[\alg{psu}(2,2|4)]$.

An $n$-point tree-level scattering amplitude can be expressed as a sum of
Yangian invariants.  A suitable way to compute such a decomposition is provided
by the supersymmetric version of the Britto--Cachazo--Feng--Witten (BCFW)
recursion relations
\cite{Britto:2004ap,Britto:2005fq,ArkaniHamed:2008gz,Brandhuber:2008pf}.  These
relations express a tree-level amplitude as a sum of terms constructed out of
lower-multiplicity on-shell amplitudes.  In these terms, the $n$-point
$\mathrm{N}^{k-2}\mathrm{MHV}$ superamplitude reads
\begin{equation}
  \label{eq:bcfw_super}
  \begin{aligned}
    \mathcal{A}_{n;k} (1,\dots,n) = \sum_{\substack{n_L + n_R = n+2 \\ k_L + k_R
        = k+1}} \int \mathrm{d}^4 P\, \mathrm{d}^4\tilde{\eta} \;
    \mathcal{A}_L \Bigl(
    \{p_1,\hat{\tilde{\eta}}_1\},\dots,\{p_{n_L-1},\tilde{\eta}_{n_L-1}\},\{p,\tilde{\eta}\}
    \Bigr)\times\,\\
    \times \frac{1}{P^2} \mathcal{A}_R \Bigl(
    \{-p,\tilde{\eta}\},\{p_{n_R+1},\tilde{\eta}_{n_R+1}\},\dots,\{p_n,\hat{\tilde{\eta}}_n\}
    \Bigr) \ ,
  \end{aligned}
\end{equation}
where the superamplitudes $\mathcal{A}_{L,R}$ include the delta functions,
$p = P + z_{P_L} \lambda_1 \tilde{\lambda}_n$ and $z_{P_L} = P_L^2
/\sbra{1}P_L\cket{n} $. In this sum each term is Yangian invariant.  %
\begin{figure}
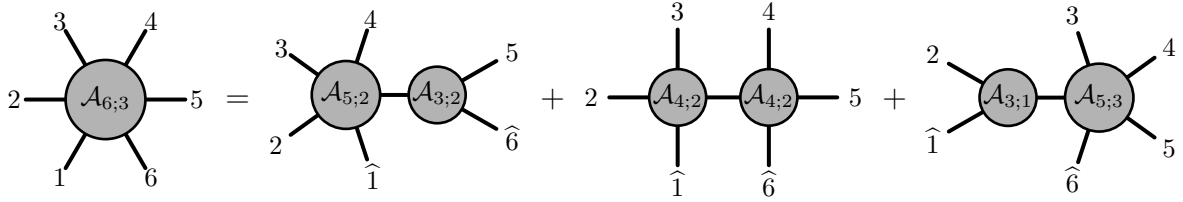

  \centering
  \includegraphicsboxex{Fig6pts_gen.mps}
  $\;=\;$
  \includegraphicsboxex{Fig6pts_ch2.mps}
  $\;+\;$
  \includegraphicsboxex{Fig6pts_ch1.mps}
  $\;+\;$
  \includegraphicsboxex{Fig6pts_ch3.mps}
  \caption{BCFW decomposition of the six-point NMHV amplitude.}
  \label{fig:6nmhvdec}
\end{figure}
%


\subsection{On-shell graphs and permutations}
\label{subsec:onshell}

In ref.~\cite{ArkaniHamed:2012nw} the authors introduced the formalism of
so-called on-shell diagrams (or on-shell graphs) to analyse the properties of
Yangian invariants and scattering amplitudes in \msYM\@. These diagrams are
constructed by gluing two basic trivalent vertices -- ``black'' and ``white''
vertices, which correspond to the three-point $\mathrm{MHV}$ and
$\overline{\mathrm{MHV}}$ amplitudes, respectively\footnote{%
  The gluing procedure amounts to the identification of legs shared by vertices
  and subsequent integration over the on-shell phase space of that internal leg.
}.
\begin{figure}
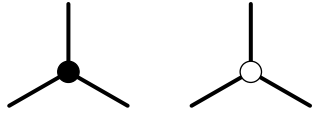

  \centering
  \includegraphicsboxex{Fig3verts.mps}
  \caption{Trivalent building blocks for on-shell graphs.}
  \label{fig:3verts}
\end{figure}
The authors show that these diagrams correspond to Yangian invariants; this
correspondence is encoded in the map from on-shell graphs to integrals over a
suitable Grassmannian manifold $G(k,n)$. This formalism is an extension of the
Grassmannian formulation for scattering amplitudes previously developed in
refs.~\cite{ArkaniHamed:2009dn,ArkaniHamed:2009vw,ArkaniHamed:2010kv,ArkaniHamed:2009dg}.
The pair $(k,n)$ consists of the $\mathrm{MHV}$ level $k$ and the multiplicity
$n$ of the amplitude the on-shell graph is related to; they are linked to
$n_\text{w},\,n_\text{b},\,n_\text{i}$ (the number of white vertices, black
vertices and internal lines of the graph, respectively) via
\begin{equation}
  \label{eq:nk}
  n=3 (n_\text{w}+n_\text{b})-2 n_\text{i} \ ,\qquad k =  n_\text{w} + 2 n_\text{b} - n_\text{i} \ .
\end{equation}

\begin{figure}
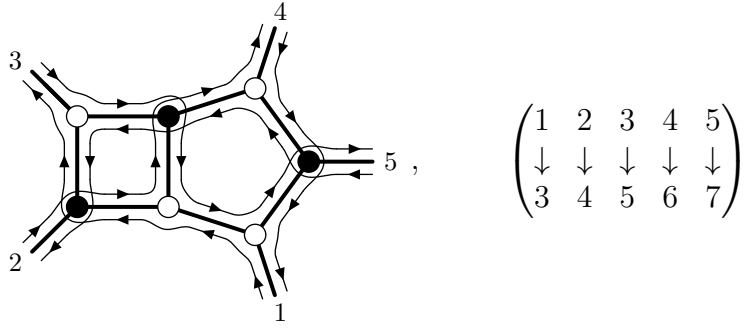

  \centering
  \includegraphicsboxex{Fig_5ptperm.mps}\ ,\hspace{1cm}
  $
  \begin{pmatrix}
    1 & 2 & 3 & 4 & 5\\
    \downarrow & \downarrow & \downarrow & \downarrow & \downarrow \\
    3 & 4 & 5 & 6 & 7 
  \end{pmatrix}
  $
  \caption{Double-line notation and decorated permutation for the five-point
    MHV on-shell diagram.}
  \label{fig:snpath}
\end{figure}

\begin{figure}
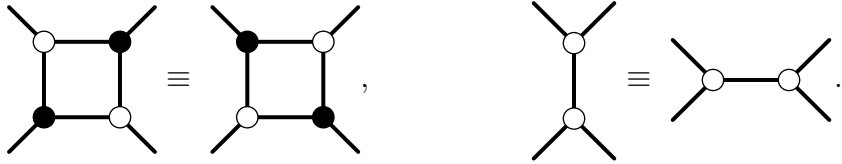

  \centering
  \includegraphicsboxex{Fig_sq1.mps}$\;\equiv\;$
  \includegraphicsboxex{Fig_sq2.mps},\hspace{2cm}
  \includegraphicsboxex{Fig_merg1.mps}$\;\equiv\;$
  \includegraphicsboxex{Fig_merg2.mps}.
  \caption{Square move and merger.}
  \label{fig:moves}
\end{figure}

One of the most important results in~\cite{ArkaniHamed:2012nw} is the
construction of a map between a subset of on-shell graphs (so-called
\emph{reduced}, related to tree-level amplitudes) and decorated
permutations\footnote{%
  To be precise, reduced on-shell graphs correspond to cells in the
  Grassmannian, and on-shell graphs related to tree-level amplitudes are always
  reduced.
}. %
A decorated permutation is an injective map
\begin{equation}
  \label{eq:perm}
  \sigma\; : \; \{1,\dots,n \} \; \; \to \; \; \{1,\dots,2n\}
\end{equation}
such that $i\leq \sigma(i)\leq i+n$ and $\sigma\; \mathrm{mod}\,n$ is an
ordinary permutation. The map is constructed starting from the on-shell graph as
follows: starting from the $i$-th leg, one follows the internal lines turning
\emph{right} at each black vertex and \emph{left} at each white vertex; the
external leg $j$ this path ends on yields $\sigma(i)$, with the
identification\footnote{%
  We will use the ``double line'' graphical notation to determine the
  permutation, as in \figref{fig:snpath}. Moreover, for self-identified legs,
  one should pay particular attention in choosing $\sigma(i)=i$ or
  $\sigma(i)=i+n$.  }
\begin{equation}
  \label{eq:decor}
  \sigma(i) = j\quad\text{if}\; j>i,\qquad \sigma(i) = j+n\quad\text{if}\;j<i \ .
\end{equation}
There are different on-shell graphs that correspond to the same decorated
permutation; however, all the on-shell graphs that correspond to a given
permutation can be mapped one into the other via two actions: merger and square
move, depicted in \figref{fig:moves}.
All diagrams that can be related via these two transformations are associated to
the \emph{same} Yangian invariant. The authors of \cite{ArkaniHamed:2012nw}
therefore conclude that there is a one-to-one map between Yangian invariants
(appearing in the tree-level amplitudes of \msYM) and decorated permutations or
that, equivalently, the invariant information of a (reduced) on-shell graph is
encoded in the associated permutation.

It is possible to derive the BCFW recursion relation in terms of on-shell
graphs. For a $(k,n)$ tree-level amplitude 
$\mathcal{A}_{n;k}$ (or $\mathcal{A}_n^{(k)}$), all BCFW channels can be obtained
starting from a single on-shell graph (the \emph{top-cell} graph\footnote{%
  It corresponds to the top-cell in the positive Grassmannian. }) %
corresponding to a permutation which is a cyclic shift by $k$. 

\begin{figure}[ht]
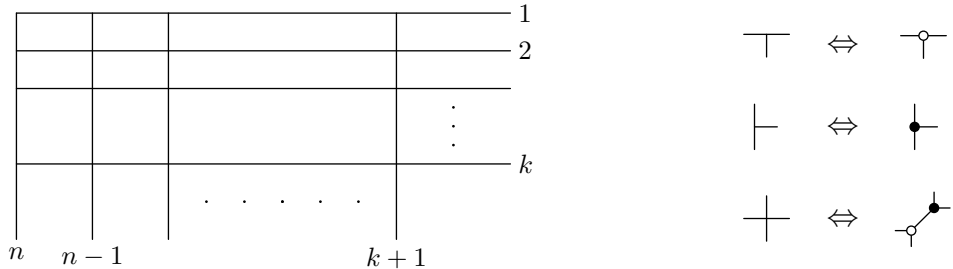

  \centering
  \includegraphicsboxex{Fig_tcrepr.mps}\hspace{.8cm}
  \begin{minipage}[r]{0.4\linewidth}
    \[
    \begin{array}{ccc}
      \includegraphicsboxex{Fig_plabicwv1.mps} &
      \;\Leftrightarrow\; &
      \includegraphicsboxex{Fig_plabicwv2.mps}\\
      &&\\
      \includegraphicsboxex{Fig_plabicbv1.mps} &
      \;\Leftrightarrow\; &
      \includegraphicsboxex{Fig_plabicbv2.mps}\\
      &&\\
      \includegraphicsboxex{Fig_plabicdv1.mps} &
      \;\Leftrightarrow\; &
      \includegraphicsboxex{Fig_plabicdv2.mps}
    \end{array}
    \nn \]
  \end{minipage}
  \caption{Construction of a representative on-shell graph for the top-cell of a
    $(k,n)$ amplitude.}
\label{fig:tcrepr}
\end{figure}

This fact allows to construct a representative on-shell graph for the top-cell
easily (as first shown in ref. \cite{Postnikov:2006kva} and reviewed in
ref. \cite{Ferro:2012xw}). It is done as follows: for the top-cell graph of a
$(k,n)$ amplitude (corresponding to the top-cell of the positive Grassmannian
$G_+ (k,n)$) draw $k$ horizontal lines, $(n-k)$ vertical lines so that the
leftmost and topmost are boundaries, then substitute the three-crossings and
four-crossings as in \figref{fig:tcrepr}.

The on-shell graphs corresponding to the BCFW channels are then obtained by
removing $(k-2)(n-k-2)$ edges from the top-cell graph. Note that not all edges
are removable, and the removable ones can be identified with a purely
combinatorial procedure.  The removal of an edge can be interpreted in terms of
the Grassmannian integral as the residue around a singularity of the integrand.

\begin{figure}
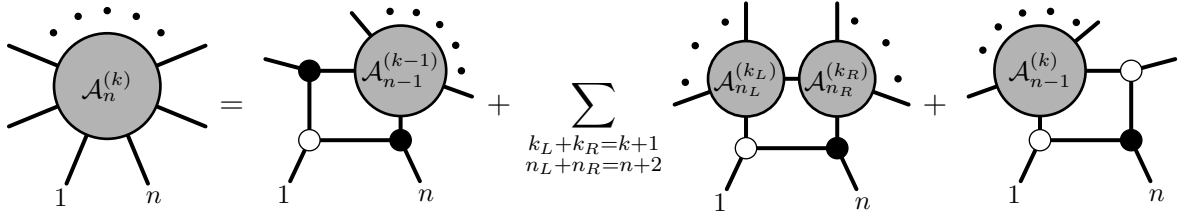

  \centering
  \includegraphicsboxex{Fig_bcfwOS1.mps}$\;=\;$
  \includegraphicsboxex{Fig_bcfwOS2.mps}$\displaystyle\;+\;\sum_{\substack{k_L+k_R=k+1\\ n_L+n_R=n+2}}$
  \includegraphicsboxex{Fig_bcfwOS3.mps}$\;+\;$
  \includegraphicsboxex{Fig_bcfwOS4.mps}
  \caption{Diagrammatic solution to the BCFW recursion relations in terms of on-shell graphs.}
  \label{fig:bcfw}
\end{figure}

The choice of which on-shell graphs obtained this way correspond to a BCFW
decomposition of the amplitude relies on the imposition of the correct unitarity
constraint and collinear limits; one of the recursive diagrammatic solution to
the BFCW recursion relations in terms of on-shell graphs is depicted in
\figref{fig:bcfw}.

\begin{figure}
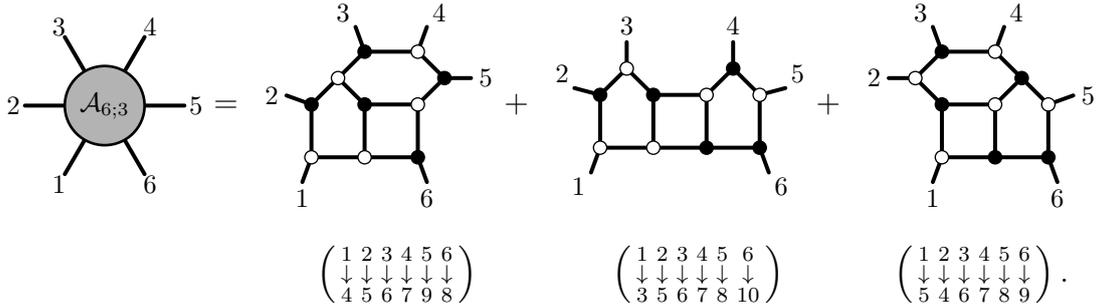

  \centering
  \[
  \begin{array}{cccc}
    \includegraphicsboxex{Fig6pts_gen.mps}$\;=$ &
    \includegraphicsboxex{Fig_6pt2.mps}$\;+$ &
    \includegraphicsboxex{Fig_6pt1.mps}$\;+$ &
    \includegraphicsboxex{Fig_6pt3.mps} \\
    &&&\\
    &
    \begin{permutation}
     1 & 2 & 3 & 4 & 5 & 6\\
     \downarrow & \downarrow & \downarrow & \downarrow & \downarrow & \downarrow\\
     4 & 5 & 6 & 7 & 9 & 8 
    \end{permutation}&
    \begin{permutation}
     1 & 2 & 3 & 4 & 5 & 6\\
     \downarrow & \downarrow & \downarrow & \downarrow & \downarrow & \downarrow\\
     3 & 5 & 6 & 7 & 8 & 10
    \end{permutation}&
    \begin{permutation}
     1 & 2 & 3 & 4 & 5 & 6\\
     \downarrow & \downarrow & \downarrow & \downarrow & \downarrow & \downarrow\\
     5 & 4 & 6 & 7 & 8 & 9 
    \end{permutation}\,.
  \end{array}
  \nn
  \]
  \caption{BCFW decomposition of the six-point NMHV amplitude, with the
    decorated permutations corresponding to the on-shell graphs.}
  \label{fig:6decomp}
\end{figure}

From the above discussion, we can infer that for MHV amplitudes there is one
single on-shell graph that corresponds to the amplitude -- the top-cell graph --
and the corresponding permutation is just a cyclic shift by two. The first
nontrivial example of a BCFW decomposition is the six-point NMHV amplitude,
which is obtained as a sum of three on-shell graphs corresponding to the
permutations reported in \figref{fig:6decomp}.
%


\subsection{Deformed on-shell graphs and Yangian invariants}
\label{sec:DefOnshell}

In refs.~\cite{Ferro:2012xw,Ferro:2013dga}, the authors studied a generalisation
of on-shell graphs (and related Yangian invariants). This generalisation relaxes
the condition of physicality of the external legs of the on-shell graph. Stated
explicitly, the deformed $n$-point Yangian invariant $\mathcal{Y}$ associated
with a given on-shell graph will satisfy\footnote{%
  We follow the sign conventions for central charges of
  ref.~\cite{Beisert:2014qba}.}
\begin{equation}
  \label{eq:ccY}
  \gen{C}_i\cdot\mathcal{Y}(1,\dots,n) = c_i \,\mathcal{Y}(1,\dots,n)
\end{equation}
with $c_i\neq 0$, in general\footnote{%
  Superconformal invariance still implies that $\sum_{i=1}^n c_i =0$.  }. %
A deformed on-shell graph can be built by gluing together two trivalent building
blocks, that correspond just to the deformation of the three-point
$\mathrm{MHV}$ and $\overline{\mathrm{MHV}}$ amplitudes. As this deformation is
switched on, one must pay attention to what happens to the Yangian generators:
in fact the representation that annihilates this deformed objects is the
evaluation representation with evaluation parameters $u_i$ different at each
site. The level-one generators annihilating an $n$-point invariant will then
read
\cite{MacKay:2004tc}
\begin{equation}
  \label{eq:lv1}
  \copro^{n-1} (\widehat{\gen{J}}^a) = \half f^a{}_{b\,c} \sum_{1\leq i < j \leq n} \,\gen{J}^b_i\,\gen{J}^c_j
  \,+ \sum_{i=1}^n (\half c_i -  u_i) \,\gen{J}^a_i \ .
\end{equation}

A deformed Yangian invariant would then naively depend on $2n$ additional
variables $(c_i,u_i)$, but this is not true, as studied in
ref.\cite{Beisert:2014qba}.  Particular attention must be paid for the gluing
procedure to preserve Yangian invariance. In order for this to be true, one must
match the parameters $u,c$ of the legs $a,b$ that are being glued as
\[
c_a = -c_b \ ,\qquad u_a - \half c_a = u_b - \half c_b \ .
\label{eq:matching}
\]
These gluing conditions impose a system of linear constraints on the parameters
$(u_i,c_i)$ attached to the legs of the diagram, and
these constraints imply that there are only $n$ independent ones. It is possible
to choose the $n$ independent parameters to be the $u$'s.  Yangian invariance
then implies the simple identification
\begin{equation}
  \label{eq:upmsys}
  c_i = u_i - u_{\sigma(i)} \ ,
\end{equation}
where $\sigma$ is the permutation associated with the graph. Note that in
\subsecref{subsec:summarysym} we will find the same condition arising from the
algebraic point of view. Therefore, a deformed on-shell graph can be associated
with a deformed Yangian invariant depending on $n$ sets of external data
$(\lambda_i,\tilde{\lambda}_i,\tilde{\eta}_i,u_i)$.

Notice that our present conventions differ from the ones used in
ref.~\cite{Beisert:2014qba}, where the iterated coproduct is defined as
\begin{equation}
  \label{eq:lv1_old}
  \copro^{n-1} (\widehat{\gen{J}}^a) =  f^a{}_{b\,c} \sum_{1\leq i < j \leq n} \,\gen{J}^b_i\,\gen{J}^c_j
  \,+ \sum_{i=1}^n w_i \,\gen{J}^a_i \ .
\end{equation}
without the factor of $\half$. The evaluation parameters
in the two different conventions ($w$'s in ref.~\cite{Beisert:2014qba}, $u$'s in
this article) are identified as
\begin{equation}
  \label{eq:eval_eq}
  w_i = - 2 u_i + c_i \ .
\end{equation}
If we solve for $u_i$ and substitute in~\eqref{eq:upmsys}, we find that
\begin{equation}
  \label{eq:pm_old}
  w_{\sigma(i)} - c_{\sigma(i)} = w_i + c_i \ ,
\end{equation}
which is the condition for Yangian invariance found in
ref.~\cite{Beisert:2014qba}.


\section{Algebraic approach}
\label{sec:algebra}

After we have introduced scattering amplitudes and on-shell diagrams we will
take a different point of view on amplitudes. In this section we will discuss
an algebraic approach to amplitudes along the lines of
refs.~\cite{Chicherin:2013ora} and \cite{Chicherin:2013sqa} to describe Yangian
invariants. 

The symmetry algebra that underlies this construction will be the Yangian of
$\alg{gl}(4|4)$ even though the amplitudes in $\superN=4$ sYM theory exhibit
$\yang[\alg{psu}(2,2|4)]$ symmetry. The reason for this is simple: we can extend
$\alg{psu}(2,2|4)$ to $\alg{su}(2,2|4)$ by adding the central element
$\gen{C}$. Moreover, although not being a symmetry of the amplitudes, the
hypercharge $\gen{B}$ can also be used as a further extension. Actually, the
hypercharge is a symmetry at the Yangian level \cite{Beisert:2011pn}.  At this
point we are effectively considering the algebra $\alg{u}(2,2|4)$, which is
equal to $\alg{gl}(4|4)$ for our purposes, since we ignore issues related to
reality conditions. We will follow ref.~\cite{Beisert:2014hya} closely.

\subsection{The Yangian algebra of \texorpdfstring{$\alg{gl}(M|N)$}{gl(M|N)}}
\label{sec:GLmn}

Consider the $\Integer_2$-graded vector space $\Complex^{M|N}$ spanned by $M$
even and $N$ odd basis vectors $E^A$ with indices $A,B,\ldots=1,\ldots, M+N$.
We define their $\Integer_2$-grading by
\[
|E^A|=|A|:=
\begin{cases}
  0 &\text{for }A\leq M \ ,\\
  1 &\text{for }A > M \ .
\end{cases}
\]
Correspondingly, we introduce a basis of canonical covectors $E_A$ of grading
$|E_A|=|A|$ as follows
\[
  E_A E^B := (-1)^{|A|}\delta^B{}_A \ .
\]
A basis for the endomorphisms $\End(\Complex^{M|N})$ is given by the matrices
$E^A{}_B$
\[\label{eq:matrixbasis}
E^A{}_B:=E^AE_B \ ,
\]
whose elements are defined to be zero except for a $(-1)^{|B|}$ in row $A$ and
column $B$.  These matrices obey the algebra
\[\label{eq:FundCommRel}
E^A{}_B E^C{}_D = (-1)^{|B|}\delta^C{}_B\,E^A{}_D \ ,
\qquad
|E^A{}_B| = |A|+|B| \ .
\]
For future reference, let us also introduce the supertransposition
\begin{align}\label{eq:supertranspose}
&(E^A{}_B)^t :=(-1)^{|A|(|B|+1)}E^B{}_A \ ,
\end{align}
such that applying the supertranspose four times is the identity
$(E^A{}_B)^{t,t,t,t} = E^A{}_B$.

The Lie superalgebra $\alg{gl}(M|N)$ is equivalent to $\End(\Complex^{M|N})$ as
a vector space. It is generated by generators
$\{\gen{J}^A{}_B\}_{A,B=1,\ldots,M+N}$ that satisfy the following commutation
relations
\begin{align}\label{eq:defGLmn}
[\gen{J}^A{}_B,\gen{J}^C{}_D\} = 
(-1)^{|B|}\delta^A_D\, \gen{J}^C{}_B - (-1)^{|B||C|+|B||D|+|C||D|} \delta^B_C\, \gen{J}^A{}_D \ ,
\end{align}
where $[A,B\} := AB-(-1)^{|A||B|}BA$ is the usual graded commutator. Let us
point out the central element $\gen{C}$ and the hypercharge $\gen{B}$ that extend
$\alg{psu}(2,2|4)$ to $\mathfrak{gl}(4|4)$ for $M=N=4$:
\begin{align}\label{eq:algCandB}
&\gen{C} = (-1)^{|A|}\gen{J}^A{}_A \ ,
&&&\gen{B} = \gen{J}^A{}_A \ .
\end{align}
In what follows, we will make use of two different representation of this
algebra, namely the fundamental and a functional (or oscillator)
representation. 

\paragraph{Representations.} 

The \emph{fundamental} representation $\repfund$ has dimension $M+N$ and the generators
$\gen{J}^A{}_B$ are simply represented by the matrices $E^A{}_B$ introduced
above
\[\label{eq:EinFundRep}
\repfund(\gen{J}^A{}_B):=E^A{}_B \ .
\]
The fact that this forms a representation is a direct consequence of
\eqref{eq:FundCommRel}. In this paper we will use the convention that algebra
generators are denoted by the fraktur font, while generators evaluated in an
explicit representation are represented by Roman letters. 

In order to introduce the \emph{functional} representation, we consider a set
of canonical conjugate (super)variables
$\mathbf{x}:=x^{A}$ and $\mathbf{p}:=p_{A}$, where $A= 1,\ldots, M+N$, such that
\begin{align}\label{eq:FCRxp} 
[x^A,p_{B}\} = - \delta^A{}_B \ .
\end{align}
These variables generate another representation $\rho^x$ of the $\alg{gl}(M|N)$
Lie superalgebra if we identify the generators as
\[\label{eq:XRep}
\rho^x(\gen{J}^A{}_B)=x^A p_B \ .
\]
The functional representation is infinite dimensional since $x$ and $p$ act on
the space of functions in the corresponding variables. Finally, let us
introduce the operators 
\begin{align}\label{eq:DefC}
&C := (-1)^{|A|} x^Ap_A \ ,
&&B := x^Ap_A \ ,
\end{align}
corresponding to the central element and hypercharge \eqref{eq:algCandB}. 

\paragraph{Yangian.}

The Yangian of a Lie (super)algebra is a Hopf algebra that is a deformation of
the universal enveloping algebra of the loop algebra. We will discuss the
Yangian of $\alg{gl}(M|N)$ in the Drinfeld realisation
\cite{Drinfeld:1985rx,Drinfeld:1986in}. 

By definition, the Yangian is generated by two
sets of generators $\gen{J}^A{}_B$ and $\genyang{J}^A{}_B$ that satisfy the
following commutation relations
\begin{align}
\label{eq:GLmnLieI}
\gcomm{\gen{J}^A{}_B}{\gen{J}^C{}_D} = (-1)^{|B|}\delta^C{}_B\, \gen{J}^A{}_D
-(-1)^{|B||C|+|B||D|+|C||D|}  \delta^A{}_D\, \gen{J}^C{}_B \ ,
\\
\label{eq:GLmnYangI}
\gcomm{\gen{J}^A{}_B}{\genyang{J}^C{}_D} = (-1)^{|B|}\delta^C{}_B\, \genyang{J}^A{}_D
- (-1)^{|B||C|+|B||D|+|C||D|} \delta^A{}_D\, \genyang{J}^C{}_D \ .
\end{align}
They must be supplemented by the Serre relations, which are the analogue of
the Jacobi identity in the Yangian. The coalgebra structure is determined by the
coproducts (\textit{cf.}~\eqref{eq:lv1})
\begin{align}
\label{eq:GLmnCoproI}
\copro(\gen{J}^A{}_B) &= \gen{J}^A{}_B\otimes 1 + 1 \otimes \gen{J}^A{}_B \ ,
\\
\copro(\genyang{J}^A{}_B) &= \genyang{J}^A{}_B\otimes 1 + 1 \otimes \genyang{J}^A{}_B
+ 
\half\bigbrk{\gen{J}^A{}_C\otimes \gen{J}^C{}_B
-(-1)^{(|A|+|C|)(|B|+|C|)}\gen{J}^C{}_B\otimes \gen{J}^A{}_C}
 \ . \nonumber
\end{align}
As usual, the Yangian can also be equipped with an opposite coproduct defined as
$P\circ\copro$ where $P$ is the (graded) permutation operator.

There is a special class of representations of Yangian algebras that is related
to representations of the underlying Lie algebra: the evaluation
representations. The evaluation representation $\rep_u$ corresponding to a
particular representation $\rep$ of the algebra $\alg{gl}(M|N)$ is defined as
\begin{align}
&\rep_u(\gen{J}^A{}_B) := \rep(\gen{J}^A{}_B) \ ,
&&\rep_u(\genyang{J}^A{}_B) := u\, \rep(\gen{J}^A{}_B) \ ,
\end{align} 
where the complex parameter $u$ is called the spectral parameter. We will be
particularly interested in the evaluation representations corresponding to the
fundamental representation $\repfund_u$ and the functional representation
$\rep^x_u$.

\paragraph{Quasi-triangular Hopf algebras.}

On the level of representations, Yangian algebras behave like quasi-triangular
Hopf algebras%
\footnote{To really achieve quasi-triangularity of the Yangian algebra, it has
  to be extended to a so-called Yangian double \cite{Khoroshkin:1994uk}.}. 
Quasi-triangular Hopf algebras constitute a special class of Hopf algebras for
which the coproduct and opposite coproduct are related by a similarity
transformation. Let $\hopf{H}$ be a quasi-triangular Hopf algebra, then there
is an invertible element $\mathcal{S}\in \hopf{H}\otimes \hopf{H}$, which is
called the universal R-matrix. It intertwines between the coproduct and the
opposite coproduct in the following way
\[\label{eq:intertwine}
\Delta^{op}(X) \mathcal{S}=\mathcal{S} \Delta(X) \ .
\]
In the remainder of this paper we shall use the term `R-matrix' for some other
object, and henceforth we will refer to the universal R-matrix $\mathcal{S}$ as
the `S-matrix'. The S-matrix must obey the so-called fusion relations
\[\label{eq:fusion}
\Delta_1(\mathcal{S}) = \mathcal{S}_{13}\mathcal{S}_{23} \ , \qquad
\Delta_2(\mathcal{S}) = \mathcal{S}_{13}\mathcal{S}_{12} \ .
\]
The two axioms above directly imply the Yang--Baxter equation
\begin{align}\label{eq:YBE}
  \mathcal{S}_{12}\mathcal{S}_{13}\mathcal{S}_{23} &=
  \mathcal{S}_{23}\mathcal{S}_{13}\mathcal{S}_{12} \ .
\end{align}
which is of central importance within integrable systems.

Furthermore, let us introduce the R-matrix as the S-matrix combined with the
permutation operator
\begin{align}
  \mathcal{R}_{12} := \mathcal{S}_{12} P_{12} \ .
\end{align}
The R-matrix is clearly equivalent to the S-matrix, but satisfies a permuted
Yang--Baxter equation
\begin{align}
  \mathcal{R}_{12}\mathcal{R}_{23}\mathcal{R}_{12} &=
  \mathcal{R}_{23}\mathcal{R}_{12}\mathcal{R}_{23} \ .
\end{align}
Even though our Yangian algebra is not quasi-triangular, on the level of
representations it behaves as if it were, thus on that level
there exists S-matrices satisfying the above properties \eqref{eq:intertwine}
and \eqref{eq:YBE}.

\paragraph{RLL-realisation.} An alternative realisation of a Yangian algebra
inspired by quasi-triangular algebras is given by the so-called RLL realisation
\cite{Drinfeld:1986in,Takhtajan:1979iv,Kulish:1980ii,Faddeev:1987ih}
\footnote{Also called RTT-realisation since sometimes the notation
  $\mathcal{T}$ is used rather than $\mathcal{L}$.}. 
The starting point for this realisation is the
Yang--Baxter equation~\eqref{eq:YBE} together with an explicit evaluation
representation $\rep_u$ (usually corresponding to the fundamental
representation).

The idea is to evaluate \eqref{eq:YBE} partially in this distinguished
representation. Defining
\begin{align}
&\mathcal{S}^\rho(u,v) := (\rho_u\otimes\rho_v) (\mathcal{S}) \ ,
&&\mathcal{L}^\rho(u) := (\rho_u\otimes 1) (\mathcal{L}) \ ,
\end{align}
and evaluating \eqref{eq:YBE} in the representation $\rho_u\otimes\rho_v\otimes
1$ yields
\begin{align}
\mathcal{S}^{\rho}_{12}(u,v)\mathcal{L}^{\rho}_1(u)\mathcal{L}^{\rho}_2(v) = 
\mathcal{L}^{\rho}_2(v)\mathcal{L}^{\rho}_1(u)\mathcal{S}^{\rho}_{12}(u,v) \ .
\end{align}
Since one leg of $\mathcal{L}^{\rho}$ lives in the Yangian algebra, the above
relation can be used as a defining relation for the algebra generated by the
elements of $\mathcal{L}^{\rho}$. The Hopf algebra structure follows from
eqn.~\eqref{eq:fusion}
\begin{align}\label{eq:coprodRLL}
  \Delta_2\mathcal{L}^{\rho}_{12}(u) =
  \mathcal{L}^{\rho}_{13}(u)\mathcal{L}^{\rho}_{12}(u) \ .
\end{align}
The Yangian generators in the Drinfeld realisation are encoded in
$\mathcal{L}^\rho(u)$ and can be extracted by expanding around $u=\infty$. For
instance, for $\alg{gl}(M|N)$ in the fundamental representation this relation
looks like
\begin{align}\label{eq:RLLandDI}
  \mathcal{L}^{\repfund}(u) = \exp\left[ (-1)^{|B|}E^B{}_A\, \gen{J}^A{}_B
    u^{-1} + (-1)^{|B|}E^B{}_A\, \genyang{J}^A{}_B u^{-2} + \ldots \right] \ .
\end{align}
This realisation offers a compact formulation of the Yangian algebra as the
object $\mathcal{L}^{\repfund}(u)$ encompasses all the Yangian levels.

\paragraph{S-matrices and R-matrices}

As mentioned above, Yangian algebras behave as quasi-triangular Hopf algebras on
the level of representations. In other words, there should be S-matrices
satisfying the Yang--Baxter eqn.~\eqref{eq:YBE} and the symmetry property
\eqref{eq:intertwine} in the representations that we introduced in the beginning
of this section. All the S- and R-matrices that we will introduce in this
section are summarised in Table~\ref{tab:SRmatrices}.

\begin{table}
  \begin{center}
    \begin{tabular}{|c|c|}
      \hline
      Representation & S-matrix\\
      \hline
      $\repfund_u\otimes\repfund_v$ & $\mathrm{S}(u-v)$\\
      \hline
      $\rep^x_u\otimes\repfund_v$ & $\lop(u-v)$\\
      \hline
      $\repfund_u\otimes\rep^x_v$ & $\lop(v-u)$\\
      \hline
      $\rep^x_{u+c/2}\otimes\rep^x_{v+c/2}$ & $\rop(u-v)P$\\
      \hline
    \end{tabular}
    \caption{S-matrices intertwining the coproduct and the opposite coproduct
      in different representations. $P$ is the graded permutation operator and
      $c$ is the value of the central operator $C$. It turns out to be
      convenient to consider the spectral parameters shifted by the central
      element for $\rop$. The explicit formulas are provided in
      eqns.~\protect\eqref{eq:Sfund,eq:Ldef} and \protect\eqref{eq:Rdef}. }
      \label{tab:SRmatrices}
  \end{center}
\end{table}

First, let us consider the fundamental S-matrix $\mathrm{S}$ that corresponds to
the intertwining operator in the tensor product of two fundamental
representations $\repfund_u\otimes\repfund_v$. It is easy to show that the usual
rational S-matrix
\begin{align}\label{eq:Sfund}
  \mathrm{S}(u) := u\,(-1)^{|A|+|B|} E^A{}_A\otimes E^B{}_B +
  (-1)^{|A|}E^A{}_B\otimes E^B{}_A
\end{align}
satisfies all the required relations, that is, it obeys the Yang--Baxter
equation and it intertwines the coproduct and opposite coproduct in the
fundamental representation.

More interesting is the mixed representation $\rep^x_v\otimes\repfund_u\simeq
\repfund_u\otimes\rep^x_v$. Consider the operator
\begin{align}\label{eq:Ldef}
  &\lop(u) := (u+\half(1-C)) \, (-1)^{|A|}E^A{}_A + (-1)^{|A|} x^A p_B\,
  E^B{}_A \ .
\end{align}
It can be shown that this operator intertwines the coproduct and the opposite
coproduct generators in the $\rep^x_v\otimes\repfund_u\simeq
\repfund_u\otimes\rep^x_v$ representation according to \eqref{eq:intertwine}.

In order to finish the discussion we would need the S-matrix in the functional
representation. However, for the remainder of this paper it turns out to be
more convenient to work with the R-matrix. It will satisfy the required
symmetry properties if we shift its arguments with the central charges of the
representations involved. This is exactly the type of twist that appears when a
Yangian is extended by a central element
\cite{Khoroshkin:1996fy,Khoroshkin:1996fz}. We denote the R-matrix in this
representation by $\rop$ and since it is an operator rather than a matrix we
will refer to it as the $\rop$-operator. 

The $\rop$-operator reads
\footnote{Notice that we have a different sign of $u$ compared to
  \cite{Chicherin:2013ora}.}
\begin{align}\label{eq:Rdef}
  \rop_{ab}(u) := \Gamma(-u)\, (\mathbf{p_a}\cdot \mathbf{x_b})^{u} = \int
  \frac{\mathrm{d} z}{z^{1+u}}e^{- z(\mathbf{p_a}\cdot \mathbf{x_b})} \ .
\end{align}
Since the symmetry properties of $\rop$ are non-trivial, let us spell
out the analogue of \eqref{eq:intertwine} which is of satisfied. It is
straightforward to show that
\begin{align}\label{eq:SymmRop}
  \left[\left(\rep^x\otimes\rep^x\right)\Delta^{op}(\gen{J}^A{}_B)\right]\rop_{12}(u_{12})
  &=
  \rop_{12}(u_{12})\left[\left(\rep^x\otimes\rep^x\right)\Delta^{op}(\gen{J}^A{}_B)\right] \ ,\\
  \left[\left(\rep^x_{u_2 + \sfrac{c_1}{2}}\otimes\rep^x_{u_1 +
        \sfrac{c_2}{2}}\right)\Delta^{op}(\genyang{J}^A{}_B)\right]\rop_{12}(u_{12})
  &= \rop_{12}(u_{12})\left[\left(\rep^x_{u_1 +
        \sfrac{c_1}{2}}\otimes\rep^x_{u_2 +
        \sfrac{c_2}{2}}\right)\Delta^{op}(\genyang{J}^A{}_B)\right] \ ,\nonumber
\end{align}
where the coproducts are evaluated in the tensor product of two functional
evaluation representations with the indicated parameters. Explicitly, from
eqn.~\eqref{eq:GLmnCoproI} we obtain
\begin{align}
  \left(\rep^x_{u_1 + \sfrac{c_1}{2}}\otimes\rep^x_{u_2 +
      \sfrac{c_2}{2}}\right)\Delta^{op}(\genyang{J}^A{}_B) = (u_1 +
  \sfrac{c_1}{2})(x_1)^A(p_1)_B + (u_2 + \sfrac{c_2}{2})(x_2)^A(p_2)_B - \nln
  -\half\left[(x_1)^A(p_1)_C(x_2)^C(p_2)_B -(-1)^{(|A|+|C|)(|B|+|C|)}
    (x_1)^C(p_1)_B(x_2)^A(p_2)_C \right] \ .
\end{align}
The permutation in the $u$'s in the second line of eqn.~\eqref{eq:SymmRop} is
due to the fact that we are working with R-operator rather than the S-matrix.

What remains to be shown are the different versions of the Yang--Baxter
equation that all these objects should satisfy. They are obtained by evaluating
\eqref{eq:YBE} in different representations. For simplicity, let us introduce
the short-hand notation $u_{ij}:=u_i-u_j$. 

On the one hand, there are two Yang--Baxter equations that do not involve the
R-operator.  While the first one is purely in the fundamental representation,
the second one has the third leg in the functional representation
\begin{align}
  \mathrm{S}_{12}(u_{12})\mathrm{S}_{13}(u_{13})\mathrm{S}_{23}(u_{23}) &=
  \mathrm{S}_{23}(u_{23})\mathrm{S}_{13}(u_{13})\mathrm{S}_{12}(u_{12}),\\
  \mathrm{S}_{12}(u_{12})\lop_{1}(u_{13})\lop_{2}(u_{23}) &=
  \lop_{2}(u_{23})\lop_{1}(u_{13})\mathrm{S}_{12}(u_{12}) \ .
\end{align}
On the other hand, there are two Yang--Baxter relations that involve the
$\rop$-operator. Consequently, those are permuted
\begin{align}
  \rop_{21}(u_1 - u_2)\lop_1(u_1 + \half C_1)\lop_2(u_2 + \half C_2) &= \lop_1(u_2 + \half
  C_1)\lop_2(u_1 + \half C_2) \rop_{21}(u_1-u_2) \ ,
  \label{eq:RLYBE}\\
  \rop_{12}(u_{12})\rop_{23}(u_{13})\rop_{12}(u_{23}) &=
  \rop_{23}(u_{23})\rop_{12}(u_{13})\rop_{23}(u_{12}) \ .\label{eq:RRYBE}
\end{align}
Remarkably, all these different instances of the Yang--Baxter equation can be
shown to hold directly from the relevant commutation relations
\eqref{eq:FundCommRel} and \eqref{eq:XRep}. The last equation
\eqref{eq:RRYBE} can be verified using the integral representation in
\eqref{eq:Rdef}.

\paragraph{Relations.}

We finish this section by deriving a few useful relations between the operators
$\rop$, $\lop$ and $\mathrm{S}$ introduced above.

To begin, let us look at some commutation relations. In particular, we can
observe that the operators $\rop$ and $\lop$ clearly commute if their indices
do not coincide. Similarly, it follows for the $\rop$-operators that
\begin{align}\label{eq:Rpermute}
  &\rop_{ab}(u)\rop_{cd}(v) = \rop_{cd}(v)\rop_{ab}(u) \ , &&~ \mathrm{if}~ a\neq d
  ~\mathrm{and}~ b \neq c \ .
\end{align}
Furthermore, the central element \eqref{eq:DefC} commutes with the Lax operator
by definition
\begin{align} 
  [\lop_a(u),C_a]=0 \ ,
\end{align}
but has a non-trivial commutation relation with the $\rop$-operator
\begin{align}\label{eq:RcComm}
  &[C_a,\rop_{ab}(u)] = -u\, \rop_{ab}(u), &&[C_b,\rop_{ab}(u)] = u\,
  \rop_{ab}(u) \ .
\end{align}
Notice, however, that as required by symmetry, $\rop_{ab}$ commutes with $C_a+C_b$.

In what follows, we will employ a representation of $\alg{gl}(M|N)$ in which
$\mathbf{x}$ and $\mathbf{p}$ are identified with the spinor-helicity variables
introduced in \secref{sec:ampl}. In such a representation the element $C$ is
central and its value $c$ characterises the representation. We can use $C$ to
invert the Lax operator at the level of representations. Indeed \eqref{eq:FCRxp}
yields (if L-operators act in the same space we suppress the indices)
\begin{align}\label{eq:Linverse}
  \lop(u+\sfrac{\alpha}{2}\, C) \lop(-u-\sfrac{\alpha}{2}\, C) =
  \lop(-u-\sfrac{\alpha}{2}\, C) \lop(u+\sfrac{\alpha}{2}\, C) =
  [\half+\sfrac{1+\alpha}{2}C+u][\half+\sfrac{1-\alpha}{2}C-u] \ ,
\end{align}
for any complex parameter $\alpha$. By using \eqref{eq:Linverse} on spaces $1$
and $2$ in \eqref{eq:RLYBE} we derive
\begin{align}
  &\rop_{12}(u_{12})\lop_{1}(u_{1}-\half C_1)\lop_{2}(u_{2}-\half C_2) =
  \lop_{1}(u_{2}-\half C_1)\lop_{2}(u_{1}-\half C_2)\rop_{12}(u_{12}) \ ,
  \label{eq:RLLinverse}
\end{align}

Moreover, we can define the transpose of $\lop$ by acting with the
supertranspose \eqref{eq:supertranspose} in the fundamental representation
\begin{align}\label{eq:Ltranspose}
  &\lop^t(u) = u (-1)^{|A|} E^A{}_A + (-1)^{|A||B|}x^A p_B E^A{}_B \ ,
\end{align}
It is easy to show that the inversion \eqref{eq:Linverse} and transposition
almost commute. In particular, we find
\begin{align}
  \lop^t(u+\sfrac{\alpha}{2}\,C) \lop^t(-u-\sfrac{\alpha}{2}\,C) =
  [\sfrac{1+\alpha}{2}C+u-\half][\sfrac{1-\alpha}{2}C-u-\half] \ ,
\end{align}
such that (we identify $C$ with $c$ on the right-hand side)
\begin{align}\label{eq:Ltransposeinverse}
  \lop^{t,-1}(u+\sfrac{\alpha}{2}\, C) =
  \frac{[\sfrac{1+\alpha}{2}c+u-\half][\sfrac{1-\alpha}{2}c-u-\half]}{[\sfrac{1+\alpha}{2}c+u+\half][\sfrac{1-\alpha}{2}c-u+\half]}
  \lop^{-1,t}(u+\sfrac{\alpha}{2}\, C) \ .
\end{align}

Finally, let us consider how the Yangian symmetry is generated in the functional
representation by the Lax-operator $\lop$. Similar to the RLL realisation, we
should be able to derive the Yangian symmetry generators in the functional
representation from $\lop$. Explicitly, the generators in the fundamental
representation can be obtained by expanding $\lop$ according to
\eqref{eq:RLLandDI}. We consider $\lop$ in a fundamental evaluation
representation with spectral parameter $u_0$ and in a the functional
representation with spectral parameter $u$.

Consequently we need to expand around $u_0=\infty$. Notice that the
normalisation of $\lop$ clearly plays a role in the explicit expansion. If we pick
a convenient normalisation $\lop\rightarrow L_0\lop$ and write\footnote{A
  different choice of normalisation simply rescales the generators and acts as
  the usual Yangian automorphism $\genyang{J}\rightarrow \genyang{J} + \sigma
  \gen{J}$.}
\begin{align}
  & L_0 = \frac{1}{u_0}+\frac{u-\half}{u_0^2}+\frac{\sfrac{c}{2}+(u-\half)^2}{u_0^2}+\ldots &\nln
  &\lop(u_0-u+\half c) =
  \sum_{n=-1}^\infty (-1)^{|A|}\lop^A_{(n)}{}_B\, E^B{}_A\, u_0^{-n-1} \ ,
\end{align}
we find the usual relations that follow from \eqref{eq:RLLandDI}
\begin{align}\label{eq:YangFromL2}
  &\delta^A{}_B = \lop^A_{(-1)}{}_B, &&\rep^x(\gen{J}^A{}_B) = \lop^A_{(0)}{}_B \ ,
\end{align}
and where the Yangian generator is shifted by the central element
\begin{align}\label{eq:YangFromL}
  &(u-\sfrac{c}{2})\rep^x(\gen{J}^A{}_B) = \lop^A_{(1)}{}_B - \half
  (-1)^{(|A|+|C|)(|B|+|C|)} \lop^C_{(0)}{}_B \lop^A_{(0)}{}_C \ .
\end{align}
Applying the fusion relations \eqref{eq:fusion} to the operator $\lop$ gives
rise to the correct Yangian coproduct. In other words, the coproduct directly
follows by expanding \eqref{eq:coprodRLL} around $u_0=\infty$.

\subsection{Yangian invariants}
\label{subsec:AmpAlgebra}

In this section we will explain how the $\lop$- and $\rop$-operators considered 
above can be used to define Yangian invariants and scattering amplitudes.

\subsubsection{Representation} 
We work in a representation where we identify the spinor variables
$\lambda,\tilde{\lambda}$ and Grassmann variables $\tilde{\eta}$ introduced in
\subsecref{sec:ampl} with the canonical variables as
\begin{align}
  &\mathbf{x} =
  (\lambda,\,\partial_{\tilde{\lambda}},\,\partial_{\tilde{\eta}}), &&\mathbf{p}
  = (\partial_\lambda,\,-\tilde{\lambda},\,-\tilde{\eta}) \ .
\end{align}
It is easy to check that they satisfy the fundamental commutation relations
\eqref{eq:FCRxp} and thus provide a functional representation of
$\alg{gl}(4|4)$. The variables $\mathbf{x}$ and $\mathbf{p}$ act on the infinite-
dimensional vector space $V_a$ of (generalised) functions in the
spinor-helicity variables $(\lambda^\alpha,\,\tilde{\lambda}_{\dot{\alpha}},\,\tilde{\eta}^A)$.

An $n$-point amplitude is a function in the $n$-fold tensor product
\footnote{An amplitude is a function of the spinor-helicity variables of $n$
particles that generically will not be of a factorised form. Thus, strictly
speaking, we can only see it as an element of $V$ if we express it via power
series. }
\begin{align}
  V=V_1\otimes\ldots\otimes V_n \ .
\end{align}
In this representation, the L-operator \eqref{eq:Ldef} is an eight-by-eight
matrix whose entries are differential operators, while the $\rop$-operator is
a scalar operator.

From the form of the $\rop$-operator \eqref{eq:Rdef}, we see that it can be
identified with a shift operator. Explicitly, the action of the $\rop$-operator
in spinor-helicity variables is given by
\begin{align}\label{eq:DefRop}
  \rop_{ab}(u)
  f(\lambda_a,\tilde{\lambda}_a,\tilde{\eta}_a,\lambda_b,\tilde{\lambda}_b,\tilde{\eta}_b)
  := \int\frac{\mathrm{d}
    z}{z^{1+u}}f(\lambda_a-z\lambda_b,\tilde{\lambda}_a,\tilde{\eta}_a,\lambda_b,\tilde{\lambda}_b+
  z \tilde{\lambda}_a,\tilde{\eta}_b+z \tilde{\eta}_a) \ .
\end{align}
and precisely corresponds to the BCFW-shift eqn.~\eqref{eq:bcfw_super}. 

Let us consider a specific subset of $V_a$ generated by the delta
functions
\begin{align}\label{eq:defdelta}
  &\dl{a}:= \delta^2(\lambda_a), &&\dlb{a}:=
  \delta^{2|4}(\tilde{\lambda}_a):=\delta^{2}(\tilde{\lambda}_a)
  \delta^{4}(\tilde{\eta}_a) \ .
\end{align}
For those, the central operator $C$ is identified with the helicity operator $h=1-\half
C$. The action of $C$ on the basis elements is trivial, while $B$ is only
trivial on the negative $\delta$-functions
\begin{align}
  &C_a \dl{\pm a} = 0 \ ,
  && B_a \dl{\pm a} = -(4\pm 4) \  .
\end{align}
On the other hand, the $\lop$-operator acts non-trivially on all
$\delta$-functions
\begin{align}
  \label{eq:LonDelta}
  &\lop_a(u)\dl{\pm b} = (u \mp \half) \,\delta_{ab}\, \dl{\pm b} \ ,
\end{align}
where the right-hand side is proportional to the unit matrix. From the explicit
form of the $\rop$-operator it is readily seen that
\begin{align}
  \label{eq:RonVac}
  &\rop_{ab}\,\dl{c} = \dl{c}\rop_{ab}  \quad \mathrm{for}\,\, c\neq a \ ,
  &&\rop_{ab}\,\dlb{c} = \dlb{c} \rop_{ab}  \quad \mathrm{for}\,\, c\neq b \ .
\end{align}
On the other $\delta$-functions the operator $\rop$ has a non-trivial action
according to \eqref{eq:DefRop}.  

\subsubsection{Invariants}

Let us now employ the above objects to define Yangian invariants. We saw in
\eqref{eq:YangFromL} that the operator $\lop$ generates the Yangian symmetry in the
functional representation (with a shifted spectral parameter). Using the fusion
relations \eqref{eq:fusion} and \eqref{eq:coprodRLL} we find that the Yangian
representation corresponding to $n$ particles is generated by the so-called
monodromy matrix
\begin{align}\label{eq:Monodromy}
  & \mathrm{T}_n := \lop_1(u_1-\sfrac{C_1}{2})\ldots \lop_n(u_n-\sfrac{C_n}{2}) \ .
\end{align}
A function $\mathcal{Y}$ on $V$ is called a Yangian invariant if
it is an eigenfunction of the monodromy matrix $\mathrm{T}_n$:
\begin{align}
  \label{eq:YangEigenValue}
  \mathrm{T}(\{u_i\})\, \mathcal{Y} = \Lambda(\{u_i\})\, \mathcal{Y} \ .
\end{align}
It is worthwhile to note that this immediately implies that all generators from
$\alg{gl}(4|4)$ apart from $\gen{B}$ annihilate the Yangian invariant. In fact,
the eigenvalue of $\gen{B}$ is directly related to the large-$u$ expansion of
$\Lambda$ as can be seen from \eqref{eq:YangFromL2}.

The $\delta$-functions introduced in \eqref{eq:defdelta} provide a natural
eigenfunction of the monodromy matrix. Indeed, it is easy to show that
\begin{align}\label{eq:Groundstate}
  \mathrm{T}_n\, \Omega_{s_1,\ldots,s_n}:=\mathrm{T}_n\, \dl{s_1
    1}\ldots \dl{s_n n} = \prod_{i=1}^n (u_i-\sfrac{s_i}{2})
  \Omega_{s_1,\ldots,s_n} \ ,
\end{align}
where $s_i = \pm$. If there are $k$ negative $s$'s, then we can interpret this
function as the trivial $\mathcal{A}_{n;k}$ superamplitude where all particles
are soft and there is no interaction. This state has all central charges $c_i=0$
has total hypercharge $\sum_i B_i \Omega = -4 (\sum_i 1 -s_i)\Omega$.

We can use the reference state $\Omega$ as a starting point for constructing
non-trivial Yangian-invariant functions. From the discussion on the Yangian
symmetry-algebra we can identify a natural candidate for building Yangian
invariants: the $\rop$-operator. By construction it commutes with $\mathrm{T}$
up to a permutation of the spectral parameters. Furthermore, $\Omega$ contains
$2n$ bosonic $\delta$-functions. The Yangian invariant should be proportional to
$\delta^4(\sum_i p_i)$ and each $\rop$-operator will remove one
$\delta$-function. Consequently, we postulate that the $n$-point amplitude
$\mathcal{A}_{n;k}$ with degree $4k$ is a linear combination of functions of the
form
\begin{align}\label{eq:AnsatzAmpl}
 \mathcal{Y} = \rop_{a_1b_1}(v_1)\ldots \rop_{a_{2n-4}b_{2n-4}}(v_{2n-4}) \Omega \ ,
\end{align}
where $\Omega$ is a product of $(n-k)$ $\dl{a_i}$'s and $k$ $\dlb{a_i}$'s such
that it is an eigenfunction of the monodromy matrix.

We would like to point out that any function of the form \eqref{eq:AnsatzAmpl}
automatically has zero total central charge, as is required  \subsecref{sec:DefOnshell}
\begin{align}
  \sum_i C_ i \mathcal{Y} = 0 \ ,
\end{align}
by the commutation relations \eqref{eq:RcComm}. However, the individual central
charges $C_i \mathcal{Y}$ are clearly non-zero.

Finally let us remark that functions of the form \eqref{eq:AnsatzAmpl} are
simply integrals of delta functions according to \eqref{eq:DefRop}. Clearly not
all possible combinations of the form \eqref{eq:AnsatzAmpl} will result in a
well-defined integral and we will have to restrict to the ones that can be
computed. There is a further subtlety regarding the overall sign. We will always
compute the integral of $\delta$-functions by applying a coordinate
transformation that trivialises them. However, we will not take the absolute
values of the associated Jacobian and consequently we can only define the
Yangian invariants up to a sign.

\subsubsection{Symmetries of amplitudes}
\label{subsubsec:SymAmp}

Apart from Yangian symmetry, scattering amplitudes have an additional `global'
symmetry: dihedral symmetry. Dihedral symmetry is generated by a reflection and
a shift operation
\begin{align}\label{eq:DihdralGroup}
  &\{1,2,\ldots,n\}\stackrel{r}{\longrightarrow}\{n,n-1,\ldots,1\} \ ,
  &&\{1,2,\ldots,n\}\stackrel{s}{\longrightarrow}\{n,1,\ldots,n-1\} \ .
\end{align}
This simply corresponds to a relabelling of the scattered particles. Note that
individual Yangian invariants do not necessarily exhibit this symmetry. Both the
reflection and the shift operation have a natural action on the level of the
monodromy matrix $\mathrm{T}$.

\paragraph{Reflection.} The reflection operation can be deduced from equation
\eqref{eq:Linverse}. Let $\mathcal{Y}$ be an eigenfunction of the monodromy
matrix with eigenvalue $\Lambda$, then
\begin{align}
  \lop_1(u_1-\sfrac{c_1}{2})\ldots \lop_n(u_n-\sfrac{c_n}{2})\, \mathcal{Y}(u_i) = \Lambda\,
  \mathcal{Y}(u_i) \ ,
\end{align}
which implies that
\begin{align}\label{eq:reflLambda}
  \Lambda^{-1}\prod_i (\half+u_i)(\half+c_i-u_i)\,\mathcal{Y} =
  \lop_n(\sfrac{c_n}{2}-u_n)\ldots\lop_1(\sfrac{c_1}{2}-u_1)\,\mathcal{Y} \ .
\end{align}
We apply a relabelling using the reflection operation to obtain
\begin{align}
  \lop_1(u_1-\sfrac{c_1}{2})\ldots \lop_n(u_n-\sfrac{c_n}{2}) \mathcal{Y}_r =
  \tilde{\Lambda}\, \mathcal{Y}_r \ ,
\end{align}
where $\mathcal{Y}_r$ corresponds to $\mathcal{Y}$ where all indices and
spectral parameters have been reflected according to
\begin{align}\label{eq:reflection_param}
  &\rop_{ab} \rightarrow \rop_{r(a)r(b)}, && u_i \rightarrow -u_{r(i)} +
  c_{r(i)} \ .
\end{align}
The eigenvalue $\tilde{\Lambda}$ can be read off directly from
\eqref{eq:reflLambda}. This shows that if $\mathcal{Y}$ is an eigenstate of the
monodromy matrix, its reflected version is an eigenstate as well. Clearly,
for generic Yangian invariants $\mathcal{Y}$ and $\mathcal{Y}_r$ need not
coincide. However, full scattering amplitudes should have this property.

\paragraph{Shifts.} The effect of a shift operation follows from the
transposition properties of $\lop$ in eqn.~\eqref{eq:Ltranspose}. In particular
we find that for an eigenstate $\mathcal{Y}$
\begin{align}
  \lop_1(u_1-\sfrac{c_1}{2})\ldots \lop_n(u_n-\sfrac{c_n}{2})\, \mathcal{Y} =
  \Lambda\, \mathcal{Y} \ ,
\end{align}
implies
\begin{align}
  \lop_2(u_2-\sfrac{c_2}{2})\ldots \lop_n(u_n-\sfrac{c_n}{2})
  \lop^{-1,t,-1,t,t,t}_1(u_1-\sfrac{c_1}{2})\, \mathcal{Y} =
  \Lambda\,\mathcal{Y} \ .
\end{align}
Then we can deduce from eqn.~\eqref{eq:Ltransposeinverse} that
\begin{align}\label{eq:shiftLambda}
  \lop_2(u_2-\sfrac{c_2}{2})\ldots \lop_1(u_1-\sfrac{c_1}{2})\, A =
  \frac{[u_1-\half][c_1-u_1-\half]}{[u_1+\half][c_1-u_1+\half]} \Lambda\,
  \mathcal{Y} \ .
\end{align}
Hence, after relabelling the indices we find
\begin{align}
  \lop_1(u_1)\ldots \lop_n(u_n) \mathcal{Y}_s = \Lambda^\prime \, \mathcal{Y}_s
  \ ,
\end{align}
where $\mathcal{Y}_s$ is $\mathcal{Y}$ with all indices shifted according to
the shift $s$ in eqn.~\eqref{eq:DihdralGroup}. The eigenvalue $\Lambda^\prime$ is
obtained by relabelling the right-hand side of eqn.~\eqref{eq:shiftLambda}.

\paragraph{Parity flip.} Finally there is the parity flip operation. Comparing
eqns.~\eqref{eq:RLYBE} and \eqref{eq:RLLinverse} we see that $\rop_{ab}$ and
$\rop_{ba}$ have similar commutation relations with respect to the monodromy
matrix. In particular, for eigenstates of the form eqn.~\eqref{eq:AnsatzAmpl}
the transformation
\begin{align}
  \label{eq:parityflip}
  &\rop_{ab}\rightarrow\rop_{ba} \ , &&\dl{a}\leftrightarrow\dlb{a} \ , &&
  u_i\rightarrow u_i - c_i \ ,
\end{align}
is a map between eigenstates. Notice that the central charges in the above
transformation are the ones of the parity-flipped function. The central charges
are not invariant under a parity flip. As described in
\subsecref{subsec:permutations} below, a parity flip corresponds to swapping
white and black dots in an on-shell diagram. 

In view of \eqref{eq:reflection_param} it is very natural to combine a parity
flip with a reflection. Indeed, the combined transformation takes the simple
form
\begin{align}
  \label{eq:refpar}
  &\rop_{ab}\rightarrow\rop_{r(b)r(a)} \ , &&\dl{a}\leftrightarrow\dlb{r(a)} \ , &&
  u_i\rightarrow -u_{r(i)} \ .
\end{align}
In terms of on-shell diagrams we will demonstrate that this transformation simply
corresponds to interchanging black and white dots and a simultaneous flip in the
orientation of the external legs (see \subsecref{subsec:permutations}).

\subsection{Three-point amplitudes}
\label{subsec:3ptAmps}

We are now ready to compute Yangian invariants explicitly. Let us start by
considering the three-point $\mathrm{MHV}$ and $\overline{\mathrm{MHV}}$
amplitudes $\mathcal{A}_{3;2}$ and $\mathcal{A}_{3;1}$, respectively. These
scattering amplitudes are generated by two $\rop$-operators acting on a product
of three $\delta$-functions. By direct computation one can readily find all states of
the form \eqref{eq:AnsatzAmpl} that satisfy the criteria outlined above.

\paragraph{The amplitude $\mathcal{A}_{3;2}$}

In this case, the vacuum is a product of two negative and one positive
$\delta$-function. We fix it to be
\begin{align}
  \Omega_{+--} := \dl{1}\dlb{2}\dlb{3} \ .
\end{align}
From the fundamental commutation relations \eqref{eq:RLYBE} and
\eqref{eq:LonDelta} it is easy to see that
\begin{align}\label{eq:MHVB3}
  \mathcal{A}_{3;2} := \rop_{23}(u_{32})\rop_{12}(u_{31})\,\Omega_{++-}
\end{align}
is an eigenfunction of $\mathrm{T}_3$ with eigenvalue
$(u_1+\half)(u_2+\half)(u_3-\half)$. The expression \eqref{eq:MHVB3} can
be straightforwardly evaluated to
\begin{align}\label{eq:defMHV3}
  \mathcal{A}_{3;2} = \frac{\delta^4(\sum_i
    p_i)\delta^{8}(\sum_i\lambda_i\eta_i)}{\langle
    12\rangle^{1+u_{32}}\langle23\rangle^{1+u_{13}}\langle 13\rangle^{1+u_{21}}}
  \ .
\end{align}
Upon setting all evaluation parameters equal, \textit{i.e.} $u_{ij}=0$, this
reduces to the usual three-point $\mathrm{MHV}$ amplitude
\eqref{eq:mhv_npt}. 

Notice furthermore that \eqref{eq:defMHV3} is \emph{almost}
invariant under the dihedral symmetry group. Under shifts and reflections it
picks up overall numerical factors of the form $(-1)^{u_{ij}}$. These factors
vanish in the undeformed limit and do not generate a new eigenstate anyway. As
such, we will actually identify states that differ by such factors and loosely
refer to \eqref{eq:defMHV3} as being invariant under dihedral symmetry. 

However, apparently eqn.~\eqref{eq:MHVB3} is not the only eigenfunction of the
monodromy matrix. Furthermore, starting with a different vacuum will also
\textit{a priori} lead to different solutions. We will discuss their relation at
the end of this section.

\paragraph{The amplitude $\mathcal{A}_{3;1}$}

The vacuum for this state is a product of two positive and one negative
$\delta$-functions. We choose
\begin{align}
  \Omega_{++-} := \dl{1}\dl{2}\dlb{3} \ .
\end{align}
This time we find that
\begin{align}
  \mathcal{A}_{3;1} := \rop_{12}(u_{21})\rop_{23}(u_{31})\,\Omega_{++-}
\end{align}
is an eigenfunction of the monodromy matrix with eigenvalue
$(u_1+\half)(u_2-\half) (u_3-\half)$.  It yields a well-defined integral that
can be evaluated to
\begin{align}\label{eq:defMHVB3}
  \mathcal{A}_{3;1} = \frac{\delta^4(\sum_i p_i)\delta([12]\eta_3 + [23]\eta_1 +
    [31]\eta_2)}{[12]^{1+u_{13}}[23]^{1+u_{32}}[31]^{1+u_{21}}} \ ,
\end{align}
This expression reduces to the usual three-point $\overline{\mathrm{MHV}}$
amplitude when all evaluation parameters are equal.

\paragraph{Classifying all eigenstates.}

Thus far we were able to derive eigenfunctions of the monodromy matrix that
correspond to deformed versions of the $\mathrm{MHV}$ and
$\overline{\mathrm{MHV}}$ amplitudes. However, the explicit expressions we
found are not the only eigenfunctions of the monodromy matrix.

In order to completely classify all eigenstates, let us take a more general
approach and consider all states of the form
\begin{align}
  \rop_{ab}\rop_{cd}\Omega_{s_1s_2s_3}
\end{align}
and investigate which ones are eigenstates of the monodromy matrix. Since
$a,b,c,d$ can only take values between one and three, we quickly find that there
are 24 possible non-trivial combinations\footnote{Clearly operators of the form
  $\rop_{12}\rop_{12}$ commute with the monodromy matrix as well, but we will
  discard these types of solutions for obvious reasons.}.  These are listed in
Table \ref{tab:3point}.

All permutations of $\{1,2,3\}$ can be generated by cyclic shifts and
reflections.  From the discussion in \subsecref{subsec:AmpAlgebra}, we know we
can relate eigenstates via dihedral transformations. In other words, this
implies that the 24 states split into four classes under
dihedral symmetry.

\begin{table}
  \begin{center}
    \begin{tabular}{|c|c|c|c|}
      \hline
      Class 1 & Class 2 & Class 3 & Class 4\\
      \hline
      $\rop_{12}\rop_{23}$ & $\rop_{23}\rop_{12}$  & $\rop_{12}\rop_{13}$ & $\rop_{23}\rop_{13}$\\
      \hline
      $\rop_{31}\rop_{12}$ & $\rop_{12}\rop_{31}$  & $\rop_{31}\rop_{32}$ & $\rop_{12}\rop_{32}$\\
      \hline
      $\rop_{23}\rop_{31}$ & $\rop_{31}\rop_{23}$  & $\rop_{23}\rop_{21}$ & $\rop_{31}\rop_{21}$\\
      \hline
      $\rop_{32}\rop_{21}$ & $\rop_{21}\rop_{32}$  & $\rop_{32}\rop_{31}$ & $\rop_{21}\rop_{31}$\\
      \hline
      $\rop_{21}\rop_{13}$ & $\rop_{13}\rop_{21}$  & $\rop_{21}\rop_{23}$ & $\rop_{13}\rop_{23}$\\
      \hline
      $\rop_{13}\rop_{32}$ & $\rop_{32}\rop_{13}$  & $\rop_{13}\rop_{12}$ & $\rop_{32}\rop_{12}$\\
      \hline
    \end{tabular}
    \caption{The equivalence classes of $\rop$-operators generating three-point
      amplitudes with the arguments left unspecified. Notice that the first
      three states of classes $3$ and $4$ can be identified with the last three by
      using \protect\eqref{eq:Rpermute}.}
      \label{tab:3point}
  \end{center}
\end{table}

Since the products of $\rop$-operators multiply a product
of $\delta$-functions it can be readily seen that \textit{any} of the products from
Table \ref{tab:3point} commutes with the monodromy matrix for appropriate
arguments. For instance,
\begin{align}\label{eq:Example3pt}
  \mathrm{T}\,\rop_{12}(u_{21})\rop_{13}(u_{32})\Omega_{s_1s_2s_3}
  &=
  \rop_{12}(u_{21})\mathrm{T}(u_2,u_1,u_3)\rop_{13}(u_{32})\Omega_{s_1s_2s_3}
  \nln
  &=(u_1-\sfrac{s_2}{2})\rop_{12}(u_{21})\lop_1(u_2)\lop_3(u_3)\rop_{13}(u_{32})\Omega_{s_1s_2s_3}
  \\ &=
  (u_1-\sfrac{s_2}{2})(u_2-\sfrac{s_3}{2})(u_3-\sfrac{s_1}{2})
  \rop_{12}(u_{21})\rop_{13}(u_{32})\Omega_{s_1s_2s_3} \ , \nonumber
\end{align}
where in the second step we used that $\delta_2$ commutes with $\rop_{13}$ and
\eqref{eq:LonDelta}. In other words, this function is only an eigenstate in this
explicit representation. Notice that the property of being an eigenstate of
$\mathrm{T}$ does not depend on the explicit choice of the vacuum.

The next criterion is whether the eigenstate gives rise to a well-defined
integral. It can be checked that this is the case for a \textit{unique} vacuum
state for each of the products from Table \ref{tab:3point}. This indicates that
there is a map that determines the vacuum corresponding to a given sequence of
$\rop$-operators. We will come back to this shortly in \subsecref{subsec:vacuum}.

The operators belonging to the classes 2 and 3 correspond to $\mathrm{MHV}$ amplitudes,
while the operators from classes 1 and 4 yield $\overline{\mathrm{MHV}}$
amplitudes. It is not difficult to show that all states from classes 2 and 3
evaluate to \eqref{eq:defMHV3} and states from classes 1 and 4 result in
\eqref{eq:defMHVB3}. In other words, the eigenfunctions \eqref{eq:defMHV3} and
\eqref{eq:defMHVB3} exhaust the space of eigenfunctions. Summarising, there is
a unique Yangian-invariant deformation of the $\mathrm{MHV}$ and
$\overline{\mathrm{MHV}}$ amplitude each of which can be written in the 12
different ways listed in Table \ref{tab:3pointAmps}.

Accordingly, Table \ref{tab:3pointAmps} provides us with an additional set of
relations that our R-matrices satisfy. We will refer to them as
$\rop\rop\delta$-relations in order to indicate that they are
representation-dependent -- in particular, they depend on the vacuum. They are
generated by shift-, reflection- and representation-dependent relations. Let us
spell the generating relations out
\begin{subequations}
  \label{eq:RRdelta}
  \begin{align}
  &\rop_{ab}(u)\rop_{bc}(v)\dl{a}\dl{b}\dlb{c}=
  \rop_{bc}(v-u)\rop_{ca}(-u)\dlb{a}\dl{b}\dl{c}\\
  &\rop_{ab}(u)\rop_{bc}(v)\dl{a}\dl{b}\dlb{c}=
  \rop_{cb}(-v)\rop_{ba}(-u)\dlb{a}\dl{b}\dl{c}\\
  &\rop_{ab}(u)\rop_{bc}(v)\dl{a}\dl{b}\dlb{c}=
  \rop_{bc}(v-u)\rop_{ac}(u)\dl{a}\dl{b}\dlb{c} \ ,
\end{align}
\end{subequations}
together with the parity flipped versions.

In addition, we can relate the $\overline{\mathrm{MHV}}$ and $\mathrm{MHV}$ by
the flip operation. Thus, these operations generate all eigenstates in Table
\ref{tab:3pointAmps} starting from any single choice of eigenstate.

\begin{table}
  \begin{center}
    \begin{tabular}{|c|c|}
      \hline
      $\mathrm{MHV}$ & $\mathrm{\overline{MHV}}$\\
      \hline
      $\rop_{12}(u_{21})\rop_{23}(u_{31})\Omega_{++-}$ & $\rop_{23}(u_{32})\rop_{12}(u_{31})\Omega_{+--}$\\
      \hline
      $\rop_{13}(u_{21})\rop_{23}(u_{32})\Omega_{++-}$ & $\rop_{12}(u_{21})\rop_{13}(u_{32})\Omega_{+--}$\\
      \hline
      $\rop_{21}(u_{32})\rop_{13}(u_{31})\Omega_{++-}$ & $\rop_{13}(u_{32})\rop_{12}(u_{21})\Omega_{+--}$\\
      \hline
      $\rop_{23}(u_{32})\rop_{12}(u_{21})\Omega_{++-}$ & $\rop_{32}(u_{21})\rop_{13}(u_{31})\Omega_{+--}$\\
      \hline
      $\rop_{21}(u_{32})\rop_{31}(u_{13})\Omega_{-++}$ & $\rop_{21}(u_{13})\rop_{32}(u_{23})\Omega_{--+}$\\
      \hline
      $\rop_{23}(u_{32})\rop_{31}(u_{12})\Omega_{-++}$ & $\rop_{32}(u_{21})\rop_{31}(u_{13})\Omega_{--+}$\\
      \hline
      $\rop_{31}(u_{13})\rop_{21}(u_{32})\Omega_{-++}$ & $\rop_{31}(u_{13})\rop_{32}(u_{21})\Omega_{--+}$\\
      \hline
      $\rop_{32}(u_{13})\rop_{21}(u_{12})\Omega_{-++}$ & $\rop_{12}(u_{21})\rop_{31}(u_{23})\Omega_{--+}$\\
      \hline
      $\rop_{12}(u_{21})\rop_{32}(u_{13})\Omega_{+-+}$ & $\rop_{13}(u_{32})\rop_{21}(u_{12})\Omega_{-+-}$\\
      \hline
      $\rop_{13}(u_{21})\rop_{32}(u_{23})\Omega_{+-+}$ & $\rop_{21}(u_{13})\rop_{23}(u_{32})\Omega_{-+-}$\\
      \hline
      $\rop_{31}(u_{13})\rop_{12}(u_{23})\Omega_{+-+}$ & $\rop_{23}(u_{32})\rop_{21}(u_{13})\Omega_{-+-}$\\
      \hline
      $\rop_{32}(u_{13})\rop_{12}(u_{21})\Omega_{+-+}$ & $\rop_{31}(u_{13})\rop_{23}(u_{12})\Omega_{-+-}$\\
      \hline
    \end{tabular}
    \caption{All possible ways to write the three-point amplitudes. The
    amplitudes in the same column are related via the
  $\rop\rop\delta$-relations. The expressions between the two columns are
related via the flip operation. }\label{tab:3pointAmps}
  \end{center}
\end{table}

\subsection{Four-point amplitude}\label{subsec:4pt}

Next we turn to the four-point $\mathrm{MHV}$ amplitude. Let us again provide a
convenient ansatz that gives an eigenstate of the monodromy matrix
\begin{align}
  \rop_{23}(u_{32})\rop_{34}(u_{42})\rop_{12}(u_{31})\rop_{23}(u_{41})\Omega_{++--} \ .
\end{align}
It evaluates to
\begin{align}\label{eq:defMHV4}
  \mathcal{A}_{4;2} = \frac{\delta^4(\sum_i p_i)\delta^{8}(\sum_i
    \lambda_i\eta_i)}{\langle
    12\rangle^{1+u_{32}}\langle23\rangle^{1+u_{43}}\langle
    34\rangle^{1+u_{14}}\langle 14 \rangle^{1+u_{21}}} \ ,
\end{align}
which for $u_{ij}=0$ reduces to the $\mathrm{MHV}$ amplitude
\eqref{eq:mhv_npt}. This function again respects dihedral symmetry.
Furthermore, we can as well apply the parity-flip operation, which leaves
\eqref{eq:defMHV4} invariant.

It is interesting to ask, whether eqn.~\eqref{eq:defMHV4} is the unique Yangian
invariant for four points. In order to perform a complete search of Yangian
invariants we assume that the only rules that we are allowed to apply in order
to show that a state is an eigenstate are:
\begin{enumerate}
\item The commutation relations between $\rop$ and $\lop$ \eqref{eq:RLYBE} and
  \eqref{eq:RLLinverse}.
\item The Yang--Baxter equation \eqref{eq:RRYBE} and the trivial permutation
  \eqref{eq:Rpermute}.
\item Dihedral symmetry of the invariant \textit{and} of subinvariants. For
  example, the above expression for the four-point Yangian invariant contains
  the three-point invariant as a subinvariant. This subinvariant can then be
  rewritten using the symmetries of the three-point amplitude summarised in the
  $\rop\rop\delta$ rules \eqref{eq:RRdelta}. 
\item The commutation relation of $\rop$ with $\delta$ according to
  \eqref{eq:RonVac}. This can be used similar to \eqref{eq:Example3pt} to deal
  with R-operators with non-adjacent legs. 
\end{enumerate} 
In what follows, we will assume that there are no additional relations and that
eigenstates can be permuted with the monodromy matrix using these relations
only. 

Constructing all eigenfunctions of the monodromy matrix that satisfy these
criteria, we find that there are about a thousand different of those states.
However, the ones that are well-defined all evaluate to the four-point
amplitude \eqref{eq:defMHV4}. In other words, there is a unique
four-point deformed amplitude. 

\subsection{Six-point invariants}
\label{subsec:sixpointinvariants}

Since the five-point scattering amplitude will be used as an example in great detail in
the next section, let us continue with the six-point Yangian invariants. For
six points there are three possible situations depending on the number of
negative $\delta$-functions. While there are Yangian invariants related to the
$\mathrm{MHV}$ and $\overline{\mathrm{MHV}}$ amplitudes, which have two and four negative
$\delta$-functions, there are several Yangian invariants
belonging to the $\mathrm{NMHV}$ sector. Those are built starting from a
product of three negative and three positive $\delta$-functions. We have again
constructed eigenfunctions using the set of rules discussed in
\subsecref{subsec:4pt}.

\paragraph{$\mathrm{MHV}$ and $\overline{\mathrm{MHV}}$}

In parallel to the results in \subsecref{subsec:4pt}, we find that the
$\mathrm{MHV}$ and $\overline{\mathrm{MHV}}$ amplitudes admit a unique
deformation. A particular representation of the deformed MHV amplitude reads
\begin{align}
\mathcal{A}_{6;2} = 
\rop_{23}(u_{32})\rop_{34}(u_{42})\rop_{45}(u_{52})\rop_{56}(u_{62})
\rop_{12}(u_{31})\rop_{23}(u_{41})\rop_{34}(u_{51})\rop_{45}(u_{61})
\dl{1}\dl{2}\dl{3}\dl{4}\dlb{5}\dlb{6} \ ,
\end{align}
while the $\overline{\mathrm{MHV}}$ is given by
\begin{align}
\mathcal{A}_{6;4} = 
\rop_{45}(u_{54})\rop_{34}(u_{53})\rop_{23}(u_{52})\rop_{12}(u_{51})
\rop_{56}(u_{64})\rop_{45}(u_{63})\rop_{34}(u_{62})\rop_{23}(u_{61})
\dl{1}\dl{2}\dlb{3}\dlb{4}\dlb{5}\dlb{6} \ .
\end{align}
These eigenstates are related by a combination of parity-flip and reflection.
It is again not difficult to show that both expressions reduce to the corresponding
$\superN=4$ sYM amplitudes when all spectral parameters are equal.
 
\paragraph{NMHV sector}

For the six-point NMHV channels we find that all possible eigenstates reduce to
exactly six different Yangian invariants. They can be represented by the
following expressions
\begin{align}\label{eq:NMHV6}
\mathcal{Y}^{(6)}_1 &= 
\rop_{34}(u_{43})\rop_{45}(u_{53})\rop_{23}(u_{42})\rop_{34}(u_{52})
\rop_{21}(u_{54})\rop_{31}(u_{51})\rop_{65}(u_{32})\rop_{64}(u_{62})
\dlb{1}\dl{2}\dl{3}\dlb{4}\dlb{5}\dl{6} \ ,\nln
\mathcal{Y}^{(6)}_2 &=  
\rop_{34}(u_{43})\rop_{45}(u_{53})\rop_{23}(u_{42})\rop_{34}(u_{52})
\rop_{21}(u_{54})\rop_{31}(u_{56})\rop_{65}(u_{31})\rop_{16}(u_{61})
\dl{1}\dl{2}\dl{3}\dlb{4}\dlb{5}\dlb{6} \ ,\nln
\mathcal{Y}^{(6)}_3 &=  
\rop_{34}(u_{43})\rop_{45}(u_{53})\rop_{23}(u_{42})\rop_{34}(u_{52})
\rop_{21}(u_{54})\rop_{31}(u_{56})\rop_{64}(u_{21})\rop_{16}(u_{61})
\dl{1}\dl{2}\dl{3}\dlb{4}\dlb{5}\dlb{6} \ ,\nln
\mathcal{Y}^{(6)}_4 &=  
\rop_{34}(u_{43})\rop_{45}(u_{53})\rop_{23}(u_{42})\rop_{34}(u_{52})
\rop_{21}(u_{54})\rop_{65}(u_{32})\rop_{64}(u_{21})\rop_{16}(u_{61})
\dl{1}\dl{2}\dl{3}\dlb{4}\dlb{5}\dlb{6} \ ,\nln
\mathcal{Y}^{(6)}_5 &=  
\rop_{34}(u_{43})\rop_{45}(u_{53})\rop_{23}(u_{42})\rop_{34}(u_{52})
\rop_{31}(u_{65})\rop_{65}(u_{32})\rop_{64}(u_{21})\rop_{16}(u_{61})
\dl{1}\dl{2}\dl{3}\dlb{4}\dlb{5}\dlb{6} \ ,\nln
\mathcal{Y}^{(6)}_6 &=  
\rop_{45}(u_{54})\rop_{23}(u_{32})\rop_{34}(u_{52})\rop_{21}(u_{53})
\rop_{31}(u_{65})\rop_{65}(u_{42})\rop_{64}(u_{21})\rop_{16}(u_{61})
\dl{1}\dl{2}\dl{3}\dlb{4}\dlb{5}\dlb{6} \ .
\end{align}
All these invariants $\mathcal{Y}^{(6)}_i$ can be related by the shift
operation from the dihedral group. Contrary to the situation for MHV amplitudes
that we encountered so far, we see that the dihedral symmetry does not leave
the individual Yangian invariants unaltered, but rather relates them. 

This is also the first occurrence where we do not have a unique Yangian
invariant corresponding to a scattering amplitude. Thus the question arises,
whether the invariants from \eqref{eq:NMHV6} belong to the same eigenspace. The
first thing to check are the eigenvalues with respect to the monodromy matrix. A
direct computation shows that
\begin{subequations}
\begin{align}
\mathrm{T}\,\mathcal{Y}^{(6)}_{1,2,3,4,5} &= 
(u_1+\half)(u_2+\half)(u_3+\half)
(u_4-\half)(u_5-\half)(u_6-\half)
\mathcal{Y}^{(6)}_{1,2,3,4,5} \ ,\\
\mathrm{T}\,\mathcal{Y}^{(6)}_{6} &= 
(u_1+\half)(u_2+\half)(u_3-\half)
(u_4+\half)(u_5-\half)(u_6-\half)
\mathcal{Y}^{(6)}_{6} \ .
\label{eq:normalmon}
\end{align}
\end{subequations}
Obviously, $\mathcal{Y}^{(6)}_{6}$ has a different eigenvalue. From the
discussion on dihedral symmetry in \subsecref{subsec:AmpAlgebra} we know that
each of the eigenstates should as well be an eigenstate of the monodromy matrices
with cyclically shifted L-operators. For instance, it can be shown
that 
\[
\begin{aligned}
\lop_2\lop_3\lop_4\lop_5\lop_6\lop_1\,\mathcal{Y}^{(6)}_{6} &= 
(u_1-\half)(u_2+\half)(u_3+\half)
(u_4+\half)(u_5-\half)(u_6-\half)
\mathcal{Y}^{(6)}_{6} \ ,\\
\lop_3\lop_4\lop_5\lop_6\lop_1\lop_2\,\mathcal{Y}^{(6)}_{6} &= 
(u_1-\half)(u_2-\half)(u_3+\half)
(u_4+\half)(u_5+\half)(u_6-\half)
\mathcal{Y}^{(6)}_{6} \ ,\\
\lop_4\lop_5\lop_6\lop_1\lop_2\lop_3\,\mathcal{Y}^{(6)}_{6} &= 
(u_1-\half)(u_2+\half)(u_3-\half)
(u_4+\half)(u_5+\half)(u_6+\half)
\mathcal{Y}^{(6)}_{6} \ ,\\
\lop_5\lop_6\lop_1\lop_2\lop_3\lop_4\,\mathcal{Y}^{(6)}_{6} &= 
(u_1+\half)(u_2-\half)(u_3-\half)
(u_4-\half)(u_5+\half)(u_6+\half)
\mathcal{Y}^{(6)}_{6} \ ,\\
\lop_6\lop_1\lop_2\lop_3\lop_4\lop_5\,\mathcal{Y}^{(6)}_{6} &= 
(u_1+\half)(u_2+\half)(u_3-\half)
(u_4-\half)(u_5-\half)(u_6+\half)
\mathcal{Y}^{(6)}_{6} \ .
\end{aligned}
\]
We see that all eigenvalues, apart from one distinct instance, are simply
related by shifts. For $\mathcal{Y}^{(6)}_{6}$ the unusual eigenvalue is the
eigenvalue of the canonical monodromy matrix eqn.~\eqref{eq:normalmon}. A similar
computation for the other Yangian invariants reveals that they do not belong to
the same eigenspace of the rotated monodromy matrices. For each pair there is
always a monodromy matrix for which they have different eigenvalues. 

From our explicit construction of the shift operation of Yangian invariants
\eqref{eq:shiftLambda}, we see that the eigenvalues can be expressed in terms
of the central charges. Thus, we can also simply compute the central charges
and compare those. The vector of central charges $\{c_i\}$ for the
six Yangian invariants reads
\[
\begin{aligned}
\label{eq:6ptcentralcharges}
&C_i \mathcal{Y}^{(6)}_{1} = 
\{u_{14},u_{25},u_{31},u_{46},u_{52},u_{63} \} \ ,
&&C_i \mathcal{Y}^{(6)}_{2} = 
\{u_{14},u_{25},u_{36},u_{42},u_{51},u_{63} \} \ ,\\
&C_i \mathcal{Y}^{(6)}_{3} = 
\{u_{14},u_{25},u_{36},u_{41},u_{53},u_{62} \} \ ,
&&C_i \mathcal{Y}^{(6)}_{4} = 
\{u_{14},u_{26},u_{35},u_{41},u_{52},u_{63} \} \ ,\\
&C_i \mathcal{Y}^{(6)}_{5} = 
\{u_{15},u_{24},u_{36},u_{41},u_{52},u_{63} \} \ ,
&&C_i \mathcal{Y}^{(6)}_{6} = 
\{u_{13},u_{25},u_{36},u_{41},u_{52},u_{64} \} \ .
\end{aligned}
\]
Thus, all Yangian invariants have different central charges and, consequently,
belong to different eigenspaces. The six-point NMHV amplitude in $\superN=4$
sYM is obtained by combining three Yangian invariants. Since our Yangian
invariants belong to different eigenspaces, we can add them only after
identifying spectral parameters in a way such that the eigenvalues coincide. In
particular, the linear combination
\begin{align}
  \label{eq:NMHV6b}
  \mathcal{A}_{6;3} = \mathcal{Y}^{(6)}_{1} + \mathcal{Y}^{(6)}_{3} +
  \mathcal{Y}^{(6)}_{5} \ ,
\end{align}
reduces to the correct undeformed amplitude. This linear combination is a
Yangian invariant only if we set
\begin{align}
\label{eq:identifyspectral}
&u_1 = u_6 \ , &&u_2=u_3 \ , &&u_4=u_5 \ .
\end{align}
Similarly, one can combine $\mathcal{Y}^{(6)}_{2,4,6}$ and find a different
deformation which only is well-defined if
\begin{align}
\label{eq:identifyspectral2}
&u_1 = u_2 \ ,
&&u_3=u_4 \ ,
&&u_5=u_6 \ .
\end{align}
This is in complete agreement with \cite{Beisert:2014qba}. If we want all
\emph{six} channels to belong to the same eigenspace we would have to set all
spectral parameters equal, which corresponds to the undeformed situation. We
would furthermore like to point out that $\mathcal{A}_{6;3}$ \eqref{eq:NMHV6b}
is invariant under dihedral symmetry as well as the parity-flip operation. 

\subsection{The top-cell}
\label{subsec:topcell}

Apart from the Yangian invariants computed so far, there are other
eigenfunctions of the monodromy matrix. For instance it is natural to consider
Yangian invariants for which the number of R-operators is not equal to $2n-4$,
for example
\begin{align}\label{eq:TopCell6}
\rop_{34}(u_{43}) \rop_{45}(u_{53}) \rop_{23}(u_{42}) 
\rop_{34}(u_{52}) \rop_{21}(u_{54}) \rop_{31}(u_{65}) 
\rop_{65}(u_{32}) \rop_{64}(u_{21}) \rop_{16}(u_{61})
\Omega_{+++---} \ .
\end{align}
This function corresponds to the so-called top-cell. For now, we remark that the
six different invariants from the six-point NMHV amplitude are elegantly
packaged in \eqref{eq:TopCell6}. Indeed, it can be seen that by removing the
R-operators at positions $\{1,5,6,7,8,9\}$ (from the left) we exactly reproduce
the channels \eqref{eq:NMHV6} discussed in the previous subsection.

On the other hand, eqn.~\eqref{eq:TopCell6} can be evaluated in a more direct way.
Performing all the integrals apart from one yields a rational function (with the
standard momentum conserving $\delta$-functions) which has six poles in the
complex plane. Unsurprisingly, the residues at each of those poles again produce
the Yangian invariants $\mathcal{Y}^{(6)}_i$.

While of course the same function can be expressed in multiple ways by using the
symmetries and relations derived earlier, there is one particular representation
that we would like to mention (see also \subsecref{subsec:permutations})
\begin{align}\label{eq:TopCell6Plabic}
\rop_{34}(u_{43}) \rop_{45}(u_{53}) \rop_{56}(u_{63})
\rop_{23}(u_{42}) \rop_{34}(u_{52}) \rop_{45}(u_{62})
\rop_{12}(u_{41}) \rop_{23}(u_{51}) \rop_{34}(u_{61})
\Omega_{+++---} \ .
\end{align}
Performing the integrations again yields a rational function whose poles give rise to
NMHV channels. However in contradistinction to \eqref{eq:TopCell6}, the channels
can not be obtained by simply removing R-operators from eqn.~\eqref{eq:TopCell6Plabic}.

\subsection{Symmetries and central charges}
\label{subsec:summarysym}

Let us summarise the results from this section: so far we found that each
undeformed Yangian invariant arising in $\superN=4$ sYM theory is associated
to a unique deformed Yangian invariant. These Yangian invariants can be
expressed in different ways in terms of R-operators. For instance, the
Yang--Baxter equation \eqref{eq:RRYBE}, the permutation \eqref{eq:Rpermute} and
dihedral symmetries of (sub)invariants can be used to find equivalent ways of
expressing the same Yangian invariant.

In order to get a well-defined deformed scattering amplitude beyond the MHV
sector, however, one needs to combine several Yangian invariants. Due to the
fact that they belong to different eigenspaces, we find that we can only
combine them if we identify spectral parameters in accordance with
\cite{Beisert:2014qba}.

In all the examples that we considered so far, the central elements turn out to
be a difference of $u$'s. This is actually a general feature. To be more
precise, there is a permutation $\sigma$ such that
\begin{align}\label{eq:CviaU}
  c_i = u_i - u_{\sigma(i)} \ .
\end{align}
This can be readily seen by commuting the monodromy matrix through the sequence
of R-operators using \eqref{eq:RLYBE} and \eqref{eq:RLLinverse}. The effect of
such a single commutation is a swap of spectral parameters. The permutation
$\sigma$ can be extracted from the distribution of spectral parameters in the
monodromy matrix after commuting it through the R-operators in the eigenstate.
From this result we see that to each eigenstate we can assign a unique
permutation $\sigma$ via the central charges.

Using the relation between central charges and permutations, we can consider the
transformation of the central charges under dihedral symmetry. Acting with the
shift operation on \eqref{eq:CviaU} results in
\begin{align}
  c_i \rightarrow u_i - u_{\sigma(i+1)-1} \ ,
\end{align}
while reflection yields
\begin{align}
  c_i \rightarrow u_i - u_{r(\sigma(r(i))} \ .
\end{align}
In particular if the permutation $\sigma$ corresponds to a translation, the
central charges remain invariant.

The map between central charges of Yangian invariants and permutations is
clearly injective\footnote{%
  Obviously, injectivity applies to  nontrivial cycles only.
}. %
However, we can actually prove that this map is surjective as well. Recall that
every permutation can be written as a sequence of minimal length of swaps of
neighbouring sites. However such swaps can be trivially realised in terms of
R-operators, which manifestly commute with the monodromy matrix. \textit{This
means that there is a bijection between eigenstates of the monodromy matrix and
permutations}. Moreover, via \eqref{eq:CviaU} this also means that Yangian
invariants are in one-to-one correspondence with a choice of central charges.


\section{Relations to on-shell graphs and amplitudes}
\label{sec:onshell}

Let us now relate the representation of Yangian invariants in terms of
$\rop$-operators to the previously known ways to describe them: the on-shell
graphs and permutations discussed in \subsecref{subsec:onshell}. All of those
descriptions are valid ways of representing Yangian invariants in
$\superN=4$ sYM theory, each of which with its advantages and disadvantages.

Naturally, one can translate between those three descriptions. While the
permutation associated to an on-shell graph can be readily deduced by using the
double-line formalism described in \subsecref{subsec:onshell}, relating the
other formulations depicted in \figref{fig:tripledescription} requires more
care. 

In the discussion to follow, we will usually not note the argument of the
$\rop$-operator as those are uniquely fixed by demanding Yangian invariance: in
the same way the as the linear system eqn.~\eqref{eq:upmsys} determines the
parameters $c_i$ in terms of the evaluation parameters $u_i$, requiring an
$\rop$-chain to be an eigenstate of the monodromy matrix
(\textit{cf.}~eqn.~\eqref{eq:AnsatzAmpl}) will identify the arguments of the
$\rop$-operators yielding a Yangian invariant.

All translational rules described below, preserve Yangian invariance: starting
from an on-shell diagram with parameters $c_i$ and $u_i$ satisfying the linear
system of equations ensuring Yangian invariance (see \subsecref{sec:DefOnshell})
one will obtain an eigenstate of the monodromy matrix and vice versa. Nicely,
the third way of describing a Yangian invariant, the permutation, already
incorporates these conditions naturally: only Yangian-invariant $\rop$-chains
and on-shell graphs can be translated into permutations.

\subsection{Which vacuum for which $\rop$-chain?}
\label{subsec:vacuum}

As already mentioned in \subsecref{subsec:3ptAmps}, there is a unique vacuum
associated to any chain of $\rop$-operators. This can be understood easily as
follows. As spelled out in eqn.~\eqref{eq:DefRop}, $\rop_{a b}$ acts as
\[
\rop_{a b} (u)\,\mathcal{F}(\lambda_a,\tilde{\lambda}_a,\tilde{\eta}_a ;
\lambda_b,\tilde{\lambda}_b,\tilde{\eta}_b) \,=\, \int
\frac{\mathrm{d}z}{z^{1+u}} \, \mathcal{F}(\lambda_a - z
\lambda_b,\tilde{\lambda}_a,\tilde{\eta}_a; \lambda_b,\tilde{\lambda}_b + z
\tilde{\lambda}_a ,\tilde{\eta}_b + z \tilde{\eta}_a) \ , \nn\]
that is, it performs a BCFW shift on the spinor-helicity variables. The
single-particle vacuum states are the superconformal invariant delta functions
$\dl{a}, \dlb{a}$ defined in eqn.~\eqref{eq:defdelta}. The operator $\rop_{a b}$
acts nontrivially only on $\dl{a}$ and $\dlb{b}$. The sequence of $\dl{a}$'s and
$\dlb{a}$'s associated with a R-chain is such that the rightmost $\rop$'s act
nontrivially on the vacuum.  In practice, reading the $\rop$-chain from the
right-hand side, one has to note whether the first occurrence of an index is at
position $a$ or $b$ in $\rop_{ab}$. For an index appearing in position $a$
first, the vacuum has to be $\dl{a}$ while an index appearing first at position
$b$ requires a vacuum state $\dlb{b}$.

Notice that a given $\rop$-chain acting on the correct vacuum can give rise to
an invariant with constrained kinematics, for example if the number of $\rop$'s
does not match $(2n-4)$.  This is not surprising, an easy example being the
invariant given by\footnote{%
  Another important example are the top-cell representatives discussed in
  \subsecref{subsec:topcell}.  }%
\begin{equation}
  \label{eq:kinconstr}
  \rop_{12}\, \rop_{13}\, \rop_{34} \cdot \dl{1} \, \dlb{2} \, \dl{3} \, \dlb{4} \ .
\end{equation}
It is easily checked that this is indeed an eigenfunction of the monodromy
matrix.

The $\rop$-operators in eqn.~\eqref{eq:kinconstr} give rise to three integrals,
whereas the total number of bosonic delta functions is eight. If we take into
account the momentum-conserving delta functions, we should be left with an
additional bosonic delta constraining the kinematics. A straightforward
computation shows that
\begin{equation}
  \label{eq:constr_res}
  \rop_{12}\, \rop_{13}\, \rop_{34} \cdot \dl{1} \, \dlb{2} \, \dl{3} \, \dlb{4}
  \,\propto\, \delta^4 (p)\, \delta(\langle34\rangle)\ ,
\end{equation}
where the proportionality factor involves eight fermionic delta functions and a
ratio of spinor brackets. This actually will not come as a surprise as soon as
the correspondence with on-shell graphs is spelled out, since the above
invariant corresponds to the following on-shell graph
\begin{equation}
  \label{eq:onsh_constr}
  \includegraphicsboxex{Fig4ptconstr.mps}
\end{equation}
which is easily seen to be proportional to $\delta(\langle34\rangle)$, thus a
factorisation channel where the internal propagator is on-shell.

As an aside note, notice that it is possible to obtain invariants with
constrained kinematics also by acting with a given chain of $\rop$'s on the
\emph{wrong} vacuum.

\subsection{Translating between $\rop$-operators on-shell graphs}

In order to create a representative for the class of on-shell diagrams
corresponding to an $\rop$-chain, there is a nice graphical method which -- in
a different language -- was already described in \cite{ArkaniHamed:2012nw}. Let us
start from a chain of $\rop$-operators acting on a vacuum which allows for a
general kinematical situation as described in \subsecref{subsec:vacuum}.
\begin{itemize}
  \item $\rop$-operators need to be applied in the succession of their
    appearance in the $\rop$-chain, starting from the rightmost $\rop$-operator.
  \item if none of the indices $a$ and $b$ of an operator $\rop_{ab}$ has
    appeared as an index in an operator to the right, these particles are still
    in their vacuum state and are not yet connected. In this case, the
    $\rop$-operator will connect them by a propagator.  Simultaneously, this
    requires the vacuum state to be $\dl{a}$ for the first index and $\dlb{b}$
    for the second index as discussed in \subsecref{subsec:vacuum}
    \begin{equation}
      \rop_{ab}\delta_a\dlb{b}\qquad\to\qquad\includegraphicsboxex{prop.mps}.
      \label{}
    \end{equation}
  \item if only the first (second) index has appeared before, that is, there is
    already an external line with that label, connect the other (vacuum)-index
    to this line by a black (white) dot. In doing so, lines need to be attached
    as to yield the correct \emph{clockwise} cyclical ordering of the external
    legs of the on-shell graph:
    \begin{equation}
      \rop_{cb}\rop_{ab}\delta_a\dlb{b}\delta_c=\rop_{ca}\rop_{ab}\delta_a\dlb{b}\delta_c
      \qquad\to\qquad\includegraphicsboxex{vac1.mps} \ .
      \label{eq:AttachExternal}
    \end{equation}
    The clockwise ordering is a choice which agrees with the convention in
    \cite{ArkaniHamed:2012nw}. It will become essential in interpreting the
    $\rop$-operators in terms of permutations in
    \subsecref{subsec:permutations} below. 
  \item if both indices are already connected, acting with an operator
    $\rop_{ab}$ amounts to connecting the two external lines $a$ and $b$ by an
    BCFW-bridge \cite{Chicherin:2013ora}: 
    \begin{equation}
      \rop_{ab}\,\includegraphicsboxex{bridge1.mps}\quad\to\quad
      \includegraphicsboxex{bridge2.mps}\,.
      \label{eq:BCFWbridge}
    \end{equation}
    Here, the line labelled by the first index, $a$, will be equipped with a
    black dot and the line of the second index, $b$, will gain a white dot. The
    assignment of black and white dots to the indices does \emph{not} depend on
    whether they appear in ascending or descending order.
\end{itemize}

Let us test this for the example of the five-point MHV amplitude:
\begin{equation}
  \rop_{54}\rop_{43}\rop_{51}\rop_{41}\rop_{23}\rop_{21}\,\dlb{1}\dl{2}\dlb{3}\dl{4}\dl{5} \ .
  \label{eq:fivepointexample}
\end{equation}
While $\rop_{21}$ acting on the vacuum leads to a line connecting particles $2$ and $1$
\begin{equation}
  \rop_{21}\dlb{1}\dl{2}\qquad\to\qquad\includegraphicsboxex{prop2.mps} \ ,
  \label{}
\end{equation}
the following three operators attach all other external particles to this line:
\begin{equation}
  \rop_{51}\rop_{41}\rop_{23}\big(\includegraphicsboxex{prop2.mps}\big)\dlb{3}\dl{4}\dl{5}
  \quad\to\quad\includegraphicsboxex{prop3.mps} \ .
  \label{eq:treegraph}
\end{equation}
The next two operators, $\rop_{43}$ and $\rop_{54}$ connect the external lines
$3$, $4$ and $5$ by BCFW-bridges: 
\begin{equation}
  \label{eq:buildingfivepoint}
  \rop_{54}\rop_{43}
  \includegraphicsboxex{prop3.mps}\quad\to\quad\rop_{54}
  \includegraphicsboxex{prop4.mps}\quad\to\quad
  \includegraphicsboxex{Fig_5pt.mps}
\end{equation}
yielding the expected diagram. 

While the method described above assigns a particular on-shell diagram to a
chain of $\rop$-operators unambiguously, the reverse operation can not be
formulated as straightforwardly. Nevertheless, here are some guidelines for
finding a chain of $\rop$'s corresponding to an on-shell graph:
\begin{itemize}
  \item find a tree-level subgraph of the on-shell graph, which connects all
    particles. For the five-point example above, the graph is the result of
    eqn.~\eqref{eq:treegraph}. In order to represent this graph in terms of
    $\rop$'s, select a baseline to start with ($\rop_{21}$ in the above
    example) and attach the other particles in the correct clockwise cyclical
    ordering using \eqref{eq:AttachExternal}.  
  \item successively add BCFW-bridges to the tree-level subgraph as in
    eqn.~\eqref{eq:BCFWbridge} leading to the complete on-shell graph as
    depicted in eqn.~\eqref{eq:buildingfivepoint}.
\end{itemize}
The $\rop$-chains thus obtained are by no means unique: choosing a different
tree-level subgraph will lead to another $\rop$-chain. Nevertheless, all
representations are related by the relations between $\rop$-operators, like
eqn.~\eqref{eq:RRdelta} and their higher-point analogues.

Although the method above allows to translate an on-shell diagram into a chain
of $\rop$'s, it is usually easier to first deduce the permutation and then
follow the directions for converting a permutation into a chain of
$\rop$-operators above as described in the next subsection. 

\subsubsection{Inverse soft limit construction of tree amplitudes}

Finally, let us note the on-shell diagrams corresponding to the inverse soft
limits used in ref.~\cite{Chicherin:2013ora}. Adding a particle with vacuum
$\dl{b}$ via inverse soft limits is represented by 
  \begin{equation}
    \rop_{ba}\rop_{bc}\,\includegraphicsboxex{addPlusParticle1.mps}\dl{b}
    \quad\to\quad\includegraphicsboxex{addPlusParticle2.mps}
    \label{ISLPlus}
  \end{equation}
  while adding a particle with vacuum $\dlb{b}$ with
  an inverse soft limit is pictured by
  \begin{equation}
    \rop_{ab}\rop_{cb}\,\includegraphicsboxex{addMinusParticle1.mps}\dlb{b}
    \quad\to\quad\includegraphicsboxex{addMinusParticle2.mps}\,.
    \label{ISLMinus}
  \end{equation}
One can infer from eqn.~\eqref{eq:nk} that in both cases the number of
particles is raised by one, while the MHV level $k$ remains constant in the
first case and is increased by one for adding a particle with vacuum $\dlb{b}$.
However, restricting the formalism just to $\rop$-chains constructible using
inverse soft limits would exclude many classes of $\rop$-chains and thus
Yangian invariants.

\subsection{Permutations}
\label{subsec:permutations}

\subsubsection{From $\rop$-chains to permutations}
In \subsecref{subsec:onshell} we described how to translate an
on-shell graph into a permutation by following the double lines (see
\figref{fig:snpath} for a five-point example). Let us now discuss how to relate
$\rop$-chains to permutations.

Finding the permutation encoded by a chain of $\rop$-operators starts with
recognising the vacuum as the trivial permutation. In the vacuum state,
particles are not connected by BCFW bridges: they do not interact and thus
their evaluation parameters and central charges are mapped onto themselves. 

The action of an operator $\rop_{ab}$ on a permutation becomes obvious after
equipping eqn.~\eqref{eq:BCFWbridge} with double-lines:
\begin{equation}
  \centering
  \rop_{ab}\,\includegraphicsboxex{RBridge1.mps}
  \qquad\to\qquad\includegraphicsboxex{RBridge2.mps} \ .
  \label{eq:BCFWbridgeDL}
\end{equation}
Following the double-lines, one can immediately see that the operator
$\rop_{ab}$ in eqn.~\eqref{eq:BCFWbridgeDL} modifies the flow of the parameters
$u$ defined in \subsecref{sec:DefOnshell} by swapping the lines \emph{ending} at
particles $a$ and $b$.  On the contrary, the lines originating in the external
points $a$ and $b$ remain untouched. Applying the operator $\rop_{ba}$ as in
eqn.~\eqref{eq:BCFWbridgeDL2}, however, will build another BCFW-bridge:
\begin{equation}
  \centering
  \rop_{ba}\,\includegraphicsboxex{RBridge1.mps}
  \qquad\to\qquad\includegraphicsboxex{RBridge3.mps} \ .
  \label{eq:BCFWbridgeDL2}
\end{equation}
this times the lines \emph{originating} at
external legs $a$ and $b$ are swapped, while those ending there are unaltered. 

Both of the above identifications eqns.~\eqref{eq:BCFWbridgeDL}
and~\eqref{eq:BCFWbridgeDL2}, however, are true only if the system of equations
ensuring Yangian invariance for the complete on-shell graph is satisfied (see
\subsecref{sec:DefOnshell}). This condition is equivalent to ensuring that the
RLL-relation eqn.~\eqref{eq:RLYBE} can be applied.

In order to translate the swaps of lines into a modification of the permutation, one has
to distinguish, which type of $\rop$-operator is considered:
\begin{itemize}
  \item for the operator $\rop_{ab}$ in eqn.~\eqref{eq:BCFWbridgeDL} the
    \emph{image} of the permutation is changed. This is the case if the indices
    of the operator are in the same succession as the clockwise ordered
    external legs:
  \[
  \rop_{34}
  \begin{permutation}
    1 & 2 & 3 & 4\\
    \downarrow & \downarrow & \downarrow & \downarrow\\
    4 & 3 & 1 & 2 
  \end{permutation}
  \quad\to\quad
  \begin{permutation}
    1 & 2 & 3 & 4\\
    \downarrow & \downarrow & \downarrow & \downarrow\\
    3 & 4 & 1 & 2 
  \end{permutation} \ .
  \]
  Conveniently, one can write this as the \emph{left} action of a cycle $(ab)$
  on a permutation\footnote{In \cite{ArkaniHamed:2012nw}, the notation $(ab)$
  is used for swapping the particles at \emph{positions} $a$ and $b$ in the
  image of the permutation. Here we will use this notation to denote cycles $(ab)$ and refer to the swap of particles at positions $a$ and $b$ by $\dblpar{ab}$.}:
  \[
  (34)\triangleright
  \begin{permutation}
    1 & 2 & 3 & 4\\
    \downarrow & \downarrow & \downarrow & \downarrow\\
    4 & 3 & 1 & 2  
  \end{permutation}
  =
  \begin{permutation}
    1 & 2 & 3 & 4\\
    \downarrow & \downarrow & \downarrow & \downarrow\\
    3 & 4 & 1 & 2  
  \end{permutation} \ .
  \]
  \item in the situation in eqn.~\eqref{eq:BCFWbridgeDL2} the indices of the
    operator $\rop_{ba}$ are not ordered clockwise. Thus the swap has to be
    applied to the \emph{preimage} of the permutation. However, swapping
    numbers $a$ and $b$ in the preimage is equivalent to swapping the entries
    at \emph{positions} $a$ and $b$ in the image, which we will note by
    $\dblpar{ab}$. The example below is the last step in
    eqn.~\eqref{eq:buildingfivepoint}:
  \[
  \rop_{54}
  \begin{permutation}
    1 & 2 & 3 & 4 & 5\\
    \downarrow & \downarrow & \downarrow & \downarrow & \downarrow\\
    3 & 4 & 5 & 2 & 1 
  \end{permutation}
  \quad\to\quad
  \begin{permutation}
    1 & 2 & 3 & 5 & 4\\
    \downarrow & \downarrow & \downarrow & \downarrow & \downarrow\\
    3 & 4 & 5 & 2 & 1 
  \end{permutation}
  =
  \begin{permutation}
    1 & 2 & 3 & 4 & 5\\
    \downarrow & \downarrow & \downarrow & \downarrow & \downarrow\\
    3 & 4 & 5 & 1 & 2 
  \end{permutation} \ .
  \]
  However, the left action of $\dblpar{ab}$ is equivalent to the \emph{right}
  action of $(ab)$ on the permutation: 
   \[
  \dblpar{ab}\triangleright
  \begin{permutation}
    1 & 2 & 3 & 4 & 5\\
    \downarrow & \downarrow & \downarrow & \downarrow & \downarrow\\
    3 & 4 & 5 & 2 & 1 
  \end{permutation}
   =
  \begin{permutation}
    1 & 2 & 3 & 4 & 5\\
    \downarrow & \downarrow & \downarrow & \downarrow & \downarrow\\
    3 & 4 & 5 & 2 & 1 
  \end{permutation}
  \triangleleft(45)
  =
  \begin{permutation}
    1 & 2 & 3 & 4 & 5\\
    \downarrow & \downarrow & \downarrow & \downarrow & \downarrow\\
    3 & 4 & 5 & 1 & 2 
  \end{permutation} \ .
  \]
  Thus, a swap in terms of positions applied from the left is equivalent to a
  swap in terms of actual numbers applied from the right.
  \end{itemize}

  Let us illustrate the step-by-step construction by translating the five-point
  tree-level amplitude (the representation is related to
  eqn.~\eqref{eq:fivepointexample} by square moves and mergers) into a
  permutation.
\[
\rop_{45}\rop_{43}\rop_{15}\rop_{12}\rop_{52}\rop_{35}\,\dl{1}\dlb{2}\dl{3}\dl{4}\dlb{5}
\ .
\]
The indices of the first operator, $\rop_{35}$, are canonically
ordered. Therefore this swap is of the kind depicted in
eqn.~\eqref{eq:BCFWbridgeDL}, and translates into
\[
  (35)\triangleright
  \begin{permutation}
    1 & 2 & 3 & 4 & 5\\
    \downarrow & \downarrow & \downarrow & \downarrow & \downarrow\\
    1 & 2 & 3 & 4 & 5 
  \end{permutation}
  =
  \begin{permutation}
    1 & 2 & 3 & 4 & 5\\
    \downarrow & \downarrow & \downarrow & \downarrow & \downarrow\\
    1 & 2 & 5 & 4 & 3 
  \end{permutation} \ .
\]
The next operator, $\rop_{52}$, is in canonical ordering as well: although it
does not seem so initially, one has to take into account that the legs $2$ and
$4$ are not yet connected. Thus the corresponding permutation is acting from the
left:
\[
  (52)\triangleright
  \begin{permutation}
    1 & 2 & 3 & 4 & 5\\
    \downarrow & \downarrow & \downarrow & \downarrow & \downarrow\\
    1 & 2 & 5 & 4 & 3 
  \end{permutation}
  =
  \begin{permutation}
    1 & 2 & 3 & 4 & 5\\
    \downarrow & \downarrow & \downarrow & \downarrow & \downarrow\\
    1 & 5 & 2 & 4 & 3 
  \end{permutation} \ .
\]
While $\rop_{12}$ is clearly in canonical order, $\rop_{51}$ is clearly not.
Note that the succession of these two operators is not significant: this will
be discussed in \subsecref{subsec:RRPerm} below:
\[
  (12)\triangleright
  \begin{permutation}
    1 & 2 & 3 & 4 & 5\\
    \downarrow & \downarrow & \downarrow & \downarrow & \downarrow\\
    1 & 5 & 2 & 4 & 3 
  \end{permutation}
  \triangleleft(15)
  =
  \begin{permutation}
    1 & 2 & 3 & 4 & 5\\
    \downarrow & \downarrow & \downarrow & \downarrow & \downarrow\\
    3 & 5 & 1 & 4 & 2 
  \end{permutation} \ .
\]
The last two operators act again from the left and the right
\[
  (45)\triangleright
  \begin{permutation}
    1 & 2 & 3 & 4 & 5\\
    \downarrow & \downarrow & \downarrow & \downarrow & \downarrow\\
    3 & 5 & 1 & 4 & 2 
  \end{permutation}
  \triangleleft(34)
  =
  \begin{permutation}
    1 & 2 & 3 & 4 & 5\\
    \downarrow & \downarrow & \downarrow & \downarrow & \downarrow\\
    3 & 4 & 5 & 1 & 2 
  \end{permutation} \ .
\]
and finally yield the expected permutation. 

The construction here relates the double-line formalism from
\secref{subsec:onshell} to the algebraic considerations in
\secref{sec:algebra}. A line originating at the external leg $i$ depicts the
flow of the evaluation parameter $u_i$ in the on-shell diagram. The endpoint of
the line is $u_{\sigma(i)}$, where $\sigma(i)$ is the image of $i$ under the
permutation encoded in the on-shell diagram. This nicely connects to
eqn.~\eqref{eq:CviaU}: given a set of $u$'s and a permutation allows to infer
the set of central charges $c_i$ or vice versa. Starting from a set of central
charges, one can determine the permutation. 

There are a couple of strings attached to the above method. The
first one is the fact that it works for planar on-shell graphs only. In other
words, BCFW-bridges can be built only between neighbouring external legs.  In
practice this is actually no restriction as long as we deal with tree
amplitudes exclusively. The second fact to consider is that the definition of
``neighbouring'' involves omitting states which are not yet connected, that is,
those whose indices have not been appearing and which therefore are still
in their vacuum configuration. Keeping these constraints in mind, starting from
the vacuum and applying the operators in reverse order as compared to their
appearance in the $\rop$-chain will yield the encoded permutation in general.

\subsubsection{From permutations to $\rop$-chains}
In order to translate a given permutation into a chain of $\rop$-operators, one
will have to decompose the permutation into successive pairwise swaps. As
pointed out above, swaps can be applied either to the image or the preimage of
a permutation, which corresponds to a different ordering of indices in the
$\rop$-operator.

In the same way as a whole class of on-shell graphs related by square moves and
mergers in \figref{fig:moves} represent the same permutation and thus the same
Yangian invariant, the description of a permutation in terms of successive
swaps is not unique. It is this freedom, which will be used to explain the
different rules eqns.~\eqref{eq:Rpermute} and \eqref{eq:RRdelta} for rewriting
$\rop$-chains in \subsecref{subsec:RRPerm}.

The freedom can as well be used to represent a permutation as a series of swaps
to be applied from the right. The construction was suggested in
\cite{ArkaniHamed:2012nw} and employs a lexicographic ordering prescription of
pairs ensuring that one obtains the kind of swaps in
eqn.~\eqref{eq:BCFWbridgeDL2} exclusively:
\begin{itemize}
  \item promote the permutation to the \emph{decorated
    permutation} described in \subsecref{subsec:onshell}.
  \item starting from the decorated permutation, swap the lexicographically
    first pair $\dblpar{ab}$ of the permutation $\sigma$ for which
    $\sigma(a)<\sigma(b)$ and which is -- if at all -- only separated by
    positions $c=\sigma(c)$. Repeat this step until reaching the trivial
    permutation. Reading the necessary swaps in reverse order will bring you
    from the trivial permutation to the desired one. 
\end{itemize}
Here is a short example of the method (a more elaborate one can be found in
\cite{ArkaniHamed:2012nw}): the five-point tree amplitude is MHV and
thus the corresponding permutation and its decorated version read
\begin{equation}
  \begin{permutation}
    1 & 2 & 3 & 4 & 5\\
    \downarrow & \downarrow & \downarrow & \downarrow & \downarrow\\
    3 & 4 & 5 & 1 & 2 
  \end{permutation}
  \quad\text{and}\quad
  \begin{permutation}
    1 & 2 & 3 & 4 & 5\\
    \downarrow & \downarrow & \downarrow & \downarrow & \downarrow\\
    3 & 4 & 5 & 6 & 7 
  \end{permutation}\,.
  \label{perm}
\end{equation}
The first swap to be applied to the decorated permutation is $\dblpar{12}$ as
$\sigma(1)<\sigma(2)$ and leads to $\lbrace 4,3,5,6,7\rbrace$. In the next step,
positions $2$ and $3$ are swapped yielding $\lbrace 4,5,3,6,7 \rbrace$. After
again swapping the first two positions obtaining $\lbrace 5,4,3,6,7\rbrace$ in
the next step one has to consider that particle $3$ is already at the correct
position. Thus the next swap is $\dblpar{24}$, which is then followed by $\dblpar{12}$ and
$\dblpar{25}$. Thus we end up with the following succession of swaps of \emph{positions}:
\begin{equation}
  \begin{permutation}
    1 & 2 & 3 & 4 & 5\\
    \downarrow & \downarrow & \downarrow & \downarrow & \downarrow\\
    3 & 4 & 5 & 1 & 2 
  \end{permutation}
  =
  \dblpar{12}\triangleright\dblpar{23}\triangleright\dblpar{12}\triangleright
  \dblpar{24}\triangleright\dblpar{12}\triangleright\dblpar{25}\triangleright
  \begin{permutation}
     1 & 2 & 3 & 4 & 5\\
     \downarrow & \downarrow & \downarrow & \downarrow & \downarrow\\
     1 & 2 & 3 & 4 & 5 
  \end{permutation}
\end{equation}
\begin{equation}
  \begin{permutation}
    1 & 2 & 3 & 4 & 5\\
    \downarrow & \downarrow & \downarrow & \downarrow & \downarrow\\
    3 & 4 & 5 & 1 & 2 
  \end{permutation}
  =
  \begin{permutation}
   1 & 2 & 3 & 4 & 5\\
   \downarrow & \downarrow & \downarrow & \downarrow & \downarrow\\
   1 & 2 & 3 & 4 & 5 
  \end{permutation}
  \triangleleft(25)\triangleleft(12)\triangleleft(24)
  \triangleleft(12)\triangleleft(23)\triangleleft(12) \ .
\end{equation}

As explained near eqn.~\eqref{eq:BCFWbridgeDL2}, a swap of positions can be
identified with an $\rop$-operator via
\begin{equation}
  \dblpar{ij}\rightarrow\rop_{ji}.
  \label{eq:bwchoice}
\end{equation}
Thus one finally obtains
\begin{equation}
  \rop_{21}\rop_{32}\rop_{21}\rop_{42}\rop_{21}\rop_{52}\,\dlb{1}\dlb{2}\delta_3\delta_4\delta_5\,,
  \label{}
\end{equation}
where we have restored the vacuum according to the discussion in
\subsecref{subsec:vacuum}.

\subsubsection{Relations between different $\rop$-chains in terms of permutations} 
\label{subsec:RRPerm}

Having mapped the action of an $\rop$-chain on the vacuum to swaps acting from
the left and from the right onto permutations in the last subsection, let us
now reexamine the relations between different combinations of $\rop$'s which
have been explored in \secref{sec:algebra} and interpret them in terms of
permutations.
\paragraph{$\rop$-swaps.} Let us start with the relations connecting different
representations of the same Yangian invariant. The most basic one is
eqn.~\eqref{eq:Rpermute}, which we repeat here for convenience: 
\begin{align}
  &\rop_{ab}(u)\rop_{cd}(v) = \rop_{cd}(v)\rop_{ab}(u) ~ ~ ~ ~\mathrm{if}~ a\neq d
  ~\mathrm{and}~ b \neq c \ .
\end{align}
If all four indices $a,b,c,d$ are different, the interpretation in terms of
permutations depends on whether the corresponding cycles act from the right or
the left. If both $\rop$-operators act from the same side, \textit{i.e.} their indices
$a,b$ and $c,d$ are in the same order, one can freely exchange the
permutations, as they do not effect each other:
\begin{align}
  (ab)\triangleright(cd)\triangleright\emptyperm=
  (cd)\triangleright(ab)\triangleright\emptyperm\\
  \emptyperm\triangleleft(ab)\triangleleft(cd)=
  \emptyperm\triangleleft(cd)\triangleleft(ab) \ .
\end{align}
If the cycles corresponding to the operators act from different sides, that is,
the indices in $\rop_{ab}$ and $\rop_{cd}$ are ordered differently, the
succession of the two neighbouring operators in the $\rop$-chain does not play
a r\^ole. The cycles will end up on the two sides in any case:  
\begin{equation}
  \begin{matrix}
    \rop_{12}\rop_{43}\,\\
    \rop_{43}\rop_{12}\,
  \end{matrix}\bigg\rbrace\quad\to\quad
  (12)\triangleright\emptyperm\triangleleft(34)\,.
\end{equation}
Considering eqn.~\eqref{eq:Rpermute} again, $\rop_{ab}\rop_{ac}$ and
$\rop_{ac}\rop_{bc}$ are the only allowed configurations where the two index
pairs of the operators share an index. As BCFW-bridges are allowed between
neighbouring legs only, it is clear that the cycles corresponding to the two
operators act from different sides.  Thus, their succession is irrelevant. Here
is an example:
\begin{equation}
  \begin{matrix}
    \rop_{12}\rop_{13}\,\\
    \rop_{13}\rop_{12}\,
  \end{matrix}\bigg\rbrace\quad\to\quad
  (12)\triangleright\emptyperm\triangleleft(13)\,.
\end{equation}
\paragraph{Dihedral symmetries for the three-point invariants.} 
Similarly, one can convince oneself that the rules in eqn.~\eqref{eq:RRdelta}
have a nice interpretation in terms of permutations.  Let us repeat them here
for convenience choosing $a,b,c=1,2,3$ as an example: 
\begin{subequations}
  \label{eq:RRdelta2}
  \begin{align}
    \label{eq:RRd1}
  &\rop_{12}(u)\rop_{23}(v)\dl{1}\dl{2}\dlb{3}=
  \rop_{23}(v-u)\rop_{31}(-u)\dlb{1}\dl{2}\dl{3} \ ,\\
    \label{eq:RRd2}
  &\rop_{12}(u)\rop_{23}(v)\dl{1}\dl{2}\dlb{3}=
  \rop_{32}(-v)\rop_{21}(-u)\dlb{1}\dl{2}\dl{3} \ ,\\
    \label{eq:RRd3}
  &\rop_{12}(u)\rop_{23}(v)\dl{1}\dl{2}\dlb{3}=
  \rop_{23}(v-u)\rop_{13}(u)\dl{1}\dl{2}\dlb{3} \ ,
\end{align}
\end{subequations}
The left-hand side of eqn.~\eqref{eq:RRdelta2} reads
\begin{align}
  \label{eq:RRdlhs}
  \rop_{12}\rop_{23}\dl{1}\dl{2}\dlb{3}\quad\to\quad(12)\triangleright(23)\triangleright
  \begin{permutation}
    1&2&3\\\downarrow&\downarrow&\downarrow\\1&2&3
  \end{permutation}=
  \begin{permutation}
    1&2&3\\\downarrow&\downarrow&\downarrow\\2&3&1
  \end{permutation}
\end{align}
in terms of permutations. The first equality, eqn.~\eqref{eq:RRd1}, is just a
cyclical shift of the external labels. Indeed, the corresponding permutation
agrees: 
\begin{align}
  \rop_{23}\rop_{31}\dlb{1}\dl{2}\dl{3}\quad\to\quad(23)\triangleright(31)\triangleright
  \begin{permutation}
    1&2&3\\\downarrow&\downarrow&\downarrow\\1&2&3
  \end{permutation}=
  \begin{permutation}
    1&2&3\\\downarrow&\downarrow&\downarrow\\2&3&1
  \end{permutation}\,.
\end{align}
The second equality, eqn.~\eqref{eq:RRd2}, is the simplest example of a
reflection: in comparison to the left-hand side (eqn.~\eqref{eq:RRdlhs}) it
just swaps right and left action and reverses the succession of operators as
well as the position of indices in each $\rop$-operator:
\begin{align}
  \rop_{32}\rop_{21}\dlb{1}\dl{2}\dl{3}\quad\to\quad
  \begin{permutation}
    1&2&3\\\downarrow&\downarrow&\downarrow\\1&2&3
  \end{permutation}\triangleleft(12)\triangleleft(23)=
  \begin{permutation}
    1&2&3\\\downarrow&\downarrow&\downarrow\\2&3&1
  \end{permutation} \ .
\end{align}
Finally, the last equality eqn.~\eqref{eq:RRd3} does not change the vacuum: it
realises the desired permutation by a different combination of swaps compared
to eqn.~\eqref{eq:RRdlhs}. Instead of acting with the cycle $(13)$ from the
left, one acts with the same cycle from the right, which is the same for the
trivial permutation:
\begin{align}
  \rop_{23}\rop_{13}\dl{1}\dl{2}\dlb{3}\quad\to\quad(23)\triangleright
  \begin{permutation}
    1&2&3\\\downarrow&\downarrow&\downarrow\\1&2&3
  \end{permutation}\triangleleft(13)=
  \begin{permutation}
    1&2&3\\\downarrow&\downarrow&\downarrow\\2&3&1
  \end{permutation}\,.
\end{align}
Similar relations for higher-point invariants encoding the dihedral symmetries
can be translated into permutations with equal ease.

\paragraph{Reflection and Parity} While the relations considered above refer to
the possibility to replace certain subchains of $\rop$-chains without changing
the Yangian invariant which is represented, the operations described in
eqns.~\eqref{eq:reflection_param} and \eqref{eq:parityflip} act on all
$\rop$-operators in the chain and implement reflection and parity-flip, which
together with the shift operation discussed in \subsecref{subsubsec:SymAmp}
constitute the dihedral symmetry of the amplitudes. Those operations map an
eigenstate of the monodromy matrix onto another eigenstate, which, however,
represents another Yangian invariant. Both parity-flip and reflection
do as well have a natural interpretation in terms of permutations: 
\begin{itemize}
  \item \textbf{Reflection} -- as described in eqn.~\eqref{eq:reflection_param}
      corresponds to writing the labels of the external legs in opposite
      direction. In terms of translating an on-shell graph into
      $\rop$-operators one will now have to replace ``clockwise'' by
      ``counterclockwise'' and vice versa where appropriate. Analogously this is
      true for translating a $\rop$-chain into permutations: in deciding,
      whether a cycle corresponding to an operator acts from the left or from
      the right, one has to swap the notions. 
      \[
      \begin{array}{cc}
	\includegraphicsboxex{Fig_5pt2.mps}&
	\includegraphicsboxex{Fig_5ptref1.mps}	
	\\[30pt]
	\rop_{21}\rop_{32}\rop_{21}\rop_{42}\rop_{21}\rop_{52}\,
	  \dlb{1}\dlb{2}\dl{3}\dl{4}\dl{5} & 
	\rop_{45}\rop_{34}\rop_{45}\rop_{24}\rop_{45}\rop_{14}\,
	\dl{1}\dl{2}\dl{3}\dlb{4}\dlb{5}
	\\[10pt]
        \begin{pmatrix}
          1 & 2 & 3 & 4 & 5\\
          \downarrow & \downarrow & \downarrow & \downarrow & \downarrow\\
          3 & 4 & 5 & 1 & 2 
        \end{pmatrix}
        &
        \begin{pmatrix}
          1 & 2 & 3 & 4 & 5\\
          \downarrow & \downarrow & \downarrow & \downarrow & \downarrow\\
          4 & 5 & 1 & 2 & 3
        \end{pmatrix}
      \end{array}
      \]
      The diagrams on the right-hand side are still the ones corresponding to a
      five-point MHV amplitude. Thus, deducing the MHV sector from the
      permutation -- which is a cyclic shift by $k=3$ here -- is possible only
      for the clockwise ordering of the legs. Translating the $\rop$-chain on
      the right-hand side back into an on-shell diagram taking care for the
      different orientation will \emph{not} lead to the first diagram on the
      right-hand side, which is, however, equal to the (expected) second
      diagram after a merger operation (see \figref{fig:moves}).
  \item \textbf{Shift} While shifting labels of external legs is a rather
    involved operation in terms of $\rop$-chains (see the discussion in
    \subsecref{subsubsec:SymAmp}), it has a straightforward interpretation in
    terms of on-shell graphs. One can easily check that the permutations
    encoded in the following two on-shell graphs are equivalent: 
      \[
      \begin{array}{cc}
	\includegraphicsboxex{Fig_5pt2.mps}&\includegraphicsboxex{Fig_5ptshift.mps}
	\\[30pt]
	\rop_{21}\rop_{32}\rop_{21}\rop_{42}\rop_{21}\rop_{52}\,
	  \dlb{1}\dlb{2}\dl{3}\dl{4}\dl{5} & 
	\rop_{32}\rop_{43}\rop_{32}\rop_{53}\rop_{32}\rop_{13}\,
	  \dl{1}\dlb{2}\dlb{3}\dl{4}\dl{5}
	\\[10pt]
        \begin{pmatrix}
          1 & 2 & 3 & 4 & 5\\
          \downarrow & \downarrow & \downarrow & \downarrow & \downarrow\\
          3 & 4 & 5 & 1 & 2 
        \end{pmatrix}
        &
        \begin{pmatrix}
          1 & 2 & 3 & 4 & 5\\
          \downarrow & \downarrow & \downarrow & \downarrow & \downarrow\\
          3 & 4 & 5 & 1 & 2
        \end{pmatrix}
      \end{array}
      \]
  \item As discussed in \subsecref{subsubsec:SymAmp}, the \textbf{parity-flip}
    operation consists of the following map: 
      \begin{align}
      &\rop_{ab}\rightarrow\rop_{ba}, &&\dl{a}\leftrightarrow\dlb{a}
      \,.
      \end{align} 
      Considering the identifications eqns.~\eqref{eq:AttachExternal} as well
      as \eqref{eq:BCFWbridge}, it is clear that parity swaps the r\^ole of
      black and white dots in an on-shell diagram. If one keeps the cyclical
      ordering of the external legs, this amounts to inverting the permutation,
      because one will now have to turn left (right) instead of right (left) at
      each vertex. 
      Naturally, we could have been arriving at the same conclusion from the
      considering the $\rop$-chain. Swapping the positions of the indices for
      each $\rop$-operator converts the a left into right action and vice
      versa, which immediately leads to a permutation describing a cyclic shift
      in the opposite direction. 

      Let us see, how this works in terms of the five-point tree-level amplitude:\\
      \[
      \begin{array}{cc}
	\includegraphicsboxex{Fig_5pt2.mps}&
	\includegraphicsboxex{Fig_5ptpar1.mps}\equiv
	\includegraphicsboxex{Fig_5ptpar2.mps}
	\\[30pt]
	\rop_{21}\rop_{32}\rop_{21}\rop_{42}\rop_{21}\rop_{52}\,
	  \dlb{1}\dlb{2}\dl{3}\dl{4}\dl{5} & 
	\rop_{12}\rop_{23}\rop_{12}\rop_{24}\rop_{12}\rop_{25}\,
	  \dl{1}\dl{2}\dlb{3}\dlb{4}\dlb{5}
	\\[10pt]
        \begin{pmatrix}
          1 & 2 & 3 & 4 & 5\\
          \downarrow & \downarrow & \downarrow & \downarrow & \downarrow\\
          3 & 4 & 5 & 1 & 2 
        \end{pmatrix}
        &
        \begin{pmatrix}
          1 & 2 & 3 & 4 & 5\\
          \downarrow & \downarrow & \downarrow & \downarrow & \downarrow\\
          4 & 5 & 1 & 2 & 3
        \end{pmatrix}
      \end{array}
      \]
      Employing eqn.~\eqref{eq:nk}, one finds $k=3$ and can thus identify
      the diagrams on the right-hand side as the ones corresponding to the
      five-point $\overline{\text{MHV}}$ amplitude.
      
      The combination of reflection and parity transformation described in
      eqn.~\eqref{eq:refpar} can be investigated in a similar fashion.
      Combining the findings from the individual transformations above, one can
      easily predict the result: one will obtain an on-shell graph with four
      black dots, three white dots and counterclockwise ordering of legs, which
      encodes the permutation for a MHV amplitude. 
  \end{itemize}

\subsubsection{A simple way to construct $\rop$-chains for top-cells}

As discussed in section \subsecref{subsec:onshell}, so called top-graphs (or
top-cells) correspond to cyclic shifts by the variable $k$ labelling the
MHV sector. Using for example $\rop$-operators of the form in
eqn.~\eqref{eq:BCFWbridgeDL} one can immediately construct a representative
$\rop$-chain by successively commuting particles to the right. For the
permutations corresponding the five-point MHV amplitude $(k=2)$ one finds:
\begin{equation}
  \begin{permutation}
   1 & 2 & 3 & 4 & 5\\
   \downarrow & \downarrow & \downarrow & \downarrow & \downarrow\\
   3 & 4 & 5 & 1 & 2 
  \end{permutation}
  =
  (23)\triangleright(34)\triangleright(45)\triangleright(12)\triangleright(23)\triangleright(34)\triangleright
  \begin{permutation}
    1 & 2 & 3 & 4 & 5\\
    \downarrow & \downarrow & \downarrow & \downarrow & \downarrow\\
    1 & 2 & 3 & 4 & 5 
   \end{permutation}\,,
\end{equation}
while the general version for the top-cell for an amplitude ${\cal A}_{n;k}$ reads
\[
  \underbrace{\rop_{k\,\,k+1}\ldots\rop_{n-1\,\,n}}\;
  \underbrace{\rop_{k-1\,\,k}\ldots\rop_{n-2\,\,n-1}}
  \;\ldots\;
  \underbrace{\rop_{12}\ldots\rop_{n-k\,\,n-k+1}}\,
  \dl{1}\ldots\dl{(n-k)}\dlb{(n-k+1)}\ldots\dlb{n} \ .
\]
The above state is a manifest eigenstate of the monodromy matrix because applying the
relations eqn.~\eqref{eq:RLYBE} is trivial: the indices of all $\rop$-operators
are in the succession suitable for permuting the monodromy matrix $T$ through
all $\rop$'s.

\subsection{Yangian invariance of the deformed six-point NMHV amplitude?}
\label{subsec:6ptNMHV}

In \cite{Beisert:2014qba} it was pointed out that the only amplitude outside
the MHV sector which can be deformed in a Yangian-invariant way is the
six-point NMHV amplitude.  Can one reproduce this result using the
$\rop$-operator formulation?  

The undeformed six-point NMHV amplitude is
composed from three BCFW-channels, which can be chosen to be represented by the
permutations (\textit{cf.} \figref{fig:6decomp}):
\begin{equation}
  \begin{permutation}
   1 & 2 & 3 & 4 & 5 & 6\\
   \downarrow & \downarrow & \downarrow & \downarrow & \downarrow & \downarrow\\
   4 & 5 & 6 & 7 & 9 & 8 
  \end{permutation}\,,\quad
  \begin{permutation}
   1 & 2 & 3 & 4 & 5 & 6\\
   \downarrow & \downarrow & \downarrow & \downarrow & \downarrow & \downarrow\\
   3 & 5 & 6 & 7 & 8 & 10
 \end{permutation}\,,\quad\text{and}\quad
  \begin{permutation}
   1 & 2 & 3 & 4 & 5 & 6\\
   \downarrow & \downarrow & \downarrow & \downarrow & \downarrow & \downarrow\\
   5 & 4 & 6 & 7 & 8 & 9 
  \end{permutation}\,.
\end{equation}
In terms of $\rop$-chains, a suitable representation reads
\begin{subequations}
\label{eq:6ptBCFWRchains}
\begin{align}
  \label{eq:BCFWchan1}
  &\rop_{34}\rop_{45}\rop_{23}\rop_{34}\rop_{21}\rop_{31}\rop_{64}\rop_{16}\cdot \dl{1}\dl{2}\dl{3}\dlb{4}\dlb{5}\dlb{6}\\
  &\rop_{34}\rop_{45}\rop_{23}\rop_{34}\rop_{21}\rop_{31}\rop_{65}\rop_{64}\cdot \dlb{1}\dl{2}\dl{3}\dlb{4}\dlb{5}\dl{6}\\
  &\rop_{34}\rop_{45}\rop_{23}\rop_{34}\rop_{31}\rop_{65}\rop_{64}\rop_{16}\cdot \dl{1}\dl{2}\dl{3}\dlb{4}\dlb{5}\dlb{6}\,.
\end{align}
\end{subequations}
These three channels can be inferred from the following top-cell
\begin{align}
  \label{eq:topcell}
  \rop_{34}\rop_{45}\rop_{23}\rop_{34}\rop_{21}\rop_{31}\rop_{65}\rop_{64}\rop_{16}\cdot
  \dl{1}\dl{2}\dl{3}\dlb{4}\dlb{5}\dlb{6}
\end{align}
by omitting the $\rop$-operators $\rop_{65}$, $\rop_{16}$ and $\rop_{21}$ at
positions $7$, $9$ and $5$ respectively. While the omissions lead to the subchains corresponding to the singularities of the top-cell in the particular representation here, this is not true for other representations. 
\begin{figure}
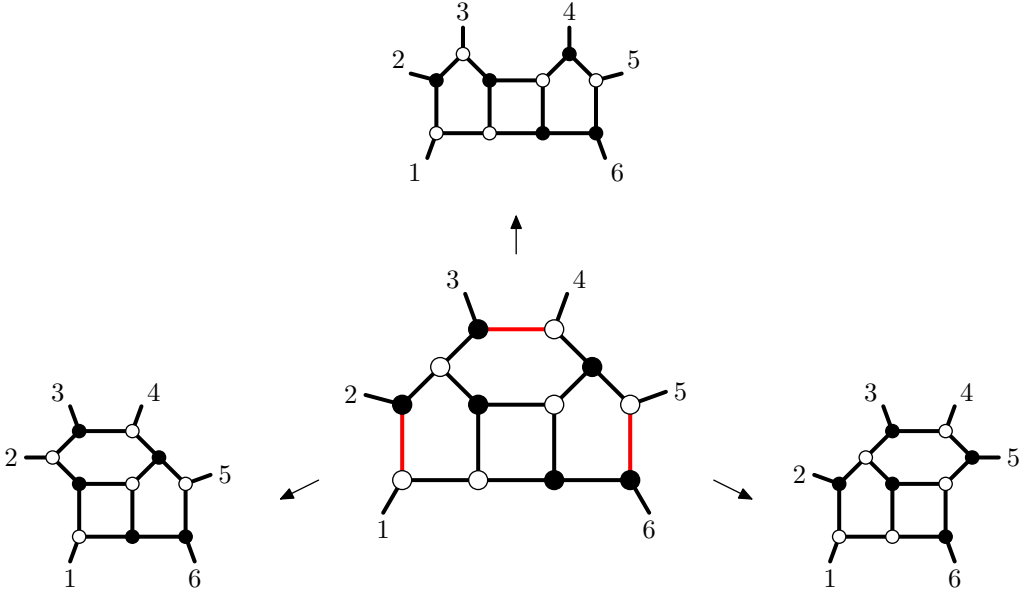

  \centering
  \includegraphicsboxex{Fig_6pttc.mps}
  \caption{Decaying the six-point NMHV top-cell into three BCFW-channels.}
\label{fig:topcell_decay}
\end{figure}
For the undeformed amplitude, it was pointed out in ref.~\cite{Beisert:2014qba} that
the conditions for Yangian invariance of the top-cell eqn.~\eqref{eq:topcell}
are equivalent to imposing Yangian invariance on the three channels
eqns.~\eqref{eq:6ptBCFWRchains}. Thus, Yangian invariance is compatible with
decaying the top-cell into individual BCFW channels as long as no deformation
is present. 

Considering a deformed amplitude, there seems to be an apparent clash: while
imposing Yangian invariance on the top-cell leaves six free deformation
parameters, demanding Yangian invariance for the three BCFW channels
\emph{simultaneously} leads to only three free deformation parameters:
\[
  \label{eq:topc_uc}
  u_5\,=u_4 \ ,\qquad u_2\,=u_3 \ , \qquad u_1\,=u_6 \ .
\]
The resolution is simple: taking a residue of the top-cell, that is ``removing
a line from the graph in \figref{fig:topcell_decay}'' is possible only if there
is no flow of central charge along the line to be removed. The three conditions
for the simultaneous vanishing of the central charges along the \emph{red}
lines delivers the three additional conditions leading to
eqn.~\eqref{eq:topc_uc}.

What is the analogue of this construction in the $\rop$-operator language?  In
\subsecref{subsec:sixpointinvariants} it was discussed that there are exactly
six different NMHV Yangian invariants with six external legs, which are listed
in eqn.~\eqref{eq:NMHV6}. As pointed out after
eqn.~\eqref{eq:6ptcentralcharges}, the three eigenfunctions corresponding to the
BCFW channels in eqn.~\eqref{eq:6ptBCFWRchains}, $\mathcal{Y}_1^{(6)}$,
$\mathcal{Y}_3^{(6)}$ and $\mathcal{Y}_5^{(6)}$, have different central charges,
and thus different eigenvalues. Since a sum of Yangian invariants is only
well-defined if all of them belong to the same eigenspace of the monodromy
matrix, one has to identify the spectral parameters as in
eqn.~\eqref{eq:identifyspectral}. This is exactly the condition in
eqn.~\eqref{eq:topc_uc}.

Thus ensuring Yangian invariance by demanding the BCFW channels to be
eigenstates of the monodromy matrix is not sufficient: if a deformed amplitude
is composed from several deformed Yangian invariants, one has to ensure that
they all are in the same eigenspace.  

\subsection{A peek at the seven-point NMHV amplitude}

In order to show the validity of the $\rop$-operator formalism, let us consider
a more difficult example: the seven-point NMHV amplitude. The on-shell graphs
derived from a standard BCFW-decomposition employing a shift of legs $1$ and $7$
correspond to the following strings of R-operators:
\begin{equation}
  \label{eq:7bcfw}
  \begin{aligned}
    &\rop_{76} \, \rop_{71} \, \rop_{32} \, \rop_{12} \, \rop_{21} \,
    \rop_{23} \, \rop_{61} \, \rop_{34} \, \rop_{53} \, \rop_{36} \,
    \cdot
    \dlb{1}\dl{2}\dl{3}\dlb{4}\dl{5}\dlb{6}\dl{7} \ ,\\
   & \rop_{76} \, \rop_{71} \, \rop_{16} \, \rop_{12} \, \rop_{21} \,
    \rop_{23} \, \rop_{61} \, \rop_{34} \, \rop_{53} \, \rop_{36} \,
    \cdot
    \dlb{1}\dl{2}\dl{3}\dlb{4}\dl{5}\dlb{6}\dl{7} \ ,\\
    &\rop_{76} \, \rop_{71} \, \rop_{16} \, \rop_{32} \, \rop_{21} \,
    \rop_{23} \, \rop_{61} \, \rop_{34} \, \rop_{53} \, \rop_{36} \,
    \cdot
    \dlb{1}\dl{2}\dl{3}\dlb{4}\dl{5}\dlb{6}\dl{7} \ ,\\
    &\rop_{43} \, \rop_{65} \, \rop_{54} \, \rop_{43} \, \rop_{56} \,
    \rop_{67} \, \rop_{16} \, \rop_{23} \, \rop_{42} \, \rop_{26} \,
    \cdot
    \dl{1}\dl{2}\dlb{3}\dl{4}\dl{5}\dlb{6}\dlb{7} \ ,\\
    &\rop_{23} \, \rop_{45} \, \rop_{34} \, \rop_{54} \, \rop_{56} \,
    \rop_{64} \, \rop_{67} \, \rop_{16} \, \rop_{32} \, \rop_{26} \,
    \cdot
    \dl{1}\dl{2}\dl{3}\dlb{4}\dl{5}\dlb{6}\dlb{7} \ ,\\
    &\rop_{45} \, \rop_{56} \, \rop_{71} \, \rop_{76} \, \rop_{43} \,
    \rop_{53} \, \rop_{36} \, \rop_{73} \, \rop_{31} \, \rop_{32} \,
    \cdot
    \dlb{1}\dlb{2}\dl{3}\dl{4}\dl{5}\dlb{6}\dl{7}\,.
  \end{aligned}
\end{equation}
We have checked that the combination of the above channels does indeed yield
the seven-point NMHV superamplitude by comparing with the explicit expressions
in \cite{Britto:2004ap}. In addition, we performed automated tests and found
complete agreement with the results obtained from the Mathematica package
\verb!GGT!  described in ref.~\cite{Dixon:2010ik}.

While the tree-amplitude is a nontrivial check, one can ask, whether it is
possible to obtain all box coefficients for the one-loop seven-point NMHV
spelled out in ref.~\cite{Bern:2004ky} from the top-cell in the
$\rop$-formalism. Finding the BCFW-channels refers to determining all
codimension-two boundaries of the top-cell which -- in turn -- amounts to
omitting \emph{two} $\rop$'s in a suitable representation of the top-cell. A
suitable representation reads 
\begin{equation}
  \label{eq:7tc_rs}
  \rop_{65}\, \rop_{76}\, \rop_{71}\, \rop_{54}\,
  \rop_{65}\, \rop_{43}\, \rop_{45}\, \rop_{61}\,
  \rop_{15}\, \rop_{41}\, \rop_{13}\, \rop_{21}\,\cdot
  \dlb{1} \, \dl{2} \, \dlb{3} \,
  \dl{4} \, \dlb{5} \, \dl{6} \, \dl{7} \,
\end{equation}
and leads to 66 $\rop$-chains of length 10. While some of them do not contain
all seven indices, and thus do not connect all particles, others do not allow
to solve the integral for general kinematics (see \subsecref{subsec:vacuum} for
a discussion). Discarding those, one is left with $20$ valid channels, which
turn out to deliver a spanning set for all box coefficients calculated in
ref.~\cite{Bern:2004ky}. 

\section{Conclusions}
\label{sec:conclusions}

In this article we have explored the $\rop$-operator formalism introduced in
ref.~\cite{Chicherin:2013ora}. We have elucidated the underlying algebraic
structures that are related to the Yangian algebra of $\alg{gl}(4|4)$. In
particular, considering the S- and R-matrix in the fundamental and functional
representation implies the forms of the Lax-operator $\lop$ and the
R-operator $\rop$ in eqns.~\eqref{eq:Ldef} and \eqref{eq:Rdef}, respectively.
Via the monodromy matrix $\mathrm{T}$ \eqref{eq:Monodromy}, the Lax operator
then generates the generators of the Yangian symmetry of $n$-particle states.
In this language, Yangian invariants are defined to be eigenfunctions of the monodromy
matrix.
 
One simple eigenfunction of the monodromy matrix is given by a product of
$\delta$-functions in the spinor-helicity variables \eqref{eq:Groundstate}.
From this ground state we can generate Yangian invariants by acting on it with
a sequence of R-operators. From the Yang--Baxter equation \eqref{eq:RLYBE}
relating $\rop$ and $\lop$, we see that functions of this form naturally give
rise to eigenfunctions.

It would be extremely interesting to derive the explicit expressions of the L-
and R-operator from the universal R-matrix. Our discussion in
\secref{sec:algebra} indicates that there is a universal R-matrix underlying
this construction of Yangian invariants. It would already be useful to
explicitly work out the constructions for $\alg{gl}(2)$ or $\alg{gl}(1|1)$ to
gain deeper insights in the algebraic structures that are involved.

We have classified all eigenstates up to six particles and we found that to each
Yangian invariant in $\superN=4$ sYM there corresponds a unique eigenfunction of
the monodromy matrix. Furthermore, each Yangian invariant is determined by its
central charges, which are of the form of a difference between spectral
parameters \eqref{eq:CviaU}. Using this identity, the central charges can be
related to permutations, which establishes a bijection between permutations and
Yangian invariants.

Subsequently, we have been identifying the maps between the $\rop$-operator
formulation of Yangian invariants and their formulations in terms on on-shell
graphs and permutations. All properties of Yangian invariants as well as those
of the amplitudes built thereof can be given an immediate interpretation: while
cyclicity, reflection and flip of parity are very natural in the $\rop$-operator
language, constraints on possible deformation parameters are realised by
demanding an Yangian-invariant object to be an eigenstate of the monodromy
matrix, which is built from Lax-operators. The result for building amplitudes
with Yangian-invariant deformations is in agreement with \cite{Beisert:2014qba}:
Yangian-invariant deformations are not compatible with the BCFW-construction of
amplitudes.

The $\rop$-operator language bridges the gap between the common
formulations of scattering amplitudes and the formalism employed in integrable
systems. The relation becomes even more evident in ref.~\cite{Kanning:2014maa}.

There is, however, a drawback of the $\rop$-operator formalism: the construction
is based on evaluating integrals on the support of delta functions. Since one
can perform the integrals in any succession, and the arguments of the delta
functions can be reversed without modifying the physical situation, there is a
sign ambiguity related to each Yangian invariant.  Although the formalism
delivers the Yangian invariants, it is not clear, in which way they need to be
combined. Naturally, this can be fixed by demanding the vanishing of spurious
poles, nevertheless, one would rather like to have a purely constructional
method. One way to realise this could be an algebraic condition singling out
BCFW-channels.

While the current article deals with tree amplitudes exclusively, there is
hope for extending the formalism to loop amplitudes.  Similar to the
forward-limit construction in \cite{ArkaniHamed:2012nw}, one will have to
identify two legs in order to build the $\rop$-chain corresponding to a
loop-amplitude. Nevertheless, the integration procedure will not be as
straightforward as in the tree-level case, because infrared divergences will
appear.

\paragraph{Acknowledgements.}

We would like to thank Niklas Beisert, James Drummond, Henrik Johansson, Nils
Kanning, Yumi Ko, Jan Plefka, Matthias Staudacher and Cristian Vergu for useful
discussions.  The work of MdL and MR is partially supported by grant no.\
200021-137616 from the Swiss National Science Foundation.

\appendix
\section{Momentum twistors and algebraic approach}

In ref.~\cite{Chicherin:2013ora}, the authors proposed an extension of their
algebraic approach to scattering amplitudes to the momentum twistor
formulation. They define the $\rop_{i j}$-operator acting on the space of
functions on momentum twistor space as\footnote{%
  We thank Cristian Vergu and James Drummond for suggesting the investigation
  of this point to us.  }
\begin{equation}
  \label{eq:rop_momtw}
  \rop_{i j} (u) \cdot \mathcal{F} \bigl( \mtw_i, \mtw_j \bigr)
  := \int \frac{\mathrm{d}z}{z^{1+u}}\, \mathcal{F} \bigl( \mtw_i - z \mtw_j, \mtw_j \bigr) \ ;
\end{equation}
in the following we will always consider $\rop$-operators whose spectral
parameter $u$ is set to zero.
Subsequently, the authors show that the following identity holds:
\begin{equation}
  \label{eq:rinv_chi}
  \mathcal{I} =  \rop_{3 4} \rop_{2 3} \rop_{1 2} \rop_{* 1} \delta^{4|4} (\mtw_* ) = [*,1,2,3,4] \ .
\end{equation}
An important remark is due here: this is \emph{not} the formulation one obtains
from translating the on-shell superspace approach directly into momentum
twistor variables. While a first proof of this is the non-locality of the BCFW
shift in momentum twistor space (whereas this $\rop$-operator acts on two sites
only), another proof is the fact that the vacuum of the on-shell superspace
approach does not translate into anything nice in momentum twistor space. While
this could be a formulation equivalent to the Grassmannian formula in momentum
twistor space, the dictionary is not direct.

The explicit computation of \eqref{eq:rinv_chi} is straightforward. From
\begin{equation}
  \label{eq:int1}
  \mathcal{I} = \int \biggl[\prod_{i=1}^4 \frac{\mathrm{d} z_i}{z_i} \biggr]
  \,\delta^{4|4}(\mtw_* - z_1 \mtw_1 + z_1 z_2 \mtw_2 - z_1 z_2 z_3 \mtw_3 + z_1 z_2 z_3 z_4 \mtw_4 )
\end{equation}
we can change variables to 
\begin{equation}
  \label{eq:changevar}
  a_i:=  (-1)^{i+1} \prod_{k=1}^i  z_k
\end{equation}
and the (inverse) Jacobian is $|\mathrm{det}\,J^{-1}| = a_1 a_2 a_3$. This leads to 
\begin{equation}
  \label{eq:int2}
  \mathcal{I} = \int\biggl[\prod_{i=1}^4 \frac{\mathrm{d} a_i}{a_i} \biggr] \delta^{4|4}(\mtw_* - \sum_i a_i \mtw_i)\ .
\end{equation}
We can now solve the integrals by localising them on the support of the bosonic
delta functions. In order to do so, we rewrite the bosonic delta functions by
dotting into the arguments\footnote{%
  Here, $\epsilon^\alpha(i,j,k,\bullet) := \epsilon_{\beta\gamma\delta\alpha}
  W_i^\beta W_j^\gamma W_k^\delta $ and $\langle ijkl\rangle =
  \epsilon_{\alpha\beta\gamma\delta} W_i^\alpha W_j^\beta W_k^\gamma W_l^\delta
  $.} %
$\epsilon^\alpha(1,2,3,\bullet),\epsilon^\alpha(1,2,4,\bullet),\epsilon^\alpha(1,3,4,\bullet),\epsilon^\alpha(2,3,4,\bullet)$:
\begin{equation}
  \label{eq:changedelta}
  \delta^4 (W_*-\sum_i W_i) = \frac{1}{\langle 1234\rangle}
  \delta\biggl(a_1 - \frac{\langle *234 \rangle}{\langle 1234 \rangle} \biggr)
  \delta\biggl(a_2 - \frac{\langle *134 \rangle}{\langle 1234  \rangle} \biggr)
  \delta\biggl(a_3 - \frac{\langle *124 \rangle}{\langle 1234  \rangle} \biggr)
  \delta\biggl(a_4 - \frac{\langle *123 \rangle}{\langle 1234  \rangle} \biggr)  \ .
\end{equation}
The result reads
\begin{equation}
  \label{eq:int3}
  \mathcal{I} = \frac{\delta^{0|4} \bigl(
      \langle 1234 \rangle \chi_* 
    + \langle 234*{} \rangle \chi_1
    + \langle 34*{}1 \rangle \chi_2
    + \langle 4* 12 \rangle \chi_3
    + \langle * 123 \rangle \chi_4
    \bigr) }
  { \langle 1234 \rangle 
    \langle 234*  \rangle 
    \langle 34* 1 \rangle 
    \langle 4* 12 \rangle 
    \langle * 123 \rangle 
  }
  \equiv
  [*,1,2,3,4] \ .
\end{equation}
This result is not unexpected, because the expression in eqn.~\eqref{eq:int2} is
-- if we allow ourselves to be not too rigorous with the integration measure --
the form of the NMHV $\mathcal{R}$-invariants as an integral over
$\mathbb{CP}^5$ first introduced by Mason and Skinner in
ref.~\cite{Mason:2009qx}. The construction of eqn.~\eqref{eq:rinv_chi} therefore
leads ``trivially'' to the correct NMHV $\mathcal{R}$-invariant.

An interesting problem is to understand whether it is possible to extend this
construction to higher-level $\mathcal{R}$-invariants. It is possible in fact to
express dual superconformal invariants in terms of residue integrals over
suitable Grassmannian manifolds (as shown in ref.~\cite{Mason:2009qx} and then
exploited in ref.~\cite{ArkaniHamed:2009vw} to show the equivalence between the
ordinary Grassmannian formulation and the dual Grassmannian formulation).
However, the problem is that there is no interpretation of the shift in
eqn.~\eqref{eq:rop_momtw}, neither in BCFW nor in Risager form. It would be
interesting to analyse whether it is possible to find a map from MHV diagrams
in momentum-twistor space to these $\rop$-operators, analogous to the
correspondence to on-shell graphs (and BCFW channels) described in the main
part of the article.

Nevertheless, it is possible to ``engineer'' a set of $\rop$-operators that lead
to the expression for $k=4$ $\mathcal{R}$-invariants\footnote{%
  Here, $k$ is the MHV level, that is the amplitude has Grassmann weight $4k$;
  in ref.\cite{Mason:2009qx} the $k$ is the ``reduced'' level, equal to our
  $k-2$. In our language, the $\mathcal{R}$-invariants thus have Grassmann
  weight $4(k-2)$.  } %
(eqn.~(55) in ref.~\cite{Mason:2009qx})
\begin{equation}
  \label{eq:rinvk4}
  \mathcal{R}_{k,n} := \frac{1}{(2\pi i)^{(k-2)(n-k-2)}} \oint_{\mathrm{\Gamma}\subset G(k-2,n)}\!\!\!\!\!
  \mathrm{d}\mu\, \prod_{i=1}^{k-2} \delta^{4|4} \biggl( \sum_{j=1}^n C_{i j} \mtw_j \biggr) \ ,
\end{equation}
where $C_{lj}$ is the $(k-2)\times n$ matrix representing a point on $G(k-2,n)$
and the measure is the ``usual'' measure for amplitude-related Grassmannian
integrals, that is
\begin{equation}
  \label{eq:meas}
  \mathrm{d}\mu = \frac{\mathrm{d}^{k\times n} c_{ab}}{\mathrm{Vol}[\grp{GL}(k)]}
  \frac{1}{ (12\dots k) (23\dots k+1) \dots (n\dots k-1)} \ .
\end{equation}

The easiest example with $k=4$ is the $6$-point $\mathcal{R}$-invariant, since
the corresponding integral is completely localised on the support of the delta
functions, and should lead to the $\mathrm{N}^2\mathrm{MHV}$ (or
$\overline{\mathrm{MHV}}$) $6$-point amplitude. We can show that the correct
expression is
\begin{equation}
  \label{eq:rstr_r2}
  \mathrm{R}^{(8)}_{34}\,\mathrm{R}^{(7)}_{23}\,
  \mathrm{R}^{(6)}_{16}\,\mathrm{R}^{(5)}_{34}\,
  \mathrm{R}^{(4)}_{63}\,\mathrm{R}^{(3)}_{56}\,
  \mathrm{R}^{(2)}_{25}\,\mathrm{R}^{(1)}_{15}\,
  \delta^{4|4}(\mtw_1) \,\delta^{4|4}(\mtw_2)\ ,
\end{equation}
where the superscript $(i)$ indicates that the integral associated to this
$\rop$-operator is over the variable $z_i$. It corresponds to the integral
\begin{equation}
  \label{eq:r2int}
  \begin{aligned}
    I = \int \biggl[\prod_{i=1}^8 \frac{\mathrm{d}z_i}{z_i} \biggr] \delta^{4|4}
    \Bigl[ \mtw_1 - z_1 \mtw_5 - z_1 z_3 z_4 \mtw_3 + (z_1 z_3 - z_6) \mtw_6 +
    z_1 z_3 z_4 (z_5+z_8)\mtw_4 \Bigr]
    \times \\
    \delta^{4|4}\Bigl[ \mtw_2 - z_2 \mtw_5 + z_2 z_3 \mtw_6 + (-z_2 z_3 z_4 -
    z_7) \mtw_3 + [z_2 z_3 z_4 (z_5+z_8) + z_7 z_8 ]\mtw_4 \Bigr] \ .
  \end{aligned}
\end{equation}
With the obvious change of variables
\[
c_{i j} := (\text{coefficient of } \mtw_j \text{ in the $i$-th delta function})
\nn
\]
one obtains
\begin{equation}
  \label{eq:res}
  I = \int \frac{\mathrm{d}^{2 \times 4}c_{a b}}
  {c_{13} c_{26} (c_{13}c_{24} - c_{14} c_{23}) (c_{15}c_{24} - c_{14} c_{25}) (c_{15}c_{26} - c_{16} c_{25})}
  \prod_{i=1,2} \delta^{4|4} (\sum_{j=1}^6 c_{i j} \mtw_j)\ ,
\end{equation}
which is exactly eqn.~\eqref{eq:rinvk4} with the $\grp{GL}(2)$ freedom used to
fix the first two columns of $c_{ab}$ to the identity, that is
\begin{equation}
  \label{eq:cmat}
  c_{ab} =
  \begin{pmatrix}
    1 & 0 & c_{13} & c_{14} & c_{15} & c_{16} \\
    0 & 1 & c_{23} & c_{24} & c_{25} & c_{26}
  \end{pmatrix} \ .
\end{equation}
This is, however, an \emph{ad hoc} construction, engineered to qualitatively
match the integral description of R-invariants of ref.~\cite{Mason:2009qx}. What
conclusions can thus be drawn from here?  

First of all, it is not clear in this momentum-twistor approach what the vacuum
state $\delta^{4|4}(\mathcal{W})$ is, even though it is pretty clear that one
such state should be associated to each off-shell leg in a given MHV diagram. It
also seems that the specific choice of a vacuum corresponds to a particular
way of fixing the $\grp{GL}(k)$ redundancy in the integral over the
Grassmannian.

Secondly, it would be nice to have a map from MHV diagrams (which have a very
natural description in momentum twistor space \cite{Bullimore2010}) to
$\rop$-chains acting on momentum-twistor space, but it is currently not clear,
whether this map exists. Moreover, the translation seems to get rather
complicated starting from the NNMHV level, since the MHV diagrams in the
expansion have an increasingly complicated topology.

\begin{bibtex}[\jobname]

@article{Kanning:2014maa,
  author         = "Kanning, Nils and Lukowski, Tomasz and Staudacher,
                    Matthias",
  title          = "{A Shortcut to General Tree-level Scattering Amplitudes
                    in N=4 SYM via Integrability}",
  year           = "2014",
  eprint         = "1403.3382",
  archivePrefix  = "arXiv",
  primaryClass   = "hep-th",
  SLACcitation   = "
}

@article{ArkaniHamed:2009dg,
  author        = {Arkani-Hamed, Nima and Bourjaily, Jacob and Cachazo, Freddy and Trnka, Jaroslav},
  title         = {{Unification of residues and Grassmannian dualities}},
  journal       = {JHEP},
  year          = {2011},
  volume        = {1101},
  pages         = {049},
  archiveprefix = {arXiv},
  doi           = {10.1007/JHEP01(2011)049},
  eprint        = {0912.4912},
  primaryclass  = {hep-th},
  slaccitation  = {
}

@article{ArkaniHamed:2010kv,
  author        = {Arkani-Hamed, Nima and Bourjaily, Jacob L. and Cachazo, Freddy and Caron-Huot, Simon and Trnka, Jaroslav},
  title         = {{The all-loop integrand for scattering amplitudes in planar N=4 SYM}},
  journal       = {JHEP},
  year          = {2011},
  volume        = {1101},
  pages         = {041},
  archiveprefix = {arXiv},
  doi           = {10.1007/JHEP01(2011)041},
  eprint        = {1008.2958},
  primaryclass  = {hep-th},
  slaccitation  = {
}

@article{ArkaniHamed:2012nw,
  author        = {Arkani-Hamed, Nima and Bourjaily, Jacob L. and Cachazo, Freddy and Goncharov, Alexander B. and Postnikov, Alexander and others},
  title         = {{Scattering amplitudes and the positive Grassmannian}},
  year          = {2012},
  archiveprefix = {arXiv},
  eprint        = {1212.5605},
  primaryclass  = {hep-th},
  slaccitation  = {
}

@article{ArkaniHamed:2009vw,
  author        = {Arkani-Hamed, Nima and Cachazo, Freddy and Cheung, Clifford},
  title         = {{The Grassmannian origin of dual superconformal invariance}},
  journal       = {JHEP},
  year          = {2010},
  volume        = {1003},
  pages         = {036},
  archiveprefix = {arXiv},
  doi           = {10.1007/JHEP03(2010)036},
  eprint        = {0909.0483},
  primaryclass  = {hep-th},
  slaccitation  = {
}

@article{ArkaniHamed:2009dn,
  author        = {Arkani-Hamed, Nima and Cachazo, Freddy and Cheung, Clifford and Kaplan, Jared},
  title         = {{A duality for the S matrix}},
  journal       = {JHEP},
  year          = {2010},
  volume        = {1003},
  pages         = {020},
  archiveprefix = {arXiv},
  doi           = {10.1007/JHEP03(2010)020},
  eprint        = {0907.5418},
  primaryclass  = {hep-th},
  slaccitation  = {
}

@article{ArkaniHamed:2008gz,
  author        = {Arkani-Hamed, Nima and Cachazo, Freddy and Kaplan, Jared},
  title         = {{What is the simplest Quantum Field Theory?}},
  journal       = {JHEP},
  year          = {2010},
  volume        = {1009},
  pages         = {016},
  archiveprefix = {arXiv},
  doi           = {10.1007/JHEP09(2010)016},
  eprint        = {0808.1446},
  primaryclass  = {hep-th},
  slaccitation  = {
}

@article{Beisert:2014qba,
  author        = {Beisert, Niklas and Broedel, Johannes and Rosso, Matteo},
  title         = {{On Yangian-invariant regularisation of deformed on-shell diagrams in N=4 super-Yang--Mills theory}},
  year          = {2014},
  archiveprefix = {arXiv},
  eprint        = {1401.7274},
  primaryclass  = {hep-th},
  slaccitation  = {
}

@article{Brandhuber:2008pf,
  author        = {Brandhuber, Andreas and Heslop, Paul and Travaglini, Gabriele},
  title         = {{A note on dual superconformal symmetry of the N=4 super Yang--Mills S-matrix}},
  journal       = {Phys.Rev.},
  year          = {2008},
  volume        = {D78},
  pages         = {125005},
  archiveprefix = {arXiv},
  doi           = {10.1103/PhysRevD.78.125005},
  eprint        = {0807.4097},
  primaryclass  = {hep-th},
  reportnumber  = {QMUL-PH-08-15},
  slaccitation  = {
}

@article{Brink:1982pd,
  author        = {Brink, Lars and Lindgren, Olof and Nilsson, Bengt E.W.},
  title         = {{N=4 Yang--Mills theory on the light cone}},
  journal       = {Nucl.Phys.},
  year          = {1983},
  volume        = {B212},
  pages         = {401},
  doi           = {10.1016/0550-3213(83)90678-8},
  reportnumber  = {GOTEBORG-82-21},
  slaccitation  = {
}

@article{Brink:1982wv,
  author        = {Brink, Lars and Lindgren, Olof and Nilsson, Bengt E.W.},
  title         = {{The ultraviolet finiteness of the N=4 Yang--Mills theory}},
  journal       = {Phys.Lett.},
  year          = {1983},
  volume        = {B123},
  pages         = {323},
  doi           = {10.1016/0370-2693(83)91210-8},
  reportnumber  = {UTTG-1-82},
  slaccitation  = {
}

@article{Brink:1976bc,
  author        = {Brink, Lars and Schwarz, John H. and Scherk, Joel},
  title         = {{Supersymmetric Yang--Mills theories}},
  journal       = {Nucl.Phys.},
  year          = {1977},
  volume        = {B121},
  pages         = {77},
  doi           = {10.1016/0550-3213(77)90328-5},
  reportnumber  = {CALT-68-574},
  slaccitation  = {
}

@article{Britto:2004ap,
  author        = {Britto, Ruth and Cachazo, Freddy and Feng, Bo},
  title         = {{New recursion relations for tree amplitudes of gluons}},
  journal       = {Nucl.Phys.},
  year          = {2005},
  volume        = {B715},
  pages         = {499-522},
  archiveprefix = {arXiv},
  doi           = {10.1016/j.nuclphysb.2005.02.030},
  eprint        = {hep-th/0412308},
  primaryclass  = {hep-th},
  slaccitation  = {
}

@article{Britto:2005fq,
  author        = {Britto, Ruth and Cachazo, Freddy and Feng, Bo and Witten, Edward},
  title         = {{Direct proof of tree-level recursion relation in Yang--Mills theory}},
  journal       = {Phys.Rev.Lett.},
  year          = {2005},
  volume        = {94},
  pages         = {181602},
  archiveprefix = {arXiv},
  doi           = {10.1103/PhysRevLett.94.181602},
  eprint        = {hep-th/0501052},
  primaryclass  = {hep-th},
  slaccitation  = {
}

@article{Chicherin:2013ora,
  author         = {Chicherin, D. and Derkachov, S. and Kirschner, R.},
  title          = {{Yang--Baxter operators and scattering amplitudes in
                       N=4 super-Yang--Mills theory}},
  year           = {2013},
  eprint         = {1309.5748},
  archivePrefix  = {arXiv},
  primaryClass   = {hep-th},
  SLACcitation   = {
}

@article{Chicherin:2013sqa,
  author        = {Chicherin, D. and Kirschner, R.},
  title         = {{Yangian symmetric correlators}},
  journal       = {Nucl.Phys.},
  year          = {2013},
  volume        = {B877},
  pages         = {484-505},
  archiveprefix = {arXiv},
  doi           = {10.1016/j.nuclphysb.2013.10.006},
  eprint        = {1306.0711},
  owner         = {mrosso},
  primaryclass  = {math-ph},
  slaccitation  = {
  timestamp     = {2014.01.24},
  url           = {http://arxiv.org/abs/arXiv:1306.0711}
}

@article{Nandan:2012rk,
  author        = {Nandan, Dhritiman and Wen, Congkao},
  title         = {{Generating All Tree Amplitudes in N=4 SYM by Inverse Soft Limit}},
  journal       = {JHEP},
  year          = {2012},
  volume        = {1208},
  pages         = {040},
  archiveprefix = {arXiv},
  doi           = {10.1007/JHEP08(2012)040},
  eprint        = {1204.4841},
  primaryclass  = {hep-th},
  reportnumber  = {QMUL-PH-12-09},
  slaccitation  = {
}

@article{Dixon:2010ik,
  author         = {Dixon, Lance J. and Henn, Johannes M. and Plefka, Jan and
                    Schuster, Theodor},
  title          = {{All tree-level amplitudes in massless QCD}},
  journal        = {JHEP},
  volume         = {1101},
  pages          = {035},
  doi            = {10.1007/JHEP01(2011)035},
  year           = {2011},
  eprint         = {1010.3991},
  archivePrefix  = {arXiv},
  primaryClass   = {hep-ph},
  reportNumber   = {CERN-PH-TH-2010-230, SLAC-PUB-14278, HU-EP-10-57},
  SLACcitation   = {
}

@article{Drummond:2008vq,
  author        = {Drummond, J.M. and Henn, J. and Korchemsky, G.P. and Sokatchev, E.},
  title         = {{Dual superconformal symmetry of scattering amplitudes in N=4 super-Yang--Mills theory}},
  journal       = {Nucl.Phys.},
  year          = {2010},
  volume        = {B828},
  pages         = {317-374},
  archiveprefix = {arXiv},
  doi           = {10.1016/j.nuclphysb.2009.11.022},
  eprint        = {0807.1095},
  primaryclass  = {hep-th},
  reportnumber  = {LAPTH-1257-08, LPT-ORSAY-08-60},
  slaccitation  = {
}

@article{Drummond:2009fd,
  author        = {Drummond, James M. and Henn, Johannes M. and Plefka, Jan},
  title         = {{Yangian symmetry of scattering amplitudes in N=4 super Yang--Mills theory}},
  journal       = {JHEP},
  year          = {2009},
  volume        = {0905},
  pages         = {046},
  archiveprefix = {arXiv},
  doi           = {10.1088/1126-6708/2009/05/046},
  eprint        = {0902.2987},
  primaryclass  = {hep-th},
  reportnumber  = {HU-EP-09-06, LAPTH-1308-09},
  slaccitation  = {
}

@article{Ferro:2012xw,
  author        = {Ferro, Livia and Lukowski, Tomasz and Meneghelli, Carlo and Plefka, Jan and Staudacher, Matthias},
  title         = {{Harmonic R-matrices for scattering amplitudes and spectral regularization}},
  journal       = {Phys.Rev.Lett.},
  year          = {2013},
  volume        = {110},
  pages         = {121602},
  number        = {12},
  archiveprefix = {arXiv},
  doi           = {10.1103/PhysRevLett.110.121602},
  eprint        = {1212.0850},
  primaryclass  = {hep-th},
  reportnumber  = {HU-EP-12-50, HU-MATHEMATIK:14-2012, DESY-12-228, ZMP-HH-12-26, AEI-2012-198, -AEI-2012-198},
  slaccitation  = {
}

@article{Ferro:2013dga,
  author        = {Ferro, Livia and Lukowski, Tomasz and Meneghelli, Carlo and Plefka, Jan and Staudacher, Matthias},
  title         = {{Spectral parameters for scattering amplitudes in N=4 super Yang--Mills theory}},
  year          = {2013},
  archiveprefix = {arXiv},
  eprint        = {1308.3494},
  primaryclass  = {hep-th},
  reportnumber  = {HU-MATHEMATIK-2013-12, HU-EP-13-33, AEI-2013-235, DESY-13-488, --ZMP-HH-13-15},
  slaccitation  = {
}

@article{Frassek:2013xza,
  author         = {Frassek, Rouven and Kanning, Nils and Ko, Yumi and Staudacher, Matthias},
  title          = {{Bethe Ansatz for Yangian Invariants: Towards Super Yang--Mills Scattering Amplitudes}},
  year           = {2013},
  eprint         = {1312.1693},
  archivePrefix  = {arXiv},
  primaryClass   = {math-ph},
  reportNumber   = {HU-MATHEMATIK-2013-14, HU-EP-13-34, AEI-2013-234, DCPT-13-47},
  SLACcitation   = "
}

@article{Gliozzi:1976qd,
  author        = {Gliozzi, F. and Scherk, Joel and Olive, David I.},
  title         = {{Supersymmetry, supergravity theories and the dual spinor model}},
  journal       = {Nucl.Phys.},
  year          = {1977},
  volume        = {B122},
  pages         = {253-290},
  doi           = {10.1016/0550-3213(77)90206-1},
  reportnumber  = {CERN-TH-2253},
  slaccitation  = {
}

@article{Howe:1983sr,
  author        = {Howe, Paul S. and Stelle, K.S. and Townsend, P.K.},
  title         = {{Miraculous ultraviolet cancellations in supersymmetry made manifest}},
  journal       = {Nucl.Phys.},
  year          = {1984},
  volume        = {B236},
  pages         = {125},
  doi           = {10.1016/0550-3213(84)90528-5},
  reportnumber  = {ICTP-82-83-20},
  slaccitation  = {
}

@article{Mandelstam1982,
  author        = {Mandelstam, S.},
  title         = {{Light cone superspace and the finiteness of the N=4 model}},
  year          = {1982},
  reportnumber  = {CERN-TH-3385},
  slaccitation  = {
}

@article{Nair:1988bq,
  author        = {Nair, V.P.},
  title         = {{A current algebra for some gauge theory amplitudes}},
  journal       = {Phys.Lett.},
  year          = {1988},
  volume        = {B214},
  pages         = {215},
  doi           = {10.1016/0370-2693(88)91471-2},
  reportnumber  = {CU-TP-408},
  slaccitation  = {
}

@article{Witten:2003nn,
  author        = {Witten, Edward},
  title         = {{Perturbative gauge theory as a string theory in twistor space}},
  journal       = {Commun.Math.Phys.},
  year          = {2004},
  volume        = {252},
  pages         = {189-258},
  archiveprefix = {arXiv},
  doi           = {10.1007/s00220-004-1187-3},
  eprint        = {hep-th/0312171},
  primaryclass  = {hep-th},
  slaccitation  = {
}

@article{Postnikov:2006kva,
  author        = {Postnikov, Alexander},
  title         = {{Total positivity, Grassmannians, and networks}},
  year          = {2006},
  archiveprefix = {arXiv},
  eprint        = {math/0609764},
  primaryclass  = {math.CO},
  slaccitation  = {
}

@article{Drinfeld:1985rx,
  author         = "Drinfeld, V. G.",
  title          = "Hopf algebras and the quantum Yang--Baxter equation",
  journal        = "Sov. Math. Dokl.",
  volume         = "32",
  year           = "1985",
  pages          = "254-258",
  SLACcitation   = "
}

@article{Drinfeld:1986in,
  author         = "Drinfeld, V. G.",
  title          = "Quantum groups",
  journal        = "J. Math. Sci.",
  volume         = "41",
  year           = "1988",
  pages          = "898",
  doi            = "10.1007/BF01247086",
  SLACcitation   = "
}

@article{Takhtajan:1979iv,
  author        = "Takhtajan, L. A. and Faddeev, L. D.",
  title         = "{The quantum method of the inverse problem and the
                    Heisenberg XYZ model}",
  journal       = "Russ.Math.Surveys",
  volume        = "34",
  pages         = "11-68",
  year          = "1979",
  SLACcitation  = "
}

@article{Kulish:1980ii,
  author        = "Kulish, P. P. and Sklyanin, E. K.",
  title         = "{On the solution of the Yang--Baxter equation}",
  journal       = "J.Sov.Math.",
  volume        = "19",
  pages         = "1596-1620",
  doi           = "10.1007/BF01091463",
  year          = "1982",
  SLACcitation  = "
}
@article{Faddeev:1987ih,
  author        = "Faddeev, L. D. and Reshetikhin, N. {\relax Yu}. and Takhtajan,
                    L.A.",
  title         = "{Quantization of Lie groups and Lie algebras}",
  journal       = "Leningrad Math.J.",
  volume        = "1",
  pages         = "193-225",
  year          = "1990",
  reportNumber  = "LOMI-E-14-87",
  SLACcitation  = "
}

@article{Khoroshkin:1996fy,
  author        = "Khoroshkin, S. and Lebedev, D. and Pakuliak, S.",
  title         = "{Intertwining operators for the central extension of the
                    Yangian double}",
  year          = "1996",
  eprint        = "q-alg/9602030",
  archivePrefix = "arXiv",
  primaryClass  = "q-alg",
  reportNumber  = "DFTUZ-95-28, ITEP-TH-15-95",
  SLACcitation  = "
}

@article{Khoroshkin:1996fz,
  author        = "Khoroshkin, S.M.",
  title         = "{Central extension of the Yangian double}",
  year          = "1996",
  eprint        = "q-alg/9602031",
  archivePrefix = "arXiv",
  primaryClass  = "q-alg",
  SLACcitation  = "
}

@article{Bern:2004ky,
  author         = "Bern, Zvi and Del Duca, Vittorio and Dixon, Lance J. and
                    Kosower, David A.",
  title          = "{All non-maximally-helicity-violating one-loop
                    seven-gluon amplitudes in N=4 super-yang-Mills theory}",
  journal        = "Phys.Rev.",
  volume         = "D71",
  pages          = "045006",
  doi            = "10.1103/PhysRevD.71.045006",
  year           = "2005",
  eprint         = "hep-th/0410224",
  archivePrefix  = "arXiv",
  primaryClass   = "hep-th",
  reportNumber   = "SLAC-PUB-10810, UCLA-04-TEP-43, DFTT-26-2004,
                    DCPT-04-136, IPPP-04-68, SACLAY-SPHT-T04-131,
                    NSF-KITP-04-114",
  SLACcitation   = "
}

@article{MacKay:2004tc,
  author         = "MacKay, N.J.",
  title          = "{Introduction to Yangian symmetry in integrable field
                    theory}",
  journal        = "Int.J.Mod.Phys.",
  volume         = "A20",
  pages          = "7189-7218",
  doi            = "10.1142/S0217751X05022317",
  year           = "2005",
  eprint         = "hep-th/0409183",
  archivePrefix  = "arXiv",
  primaryClass   = "hep-th",
  reportNumber   = "ESI-1514",
  SLACcitation   = "
}

@article{Khoroshkin:1994uk,
  author         = "Khoroshkin, S.M. and Tolstoi, V.N.",
  title          = "{Yangian double and rational R matrix}",
  year           = "1994",
  eprint         = "hep-th/9406194",
  archivePrefix  = "arXiv",
  primaryClass   = "hep-th",
  SLACcitation   = "
}

@article{Beisert:2014hya,
  author         = "Beisert, Niklas and de Leeuw, Marius",
  title          = "{The RTT-Realization for the Deformed $\mathfrak{gl}(2|2)$ Yangian}",
  year           = "2014",
  eprint         = "1401.7691",
  archivePrefix  = "arXiv",
  primaryClass   = "math-ph",
  SLACcitation   = "
}

@article{Beisert:2011pn,
  author         = "Beisert, Niklas and Schwab, Burkhard U.W.",
  title          = "{Bonus Yangian Symmetry for the Planar S-Matrix of N=4
                    Super Yang--Mills}",
  journal        = "Phys.Rev.Lett.",
  volume         = "106",
  pages          = "231602",
  doi            = "10.1103/PhysRevLett.106.231602",
  year           = "2011",
  eprint         = "1103.0646",
  archivePrefix  = "arXiv",
  primaryClass   = "hep-th",
  reportNumber   = "AEI-2011-006",
  SLACcitation   = "
}

@article{Mason:2009qx,
  author         = "Mason, L.J. and Skinner, David",
  title          = "{Dual Superconformal Invariance, Momentum Twistors and
                    Grassmannians}",
  journal        = "JHEP",
  volume         = "0911",
  pages          = "045",
  doi            = "10.1088/1126-6708/2009/11/045",
  year           = "2009",
  eprint         = "0909.0250",
  archivePrefix  = "arXiv",
  primaryClass   = "hep-th",
  SLACcitation   = "
}

@article{Bullimore2010,
  author         = "Bullimore, Mathew and Mason, L.J. and Skinner, David",
  title          = "{MHV Diagrams in Momentum Twistor Space}",
  journal        = "JHEP",
  volume         = "1012",
  pages          = "032",
  doi            = "10.1007/JHEP12(2010)032",
  year           = "2010",
  eprint         = "1009.1854",
  archivePrefix  = "arXiv",
  primaryClass   = "hep-th",
  SLACcitation   = "
}

\end{bibtex}

\bibliographystyle{nb}
\bibliography{\jobname}

\end{document}